\documentclass[fleqn,usenatbib]{mnras}

\usepackage{newtxtext,newtxmath}

\usepackage[T1]{fontenc}

\DeclareRobustCommand{\VAN}[3]{#2}
\let\VANthebibliography\thebibliography
\def\thebibliography{\DeclareRobustCommand{\VAN}[3]{##3}\VANthebibliography}

\usepackage{graphicx}	
\usepackage{amsmath}	
\usepackage{subfigure}
\usepackage{natbib}
\usepackage{xcolor}
\usepackage{tabularx}
\usepackage{longtable}
\usepackage{bm}
\usepackage{threeparttable}
\usepackage[normalem]{ulem}
\usepackage{verbatim}

\newcommand{\code}[1]{{\texttt{#1}}}
\newcommand{\tess}{{\it TESS}}
\newcommand{\kepler}{{\it Kepler}}

\newcommand{\gaia}{{\it Gaia}}
\newcommand{\wise}{{\it WISE}}
\newcommand{\jwst}{{\it JWST}}

\newcommand{\tar}{{TOI-2136}}

\title[TOI-2136\,b]{TESS discovery of a sub-Neptune orbiting a mid-M dwarf TOI-2136}

\author[T. Gan et al.]{Tianjun Gan,$^{1}$\thanks{E-mail: gtj18@mails.tsinghua.edu.cn}
Abderahmane Soubkiou,$^{2,3,4}$
Sharon X. Wang,$^{1}$
Zouhair Benkhaldoun,$^{2}$
Shude Mao,$^{1,5}$
\newauthor
\'Etienne Artigau,$^{6,7}$ 
Pascal Fouqu\'e,$^{8,9}$
Steven Giacalone,$^{10}$
Christopher A. Theissen,\thanks{NASA Sagan Fellow}$^{11}$
Christian Aganze,$^{11}$
\newauthor
Karen A.\ Collins,$^{12}$
Avi Shporer,$^{13}$
Khalid Barkaoui,$^{14,15,16}$
Mourad Ghachoui,$^{2,14}$
Steve~B.~Howell,$^{17}$
\newauthor
Claire Lamman,$^{12}$
Olivier D.\ S. Demangeon,$^{3,4}$
Artem Burdanov,$^{15}$
Charles Cadieux,$^{6}$
Jamila Chouqar,$^{2}$
\newauthor
Kevin I. Collins,$^{18}$
Neil J. Cook,$^{6}$
Laetitia Delrez,$^{14,19}$
Brice-Olivier Demory,$^{20}$
Ren\'e Doyon,$^{6,7}$
\newauthor
Georgina Dransfield,$^{21}$
Courtney D. Dressing,$^{10}$
Elsa Ducrot,$^{14,22}$
Jiahao Fan,$^{23}$
Lionel Garcia,$^{14}$
\newauthor
Holden Gill,$^{10}$
Micha{\"e}l Gillon,$^{14}$
Crystal~L.~Gnilka,$^{17,24}$
Yilen G\'omez Maqueo Chew,$^{25}$
\newauthor
Maximilian N. G{\"u}nther,\thanks{ESA Research Fellow}$^{26}$
Christopher E. Henze,$^{17}$
Chelsea X. Huang,$^{13,27}$
Emmanuel Jehin,$^{19}$
\newauthor
Eric L.\ N.\ Jensen,$^{28}$
Zitao Lin,$^{29}$
James McCormac,$^{30}$
Catriona A. Murray,$^{31}$
Prajwal Niraula,$^{15}$
\newauthor
Peter P. Pedersen,$^{31}$
Francisco J. Pozuelos,$^{14,19}$
Didier Queloz,$^{31,32}$
Benjamin V. Rackham,\thanks{51 Pegasi b Fellow}$^{15}$
Arjun B. Savel,$^{33}$
\newauthor
Nicole Schanche,$^{20}$
Richard P. Schwarz,$^{34}$
Daniel Sebastian,$^{21}$
Samantha Thompson,$^{31}$
Mathilde Timmermans,$^{14}$
\newauthor
Amaury H. M. J. Triaud,$^{21}$
Michael Vezie,$^{13}$
Robert D. Wells,$^{20}$
Julien de Wit,$^{15}$
George~R.~Ricker,$^{13}$
\newauthor
Roland~Vanderspek,$^{13}$
David~W.~Latham,$^{12}$
Sara~Seager,$^{13,15,35}$
Joshua~N.~Winn,$^{36}$
and Jon~M.~Jenkins$^{17}$
\\
Affiliations are listed at the end of the paper
}

\date{Accepted XXX. Received YYY; in original form ZZZ}

\pubyear{2022}

\begin{document}
\label{firstpage}
\pagerange{\pageref{firstpage}--\pageref{lastpage}}
\maketitle

\begin{abstract}
We present the discovery of \hbox{TOI-2136\,b}, a sub-Neptune planet transiting every 7.85 days a nearby M4.5V-type star, identified through photometric measurements from the TESS mission. The host star is located $33$ pc away with a radius of $R_{\ast} = 0.34\pm0.02\ R_{\odot}$, a mass of $0.34\pm0.02 M_{\odot}$ and an effective temperature of $3342\pm100$ K. We estimate its stellar rotation period to be $75\pm5$ days based on archival long-term photometry. We confirm and characterize the planet based on a series of ground-based multi-wavelength photometry, high-angular-resolution imaging observations, and precise radial velocities from CFHT/SPIRou. Our joint analysis reveals that the planet has a radius of $2.19\pm0.17\ R_{\oplus}$, and a mass measurement of $6.4\pm2.4\ M_{\oplus}$. The mass and radius of \hbox{TOI-2136\,b} is consistent with a broad range of compositions, from water-ice to gas-dominated worlds. \hbox{TOI-2136\,b} falls close to the radius valley for low-mass stars predicted by the thermally driven atmospheric mass loss models, making it an interesting target for future studies of its interior structure and atmospheric properties.
\end{abstract}




\section{Introduction}

The \kepler\ mission enabled the discovery of thousands of transiting exoplanets \citep{Borucki2010}, which began a new chapter in exoplanet research. One of the most important findings of \kepler\ is that super-Earths and sub-Neptunes ($1\ R_{\oplus}<R_{p}<4\ R_{\oplus}$) are abundant in close-in orbits around other stars \citep{Howard2012,Fressin2013,Petigura2013}, whereas our Solar System has no such planets. Later demographic studies based on a well-characterized sample with refined stellar properties, as part of the California-\kepler\ Survey \citep[CKS;][]{Petigura2017,Johnson2017}, revealed that the radius distribution of small planets has a bimodal profile with a valley centered at around $1.8\ R_{\oplus}$ \citep{Fulton2017,Fulton2018gap}. In particular, \cite{Van_Eylen2018} and \cite{Martinez2019} looked into the radius distribution of small planets around stars with spectral types F, G, or K in a multi-dimensional parameter space. Both of them reached the same conclusion that the location of the radius gap depends on the planet orbital period, and modeled it as a power-law function. This relation is consistent with the predictions from theoretical models on photoevaporation \citep{Owen2013,Lopez2014,Jin2014,Chen2016,Owen2017}, which proposed that the $\rm H/He$ gaseous envelopes of small planets would be stripped away by high energy stellar radiation such as X-rays during the first few Myrs of the evolution when the host stars are still active \citep{Lopez2018}. A similar trend can also be sculpted according to the core-powered mass-loss theory \citep{Ginzburg2018,Lopez2018,Gupta2019,Gupta2020,Gupta2021}. Under this hypothesis, the luminosity of cooling planetary cores offers the energy for atmosphere escape, and causes the planetary radius to shrink. 

However, the transition radius between super-Earths and sub-Neptunes around M dwarfs tends to behave differently compared with Sun-like hosts. \cite{Cloutier2020} investigated the radius valley of small planets around low-mass stars based on a composite sample from \kepler\ and {\it K2} \citep{Howell2014}, and they found that the slope of the valley likely follows a power law relation with planet orbital period but with an index of the opposite sign, compared to the trend for Sun-like stars. Though this feature is in possible disagreement with the aforementioned thermally-driven mass-loss models, it conforms to the gas-poor formation scenario \citep{Lee2014,Lee2016}, suggesting that the radius gap is a result of the superposition of two distinct populations with the rocky group forming at late times when the protoplanetary disk had mostly dissipated. A straightforward way to distinguish the dominant mechanism that results in the transition radius at the low stellar mass end is to examine the bulk compositions of small planets around low mass stars \citep{Cloutier2020}. Nevertheless, only a few small planets around M dwarfs have been confirmed with both precise radius and mass determination so far \citep[e.g.,][]{Charbonneau2009,Ment2019,Agol2021}. 


Fortunately, the Transiting Exoplanet Survey Satellite (\tess, \citealt{Ricker2015}) is performing an all-sky survey and targets bright nearby stars, providing an exciting opportunity to discover small transiting planets around M dwarfs. The \tess\ Primary Mission has already yielded the detections of several such systems \citep[e.g.,][]{Vanderspek2019,Gan2020,Wells2021,Fukui2021}. Some of those planets also have precise mass constraints through spectroscopic measurements thanks to the brightness of their host stars \citep[e.g.,][]{Luque2019,Shporer2020,Cloutier2020_1235b,Soto2021}. However, 
it is challenging to achieve a high enough signal-to-noise ratio (SNR) and obtain precise radial velocities for mid-to-late M dwarfs as they are, in general, faint at optical wavelengths. The new-generation near-infrared spectrograph SpectroPolarim\`etre InfraROUge spectrograph (SPIRou) on the Canada-France-Hawaii-Telescope (CFHT) opens a window to characterize planets around faint stars via high-precision velocimetry and spectropolarimetry \citep{Donati2020,Klein2021,Gantoi530}.

Here we report the discovery and follow-up observations of a transiting sub-Neptune around the nearby M4.5V dwarf, \tar. We present RV measurements from SPIRou along with a series of additional time-series observations including ground-based photometry and high resolution images that allow us to confirm that the \tess\ signal is due to a transiting planet. The rest of the paper is organized as follows. In Section \ref{observations}, we detail all space and ground-based observational data used in this work. Section \ref{stellar_properties} provides the stellar characterization. We present our analysis of light curves as well as the RVs in Section \ref{analysis} before we discuss the properties and the prospects of future atmospheric characterization of \hbox{TOI-2136\,b} in Section \ref{discussion}. A summary of our findings is given in Section \ref{conclusion}.

\section{Observations}\label{observations}
\subsection{\tess\ photometry}\label{tess_data}

\begin{figure*}
\centering
\includegraphics[width=0.31\textwidth]{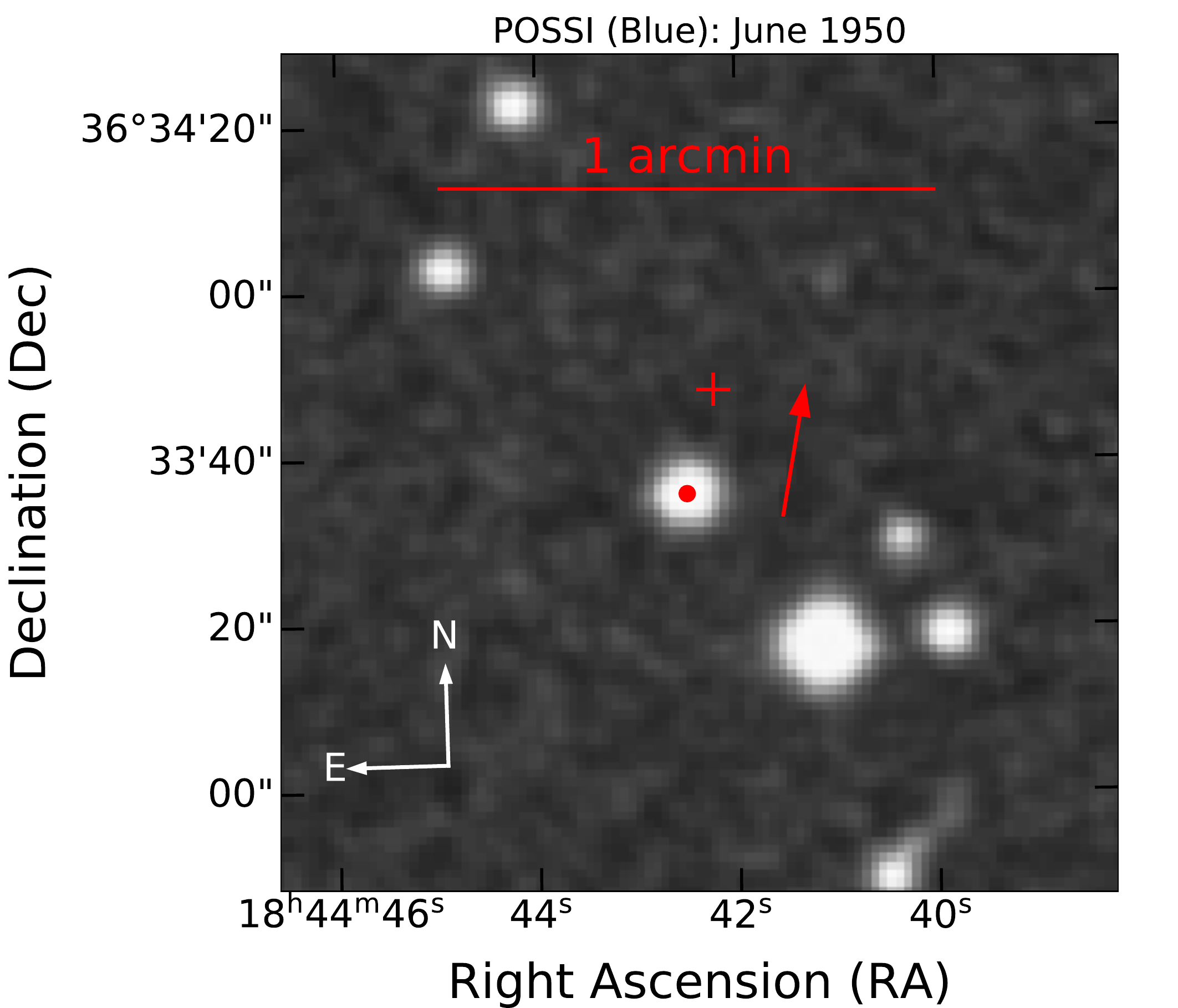}
\includegraphics[width=0.33\textwidth]{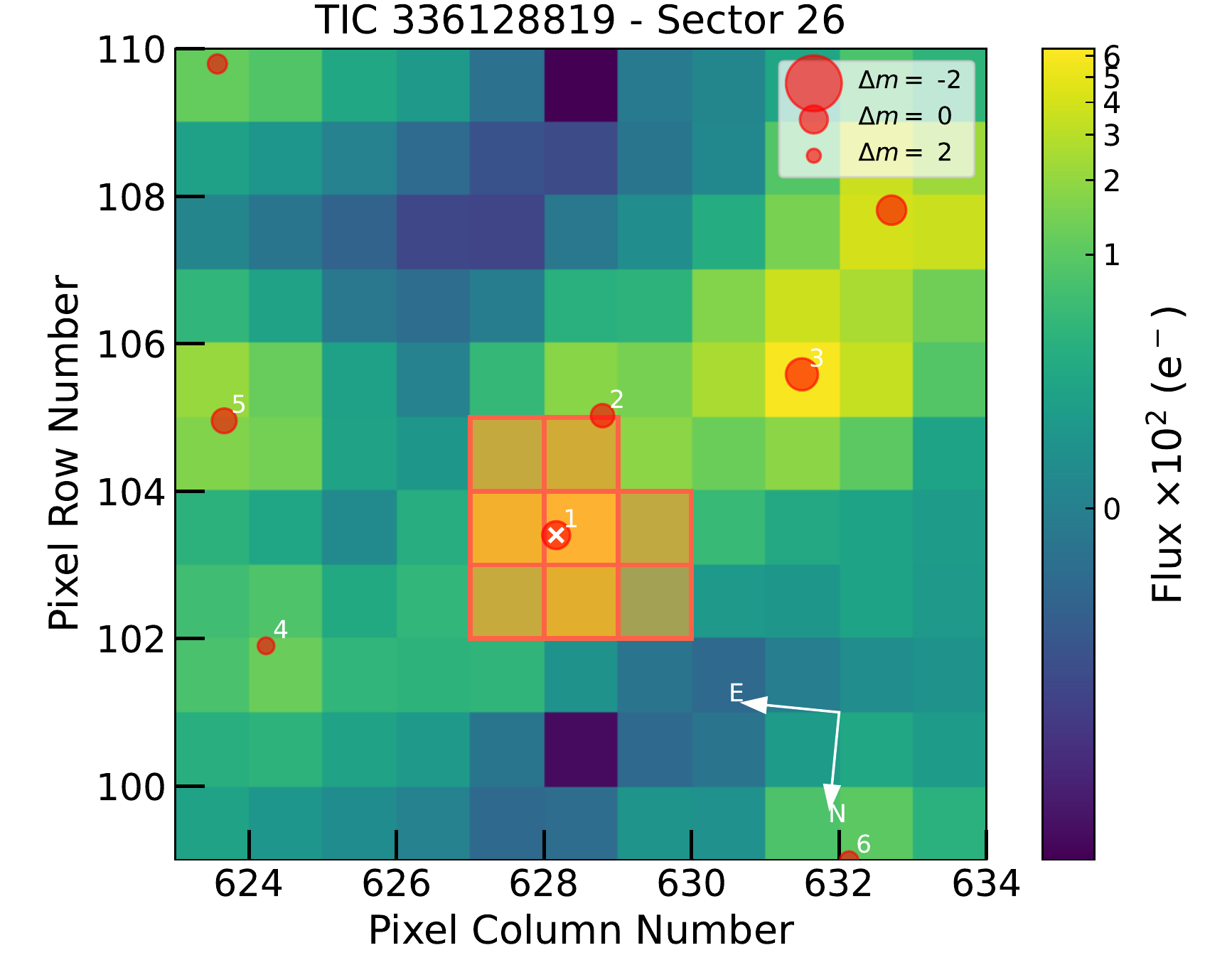}
\includegraphics[width=0.33\textwidth]{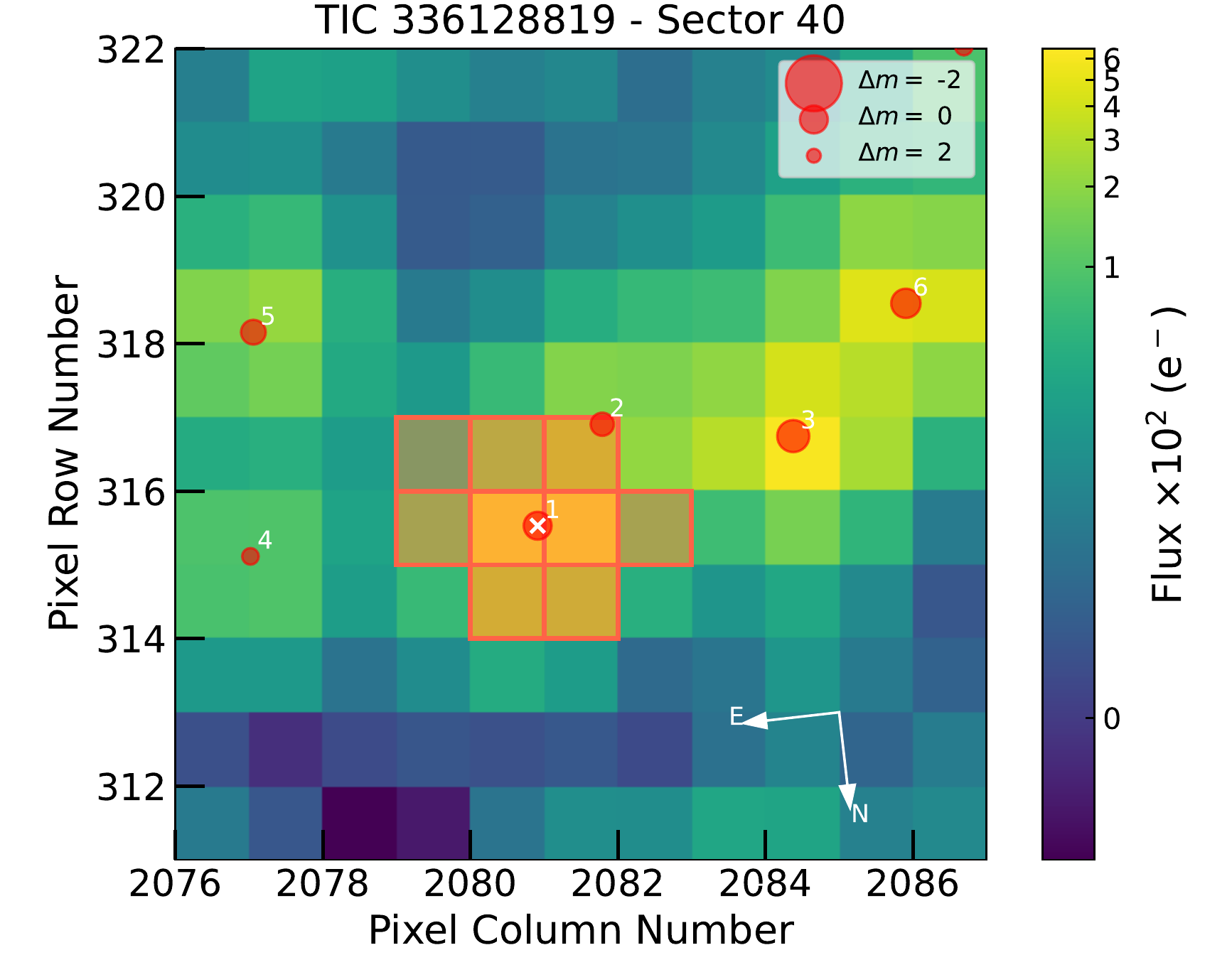}
\caption{{\it Left panel:} The POSSI blue image of \tar\ taken in 1950. The central red dot marks the position of \tar\ in this image while the red cross represents its current location. Red arrow indicates the direction of proper motion. {\it Middle and right panels:} Target pixel files (TPF) of \tar\ in \tess\ Sector 26 and 40, created with \code{tpfplotter}. The orange shaded region represents the aperture used to extract the photometry. The red circles are the \gaia\ DR2 sources. Different sizes represent different magnitudes in contrast with \tar. } 
\label{fov}
\end{figure*}

\tar\ (TIC 336128819) was first observed by \tess\ on its Camera 1 with the two-minute cadence mode in Sector 26 during the primary mission from 9th June 2020 to 4th July 2020 and it was re-observed in Sector 40 between 24$^{\rm th}$ June 2021 and 23$^{\rm th}$ July 2021 during the Extended Mission. The left panel of Figure \ref{fov} shows the POSSI image of \tar\ taken in 1950. Based on the relatively large stellar proper motion ($\sim180$ mas/yr), we rule out the possibility that the light from an unassociated distant eclipsing binary system with $V\lesssim21$ mag caused the \tess\ detection. The other panels of Figure \ref{fov} show the target pixel files (TPFs) and the Simple Aperture Photometry (SAP) apertures used in each sector
, plotted with \code{tpfplotter} \citep{Aller2020}. A nearby star (\gaia\ DR2 2096535788163295744, $T_{\rm mag}=13.23$) 33$''$ away is located at the edge of the aperture, which is expected to make only a slight contribution to the \tess\ signal. We summarize the host star properties in Table \ref{starparam}.

The \tess\ time-series data were initially processed by the Science Processing Operations Center (SPOC; \citealt{Jenkins2016}) pipeline. After correcting the instrumental and systematic effects as well as the light dilution with the Presearch Data Conditioning (PDC; \citealt{Stumpe2012,Smith2012,Stumpe2014}) module, transit signals were searched using the Transiting Planet Search \citep[TPS;][]{Jenkins2002,Jenkins2020} algorithm, which resulted in a periodic signal with an orbital period of 7.85 days and a duration of 1.61 hours. Validation tests were then conducted to confirm the transit signature \citep{Twicken2018,Li2019}, including locating the source of the transit signal to within \hbox{$1 - 3^{\prime\prime}$}  of the target star, and searching for additional transiting planet signatures in the residual light curve before \tar\ was finally alerted as a planet candidate in the TESS Object of Interest catalog (TOI-2136.01).

\begin{figure*}
\centering
\includegraphics[width=\textwidth]{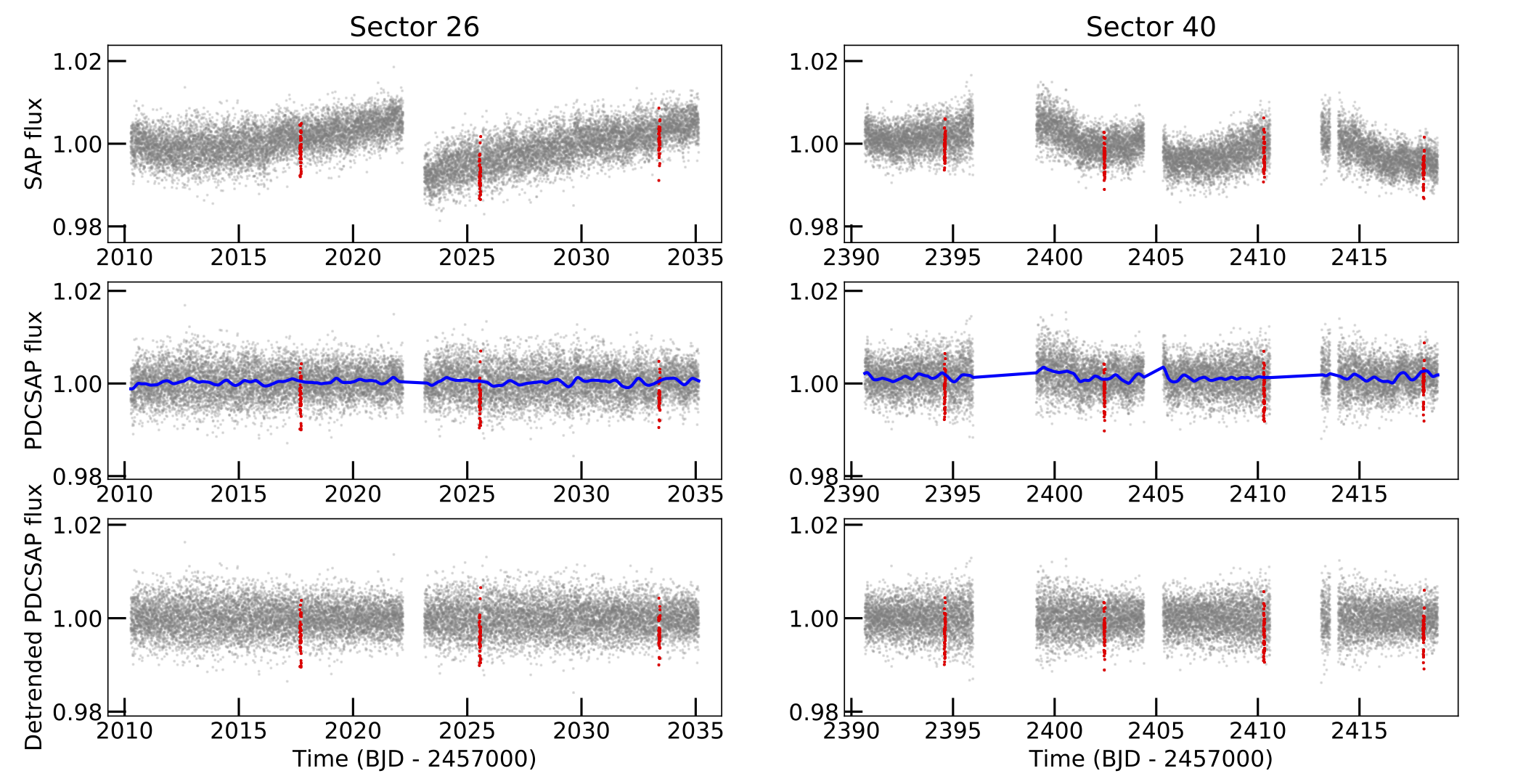}
\caption{\tess\ light curves of \tar\ from Sector 26 and 40. {\it Top panels:} The \tess\ simple aperture photometry light curves. {\it Middle panels:} The \tess\ raw PDCSAP light curves after correcting the systematic and instrumental errors. The blue curves represent the best-fit GP models used to remove the correlated noise existing in the PDCSAP light curves (Section \ref{tess_data}). {\it Bottom panels:} The final detrended \tess\ PDCSAP light curves. The red dots highlight each transit of \tar\,b.} 
\label{TESS_photometry}
\end{figure*}

We retrieved the Presearch Data Conditioning Simple Aperture Photometry (PDCSAP) light curve from the Mikulski Archive for Space Telescopes\footnote{\url{http://archive.stsci.edu/tess/}}\citep{Twicken2010,Morris2020}. We found a total of 16941 and 15319 useful measurements within the data from Sector 26 and Sector 40, respectively. We then performed our own transit search by utilizing the Transit Least Squares (TLS; \citealt{Hippke2019}) algorithm, which is an advanced version of Box Least Square (BLS; \citealt{Kovacs2002}). We confirmed the 7.85\,days signal with a signal detection efficiency (SDE) of 34 but we did not find additional significant signals existing in the light curve. To detrend the \tess\ light curve and remove the systematic trends left in the PDCSAP light curve, we fit a Gaussian Process (GP) model with a Mat\'{e}rn-3/2 kernel using the \code{celerite} package \citep{Foreman2017}, after masking out all in-transit data. We show the SAP, raw PDCSAP and detrended PDCSAP light curves in Figure \ref{TESS_photometry}. 

\begin{table}
    \caption{Basic information of \tar}
    \begin{tabular}{lll}
        \hline\hline
        Parameter       &Value       \\\hline
        \it{Main identifiers}                    \\
         TOI                     &$2136$         \\
         TIC                     &$336128819$   \\
         \gaia\ ID            &$2096535783864546944$ \\
         \it{Equatorial Coordinates} \\
         $\rm R.A.\ (J2015.5)$                    &18:44:42.32 \\
         $\rm DEC.\ (J2015.5)$                    &36:33:47.27    \\
         \it{Photometric properties}\\
         $\tess$\ (mag)           &$11.737\pm0.007$   &$\rm TIC\ V8^{[1]}$     \\
         $\gaia$\ (mag)           &$12.946\pm0.011$   &\gaia\ EDR3$^{[2]}$   \\
         \gaia\ BP\ (mag)           &$14.367\pm0.010$   &\gaia\ EDR3   \\
         \gaia\ RP\ (mag)           &$11.780\pm0.012$   &\gaia\ EDR3   \\
         $J$\ (mag)                    &$10.184\pm0.024$   &2MASS$^{[3]}$\\
         $H$\ (mag)                    &$9.604\pm0.028$   &2MASS \\
         $K$\ (mag)                    &$9.343\pm0.022$    &2MASS \\
         \wise1 (mag)                   &$9.194\pm0.022$   &\wise$^{[4]}$ \\
         \wise2 (mag)                   &$9.050\pm0.021$   &\wise \\
         \wise3 (mag)                   &$8.924\pm0.027$   &\wise \\
         \wise4 (mag)                   &$8.763\pm0.328$   &\wise \\
         \it{Astrometric properties}\\
         $\varpi$ (mas)              &$29.976\pm0.017$  &\gaia\ EDR3  \\
         $\mu_{\rm \alpha}\ ({\rm mas~yr^{-1}})$     &$-33.81\pm0.02$   &\gaia\ EDR3   \\
         $\mu_{\rm \delta}\ ({\rm mas~yr^{-1}})$     &$177.05\pm0.02$   &\gaia\ EDR3  \\
         RV\ (km~s$^{-1}$)                          &$-28.8\pm6.0$ &This work  \\
         \it{Derived parameters} \\
         Distance (pc)                &$33.36\pm0.02$  &This work     \\
         $U_{\rm LSR}$ (km~s$^{-1}$)       &$-25.15\pm2.26$     &This work\\
         $V_{\rm LSR}$ (km~s$^{-1}$)       &$-9.42\pm5.27$     &This work\\
         $W_{\rm LSR}$ (km~s$^{-1}$)       &$13.16\pm1.75$     &This work\\
         $M_{\ast}\ (M_{\odot})$ &$0.34\pm 0.02$ &This work       \\
         $R_{\ast}\ (R_{\odot})$ &$0.34\pm 0.02$ &This work       \\
         $\rho_\ast\ ({\rm g~cm^{-3}})$ &$12.20\pm 2.53$ &This work \\
         $\log g_{\ast}\ ({\rm cgs})$       &$4.91\pm 0.03$  &This work        \\
         $L_{\ast}\ (L_{\odot})$ &$0.013\pm0.003$  &This work    \\
         $T_{\rm eff}\ ({\rm K})$           &$3342\pm 100$  &This work       \\
         $\rm [Fe/H]$  &$0.03\pm 0.07$ &This work \\
         $\rm [M/H]$  &$-0.01\pm0.08$ &This work\\
         $P_{\rm rot}$\ (days) &$75\pm5$ &This work\\
         $\rm Age$ (Gyr) &$4.6\pm1.0$ &This work\\
         \hline\hline 
    \end{tabular}
    \begin{tablenotes}
    \item[1]  [1]\ \cite{Stassun2017tic,Stassun2019tic}, [2]\ \cite{GaiaEDR3}, 
    \item[2]  [3]\ \cite{Cutri:2003}, [4]\ \cite{Wright:2010}.

    \end{tablenotes}
    \label{starparam}
\end{table}

\subsection{Ground-Based photometry}\label{gbp}
Due to the large pixel scale of \tess\ (21$''$/ pixel, \citealt{Ricker2015}), the host star is likely to be blended with close stars in a single \tess\ pixel. Consequently, the transit signal of \tar\ detected in the space data could be caused by nearby eclipsing binaries. Even though the transit signal is on target, the depth might be biased to a smaller value because of light contamination. With all of the above in mind, we collected a series of ground-based observations of \tar, as part of the TESS Follow-up Observing Program (TFOP\footnote{\url{https://tess.mit.edu/followup}}), to validate the planetary nature and refine both the transit ephemeris and the radius measurement. We scheduled these photometric time-series by using the \tess\ Transit Finder (\code{TTF}) tool, which is a customized version of the \code{Tapir} software package \citep{Jensen2013}. We summarize the details in Table \ref{po} and describe individual observations below. We show the raw and detrended ground-based light curves in Figure \ref{ground_transit_detrend} (see Section \ref{ground_based_photometry}).

\subsubsection{TRAPPIST-North}\label{TN}
A total of three full transits of \hbox{TOI-2136\,b} were acquired by the 60-cm robotic TRAPPIST-North telescope on 12$^{\rm th}$ May 2021, 28$^{\rm th}$ June 2021 and 6$^{\rm th}$ July 2021. TRAPPIST-North is located at Oukaimeden Observatory in Morocco \citep{Jehin:2011,Gillon:2011,Barkaoui:2019}, which has an f/8 Ritchey-Chr\'etien optical design. It is equipped with a thermoelectrically cooled $2K\times2K$ Andor iKon-L BEX2-DD CCD camera with a pixel scale of $0.60''$ pixel$^{-1}$, resulting in a field of view of $20\arcmin\times20\arcmin$. Due to the faintness of the host star, all of the three observations were carried out in the Sloan-$z'$ filter with an exposure time of 20\,s. We took a total of 441, 548 and 334 raw images during the three visits. Data calibration and photometric measurements were performed using a custom pipeline,  \code{PROSE}\footnote{\url{https://github.com/lgrcia/prose}}, which is detailed in \cite{garcia2021}. In all observations, the transit signal is detected on target.


\subsubsection{LCOGT}\label{LCO}
We obtained two ground-based follow-up observations using the 1.0-m telescopes at Cerro Tololo Interamerican Observatory\ (CTIO), one of the southern hemisphere sites of the Las Cumbres Observatory Global Telescope\ (LCOGT\footnote{\url{https://lco.global/}}) network \citep{Brown2013}. The photometric observations were acquired in the Pan-STARRS $z$-short band ($z_{s}$) with an exposure time of 80 s on 21$^{\rm th}$ June 2021 and 22$^{\rm th}$ August 2021, and both were done with the Sinistro cameras, which have a $26' \times 26'$ field of view as well as a plate scale of $\rm 0.389''$ per pixel. The images were focused and have stellar point-spread-functions (PSF) with a full-width-half-maximum (FWHM) of $\sim 2.0\arcsec$ and $\sim 3.1\arcsec$, respectively. The raw images were first calibrated by the LCOGT standard automatic \code{BANZAI} pipeline \citep{McCully2018}. We then carried out photometric analysis using the \code{AstroImageJ} (AIJ) package \citep{Collins2017} to extract the target light curve with uncontaminated apertures of 11 and 15 pixels ($4.3\arcsec$ and $5.8\arcsec$), and examine all nearby stars within $2.5\arcmin$ to look for the sources that may caused the \tess\ signal at the periods of the planet candidate (see Figure \ref{fov}). We confirmed the transit signal on target and ruled out the nearby eclipsing binary scenario.

\subsubsection{SPECULOOS-North}

We observed a full transit of \hbox{TOI-2136\,b} with the 1.0-m SPECULOOS-North/Artemis telescope on 24$^{\rm th}$ October 2021. Artemis telescope is a Ritchey-Chr\'etien telescope equipped with a thermoelectrically cooled $2K\times2K$ Andor iKon-L BEX2-DD CCD camera with a pixel scale of 0.35\,arcsec\,pixel$^{-1}$ and a field of view of $12\arcmin\times12\arcmin$. It is a twin of the four SPECULOOS-South telescopes located at the Paranal observatory \citep{Delrez2018,Sebastian_2021AA}, optimized for detecting planetary transits around cool stars \citep[e.g.,][]{Niraula2020}. The observations were done in the Sloan-$z'$  filter in order to improve the transit SNR. The observation consisted of 514 raw images with an exposure time of 16~seconds, covering 137 minutes total. Data reduction and photometric measurements were performed using the \code{PROSE} pipeline \citep{garcia2021} with an uncontaminated aperture of 8~pixels ($2.8\arcsec$).

\begin{table*}
    \centering
    \caption{Ground-based photometric follow-up observations for \tar}
    \begin{tabular}{ccccccc}
        \hline\hline
        Telescope &Pixel Scale (arcsec) &Date (UT) &Filters &Aperture Size (pixel) &PSF FWHM (arcsec) &$\#$ of exposures  \\\hline
        Trappist-North-0.6m &0.60 &2021 May 12 &$z'$ &7.4 &1.5 &441 \\
        & &2021 Jun. 28 &$z'$ &9.2 &1.5 &548 \\
        & &2021 Jul. 6 &$z'$ &10.1 &1.4 &334 \\\hline
        LCO-CTIO-1m &0.39 &2021 Jun. 21 &$z_{s}$ &11.0 &2.1 &95  \\
        & &2021 Aug. 22 &$z_{s}$ &15.0 &3.1 &93 \\\hline
        SPECULOOS-North-1m &0.35 &2021 Oct. 24 &$z'$ &8.0 &1.3 &514\\
        \hline
    \end{tabular}
    \label{po}
\end{table*}

\begin{figure*}
\centering
\includegraphics[width=0.9\textwidth]{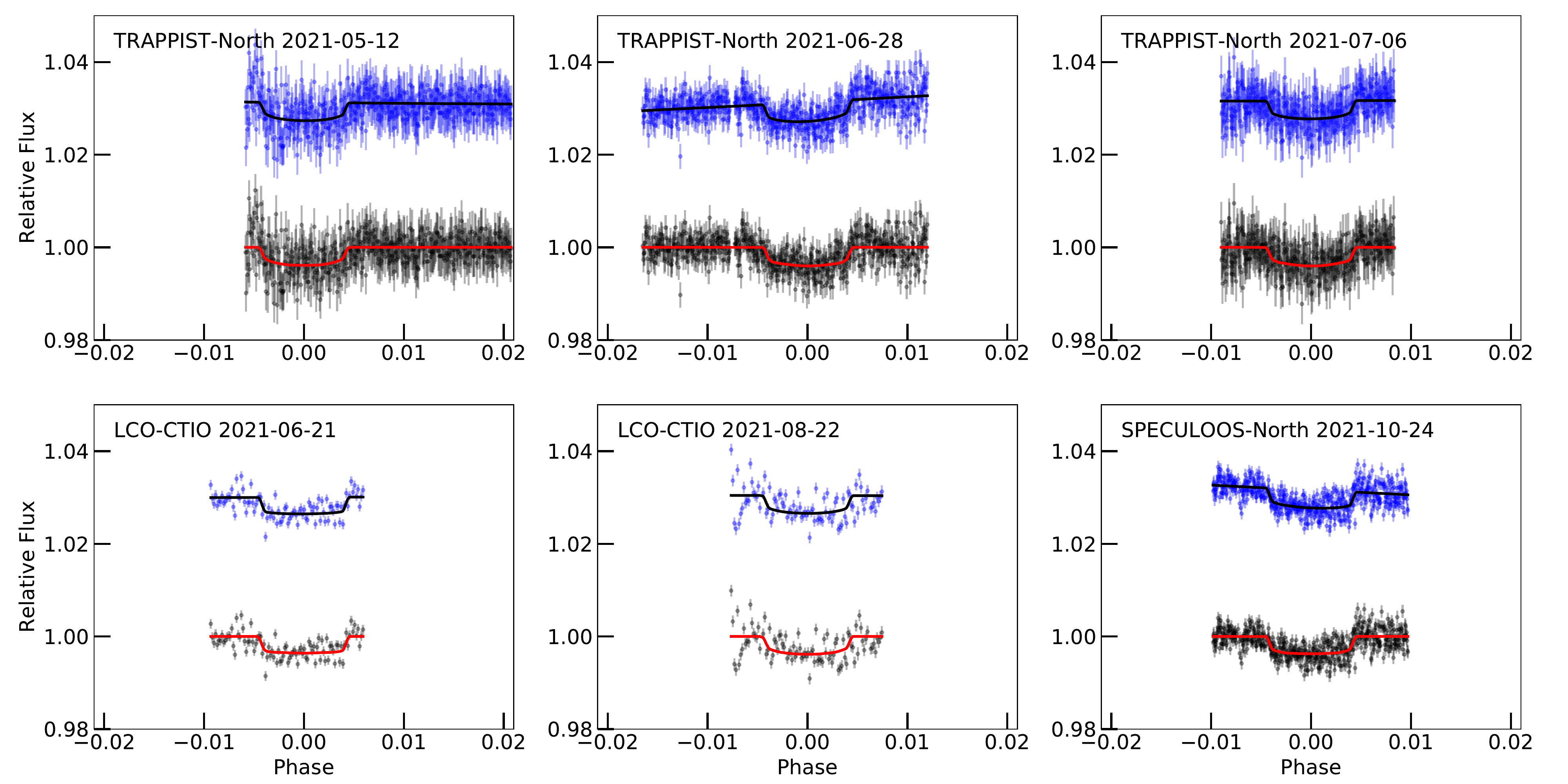}
\caption{Ground-based light curves of \tar. The blue dots are the raw data. The black solid curve represents our best-fit GP+transit model used to remove the systematic trends. The black dots are the final detrended light curves along with a best-fit transit model, shown as a red solid curve (see Section \ref{ground_based_photometry}). The facility and the observation date are listed at upper left in each panel.} 
\label{ground_transit_detrend}
\end{figure*}

\subsection{Spectroscopic Observations}
\subsubsection{IRTF/SpeX}
\begin{figure}
\centering
\includegraphics[width=0.49\textwidth]{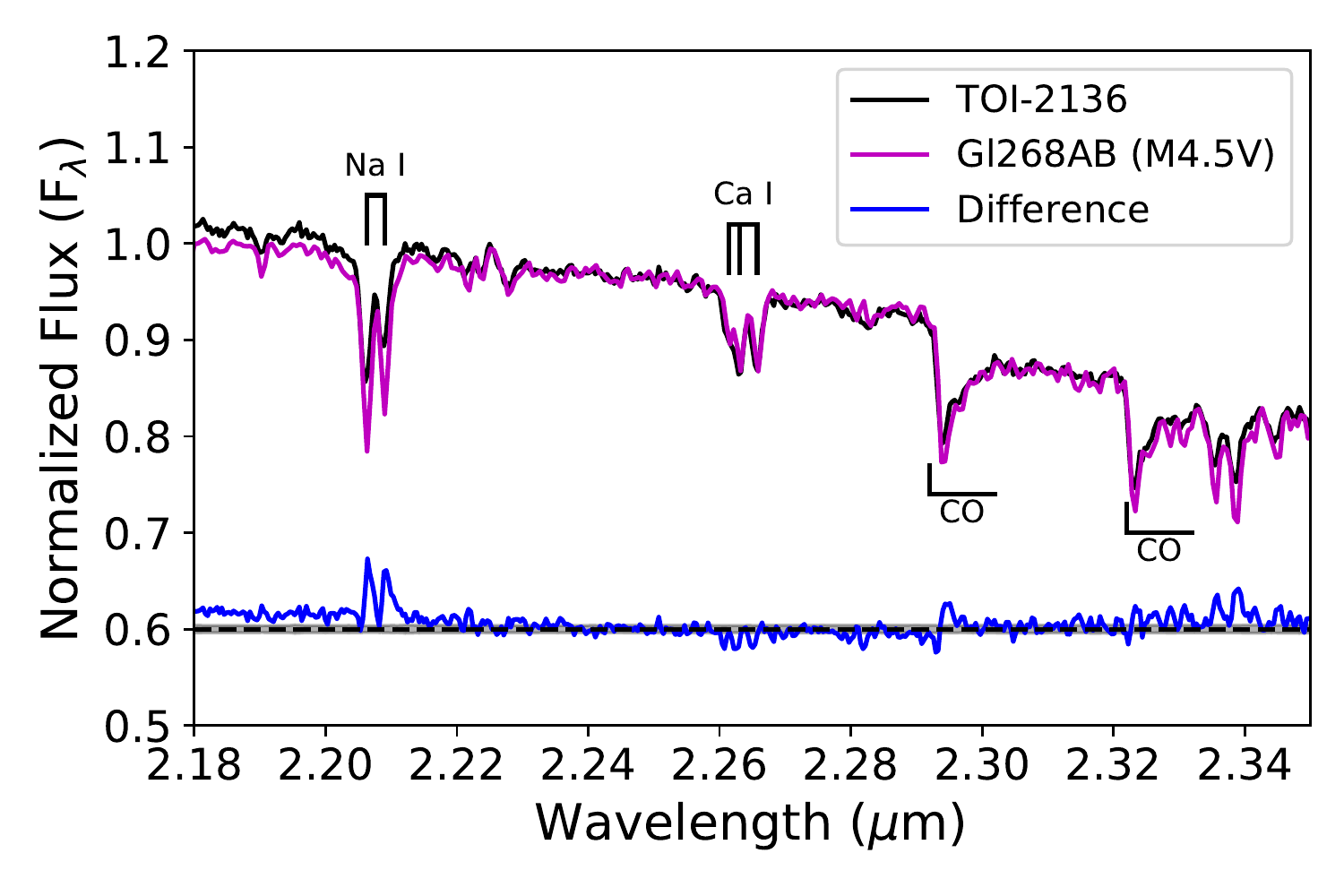}
\caption{Normalized SpeX near-infrared spectrum of \tar\ (black line) and the comparison spectrum (magenta line) taken from the IRTF library \citep{Rayner2009}. The strong atomic features are marked based on the results from \citet{Cushing2005}. The difference between these two spectra is shown below (blue line). The NIR spectrum of \tar\ is consistent with a spectral type of M4.5V.} 
\label{Spex}
\end{figure}

Infrared spectroscopy of \hbox{TOI-2136} was obtained with the SpeX spectrograph \citep{Rayner2003} on the 3.2-m NASA Infrared Telescope Facility on Maunakea, Hawaii, on 15$^{\rm th}$ September 2021 (UT). Conditions were mostly clear with thin clouds and 0.7\arcsec\ seeing. The short-wavelength cross-dispersed (SXD) mode was used with the 0.5\arcsec-wide slit to obtain a 0.7--2.5~$\mu$m spectrum in seven orders at a spectral resolving power $\lambda$/$\Delta\lambda \approx 2000$. A total of two ABBA nod sequences (8 exposures) were obtained with an integration time of 240 s per exposure with the slit aligned with the parallactic angle. The A0 V star HD 174567 (V = 6.63) was observed afterwards at an equivalent airmass for flux and telluric calibration, followed by arc lamp and flat field lamp exposures. Data were reduced using SpeXtool v4.1 \citep{Cushing2004} using standard settings. The resulting spectrum of \hbox{TOI-2136} had a median SNR of 200, with $JHK$ peaks of around 250--300 (see Figure \ref{Spex}). 

\subsubsection{CFHT/SPIRou}

\tar\ was monitored by SPIRou between 24$^{\rm th}$ April 2021 and 28$^{\rm th}$ June 2021. SPIRou has a spectral resolution of $R\approx75\,000$, covering a bandwidth from 0.98 to 2.5\,$\mu$m \citep{Moutou2020}. A total of 69 spectra were obtained. The observations were mainly done consecutively in two separate weeks, spanning roughly 50 days. We adopted an exposure time of 900s, and we repeated the observations $2\sim4$ times every night. Given the brightness of the host star in $H$ band (9.6\,mag), we opted to use the Farby-P\'erot (FP) mode to perform a simultaneous drift calibration during each observation, aiming for a RV precision better than 10\,m/s \citep{Cersullo2017}.

The SPIRou data reduction was performed using the 0.7.194 version of the \texttt{APERO} pipeline (Cook et al., in prep). Basic \texttt{APERO} steps have been described in a number of contributions \citep{Artigau2021,cristofari_estimating_2021,Martioli2022}. In brief, the major \texttt{APERO} modules are as follows: 
\begin{itemize}
    \item For all frames (science and calibrations), remove spatially correlated noise in the $4096\times4096$ images produced by the detector control software.
    \item Locate orders in nightly calibrations.
    \item Extract science and calibration frames into per-order spectra.
    \item Derive a nightly wavelength solution using the method described in \citet{hobson_spirou_2021}.
    \item Measure the instantaneous drift in individual science frame relative to the nightly wavelength solution using the simultaneous FP measurements.
    \item Apply a telluric correction to science data mainly based on a principal component analysis (PCA)-based approach  \citep{Artigau2014PCA}.
    \item Using the line-by-line method (see below) and derive a radial velocity. 
\end{itemize}

Velocity measurements were obtained with the line-by-line method (LBL; Artigau et al., in prep), which is discussed in \cite{Martioli2022}. Overall the approach of the LBL is to subdivide the spectral domain in a large number of `lines', typically 16\,000 for SPIRou, that corresponds to domain between consecutive local maxima in spectrum. Within each line, one applies the \citet{bouchy_fundamental_2001} framework to the difference between a high-SNR template and the spectrum to derive a velocity by projecting the residuals onto the first derivative of the template. This method provides a per-line velocity and the corresponding uncertainty. One then constructs a mixture model, where the mean velocity is derived simultaneously with the likelihood that a given line is valid (i.e., consistent with the mean velocity considering uncertainties) or that it belongs to a population of ``outliers'' that should be disregarded. The LBL framework fully utilizes the radial-velocity content of the spectrum and significantly out-performs the CCF in the near-infrared where numerous residuals (e.g., sky emission, telluric absorption, detector defects) plague precise RV observations.

All RVs we extracted are listed in Table \ref{spirourv}. We dropped three outliers above the $3\sigma$ limit, and a total of 66 measurements were used in the following analysis.

\subsection{High Angular Resolution Imaging}
High-resolution imaging is one of the standard follow-up observations made for exoplanet host stars. Spatially close companions, bound or line of sight, can create a false-positive transit signal and provide ``third-light'' flux leading to an underestimated planetary radius \citep{Ciardi2015}, incorrect planet and star properties \citep{Furlan2017b,Furlan2020} and can cause non-detections of small planets residing with the same exoplanetary system \citep{Lester2021}. Additionally, the discovery of close, bound companion stars provides crucial information toward our understanding of exoplanetary formation, dynamics and evolution \citep{Howell2021}. Generally, \gaia\ is not capable to recover binaries with separations smaller than $0.7\arcsec$ \citep{Ziegler2020}. Thus, to search for close-in bound companions unresolved in \gaia, \tess\ or other ground-based follow-up observations, we obtained high-resolution imaging observations of \tar.

\subsubsection{Robo AO}
As part of the M dwarf multiplicity survey \citep{Lamman2020}, a sub-arcsecond imaging of \tar\ was previously obtained from Robo-AO, an autonomous laser-guided adaptive optics system \citep{Baranec2014}, on 29$^{\rm th}$ July 2016 on the Kitt Peak 2.1-m telescope. The observation was taken with an Andor iXon DU-888 camera in the $i'$-band with a 90s exposure time. Median seeing at the telescope was $1. 44''$ which resulted in an $i'$-band Strehl ratio of 4.2\% for this observation and a full-width at half-maximum of $\sim0.12''$. The image was processed via an automatic pipeline, which shifts and adds data to optimize for both high and low SNR images \citep{Jensen2018}. \cite{Lamman2020} identified that there is no stellar companion of \tar\ with a contrast above the curve shown in Figure \ref{AO_image}.


\subsubsection{Shane}
We observed TIC 336128819 (TOI-2136) on 30$^{\rm th}$ April 2021 (UT) using the ShARCS camera on the Shane 3-meter telescope at Lick Observatory \citep{2012SPIE.8447E..3GK, 2014SPIE.9148E..05G, 2014SPIE.9148E..3AM}. Observations were taken with the Shane adaptive optics system in natural guide star mode. We refer the readers to \cite{2020AJ....160..287S} for a detailed description of the observing strategy and reduction prodecure. We collected two sequences of observations, one with a $Ks$ filter ($\lambda_0 = 2.150$ $\mu$m, $\Delta \lambda = 0.320$ $\mu$m) and one with a $J$ filter ($\lambda_0 = 1.238$ $\mu$m, $\Delta \lambda = 0.271$ $\mu$m). Our contrast curves are shown in Figure \ref{AO_image}. We find no nearby stellar companions within our detection limits.

\subsubsection{Gemini-North}
\begin{figure}
\includegraphics[width=0.5\textwidth]{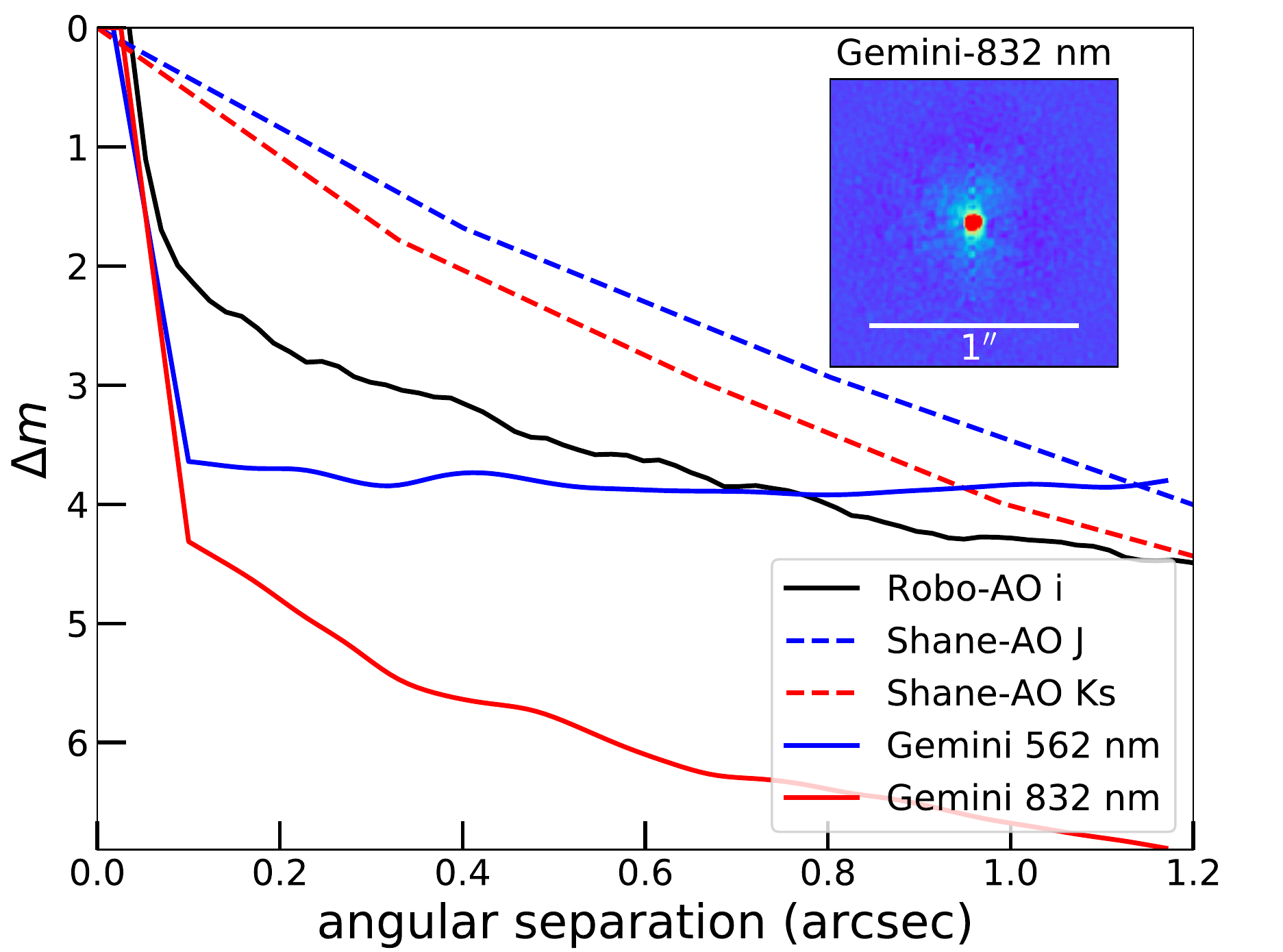}
\caption{$5\sigma$ contrast curves for \tar. Different lines represent results from different observations. The inset figure shows the reconstructed Gemini 832 nm image with a 1 arcsec scale bar. \tar\ was found to be an isolated single star within the contrast levels achieved.}
\label{AO_image}
\end{figure}

We obtain speckle imaging observation of \tar\ on 17$^{\rm th}$ October 2021 (UT) using the `Alopeke speckle instrument on the Gemini North 8-m telescope \citep{Scott2021}.  `Alopeke provides simultaneous speckle imaging in two bands (562nm and 832 nm) with output data products including a reconstructed image with robust contrast limits on companion detections (e.g., \citealt{Howell2016}). Five sets of $1000 \times 0.06$ sec exposures were collected and subjected to Fourier analysis in our standard reduction pipeline (see \citealt{Howell2011}). Figure \ref{AO_image} shows our final contrast curves and the 832 nm reconstructed speckle image. We find that \tar\ is a single star with no companion brighter than 4-7 magnitudes below that of the target star from the diffraction limit (20 mas) out to 1.2$\arcsec$. At the distance of \hbox{TOI-2136} (d=33\,pc) these angular limits correspond to spatial limits of 0.7 to 40\,au.

\section{Stellar Properties}\label{stellar_properties}
\subsection{Stellar Characterization}

We first estimate the absolute $K$ band magnitude from the 2MASS observed $m_{K}$ and the parallax from \gaia\ EDR3 \citep{GaiaEDR3}, which yields $M_{K}= 6.73 \pm 0.02$ mag. Taking use of the polynomial relation between $R_{\ast}$ and $M_{K}$ derived by \cite{Mann2015}, we obtain a stellar radius of $R_{\ast}=0.34\pm0.01\ R_{\odot}$, assuming a typical uncertainty of 3\% (see Table 1 in \citealt{Mann2015}). This is consistent with the estimation $R_{\ast}=0.34\pm0.02\ R_{\odot}$ within $1\sigma$ using the angular diameter relation in \cite{Boyajian2014}.

Based on the empirical relation between bolometric correction ${\rm BC}_{K}$ and stellar color $V-J$ found by \cite{Mann2015}, we obtain a ${\rm BC}_{K}$ of $2.73\pm0.21$ mag, leading to a bolometric magnitude $M_{\rm bol}=9.46\pm 0.22$ mag. We then calculate the bolometric luminosity to be $L_{\ast}=0.013\pm0.003\ L_{\odot}$. We further estimate the stellar effective temperature of \tar\ using two different methods. Combined with the stellar radius and bolometric luminosity, we find $T_{\rm eff}=3324\pm55$ K by utilizing the Stefan-Boltzmann law. Additionally, we also obtain $T_{\rm eff}$ following the empirical relation with stellar color $V-J$ and $J-H$ reported by \cite{Mann2015}, and we find $T_{\rm eff}=3314\pm104$ K. Both estimations agree well with the result $T_{\rm eff}=3267\pm133$ K from \cite{Pecaut2013}. 

We also evaluate that \tar\ has a mass of $M_{\ast}=0.33\pm0.02\ M_{\odot}$ using Equation 2 in \cite{Mann2019} according to the $M_{\ast}$-$M_{K}$ relation. This is consistent with the value $M_\ast = 0.35 \pm 0.02\ M_\odot$ given by the $M_{K}$-Mass empirical relation of \citet{Benedict2016}. 

\begin{figure}
\includegraphics[width=0.5\textwidth]{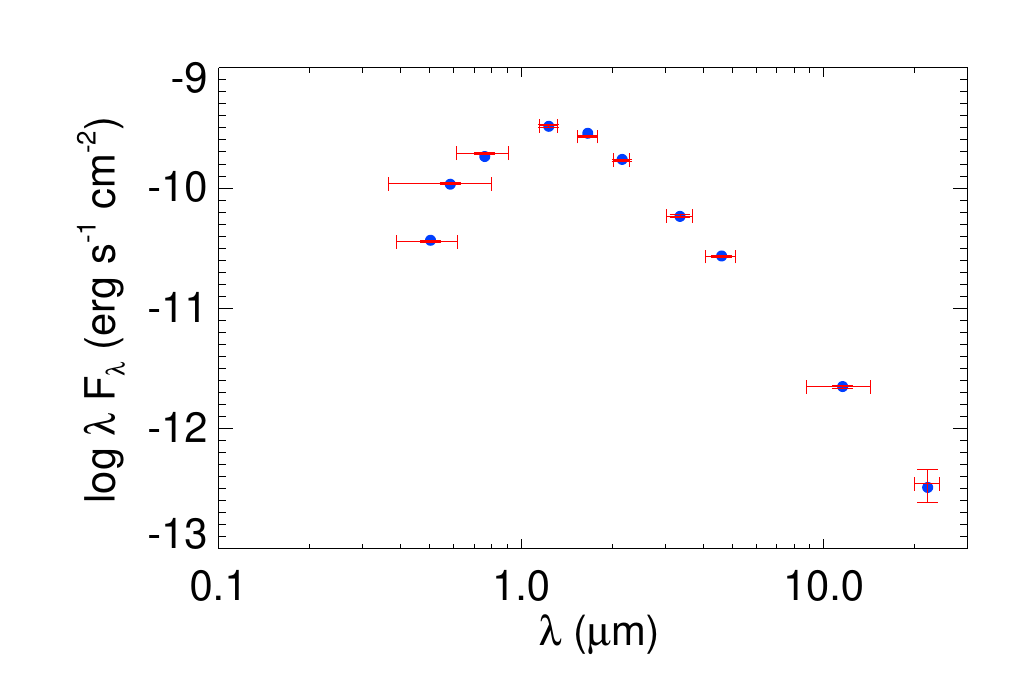}
\caption{SED model for \hbox{TOI-2136}. The red symbols are the broadband photometric measurements used in the SED analysis (provided in Table \ref{starparam}) with the horizontal uncertainty bars representing the effective width of the passband. The blue symbols are the model fluxes from the best-fit Kurucz atmosphere model.}
	\label{fig:sed_isochrone}
\end{figure}

As an independent check, we performed an analysis of the broadband Spectral Energy Distribution (SED) of \tar\ using MIST stellar models \citep{Dotter:2016, choi:2016} along with the {\it Gaia\/} EDR3 parallax \citep{GaiaEDR3} in order to derive the stellar parameters of \tar. We made use of the \code{EXOFASTv2} package \citep{Eastman:2019} to conduct the SED fit. We used the MIST method (the favored method reported in \cite{Eastman:2019}, -MISTSEDFILE) that interpolates the 4D grid of $\log g$, $T_{\rm eff}$, [Fe/H], and an extinction grid from Conroy et al., (in prep) to determine the bolometric corrections in each of the observed band. We pulled the $JHK_S$ magnitudes from 2MASS \citep{Cutri:2003}, the W1-W4 magnitudes from WISE \citep{Wright:2010}, and three \gaia\ magnitudes $G, G_{\rm BP}, G_{\rm RP}$ \citep{Gaia:2018}. Together, the available photometry spans the full stellar SED over the wavelength range 0.5-22$\mu$m (see Figure \ref{fig:sed_isochrone}). We applied an upper limit on the V-band extinction from the dust maps of \citet{Schlafly:2011} and a Gaussian prior on the [Fe/H] taken from the spectroscopic analysis. 
The \code{EXOFASTv2} analysis ran until convergence when the Gelman-Rubin statistic (GR) and the number of independent chain draws (Tz) were less than 1.01 and greater than 1000, respectively. The full results of the SED fit are provided in Table \ref{exofaststarparam}, which are in excellent agreement with our previous estimation.

Taking all the results above into account, we adopt the weighted-mean values of effective temperature $T_{\rm eff}$, stellar radius $R_{\ast}$ and stellar mass $M_{\ast}$ with conservative uncertainties as listed in Table \ref{starparam}. Combining the derived stellar radius with mass, we find a mean stellar density of $\rm \rho_\ast = 12.20 \pm 2.53$~g~cm$^{-3}$.

Finally, we also estimate the systemic radial velocity of \tar\ to be $-28.8 \pm 6.0$ km/s by RV-correcting our SpeX spectrum using \texttt{tellrv} \citep{2014AJ....147...20N}. To determine the stellar type, we further compare our SpeX spectrum with the IRTF library \citep{Rayner2009} and find that \tar\ is consistent with a star of spectral type M4.5V (Figure \ref{Spex}). Lastly, following the procedure described in \cite{Gantoi530}, we obtain the metallicity of \tar\ based on the relations defined in \cite{Mann2013} for cool dwarfs with spectral types between K5 and M5. Our analysis yield metallicities of [Fe/H] = $0.03 \pm 0.07$ and [M/H] = $-0.01 \pm 0.08$.

\begin{table}
	\caption{Median values and 68\% confidence interval for TOI-2136 from the SED fit alone.}
	\begin{tabular}{lll}
		\hline\hline
		Parameter & Units      &Value       \\ \hline
		$M_*$  & Mass ($M_{\odot}$)  &$0.350^{+0.024}_{-0.028}$\\
		$R_*$  & Radius ($R_{\odot}$)  &$0.342^{+0.011}_{-0.011}$\\
		$\rho_*$  &Density (cgs)  &$12.2^{+1.2}_{-1.1}$\\
		$\log{g}$ &Surface gravity (cgs)  &$4.912^{+0.031}_{-0.033}$\\
		$L_*$  &Luminosity ($L_{\odot}$)  &$0.01381^{+0.00048}_{-0.00043}$\\
		$F_{\rm bol}$ &Bolometric Flux (cgs 10$^{-10}$) &$3.98^{+0.14}_{-0.12}$\\
		$T_{\rm eff}$ &Effective Temperature (K) &$3383^{+52}_{-54}$\\
		$[{\rm Fe/H}]$  &Metallicity (dex)  &$0.15^{+0.10}_{-0.10}$\\
		$A_V$  & V-band extinction (mag)  &$0.070^{+0.090}_{-0.052}$\\
		$\sigma_{\rm SED}$  &SED photometry error scaling   &$1.9^{+0.74}_{-0.45}$\\
		$\varpi$  &Parallax (mas)  &$29.997^{+0.057}_{-0.057}$\\
		$d$  & Distance (pc) &$33.337^{+0.064}_{-0.063}$\\
		\hline\hline 
	\end{tabular}
	\label{exofaststarparam}
\end{table}

\subsection{Galactic Component}
Combined with the tangential velocity ($\mu_{\alpha}$, $\mu_{\delta}$) and the stellar parallax ($\varpi$) from \gaia\ EDR3 as well as the spectroscopically determined systemic RV from the SpeX spectrum, we calculate the three-dimensional space motion with respect to the LSR based on the methodology described in \cite{Johnson1987}. We obtain three-dimensional space velocities $U_{\rm LSR}=-25.15\pm2.26$ km s$^{-1}$, $V_{\rm LSR}=-9.42\pm5.27$ km s$^{-1}$, $W_{\rm LSR}=13.16\pm1.75$ km s$^{-1}$, respectively. We further identify the Galactic population membership of \tar\ following the criterion first used in \cite{Bensby2003}. We compute the relative probability $P_{\rm thick}/P_{\rm thin}=0.01$ of \tar\ to be in the thick and thin disks by taking use of the recent kinematic values from \cite{Bensby2014}, indicating a thin-disk origin. Finally, we integrate the stellar orbit with the ``MWPotential2014'' Galactic potential using \code{galpy} \citep{Bovy2015} following the procedure described in \cite{Gan2020}, and we estimate that the maximal height $Z_{\rm max}$ of \tar\ above the Galactic plane is about $206$ pc. Therefore, we conclude that \tar\ belongs to the thin-disk population, which is also consistent with its solar-like metallicity. 

\subsection{Stellar activity and rotation period}

\begin{figure*}
\centering
\includegraphics[width=\textwidth]{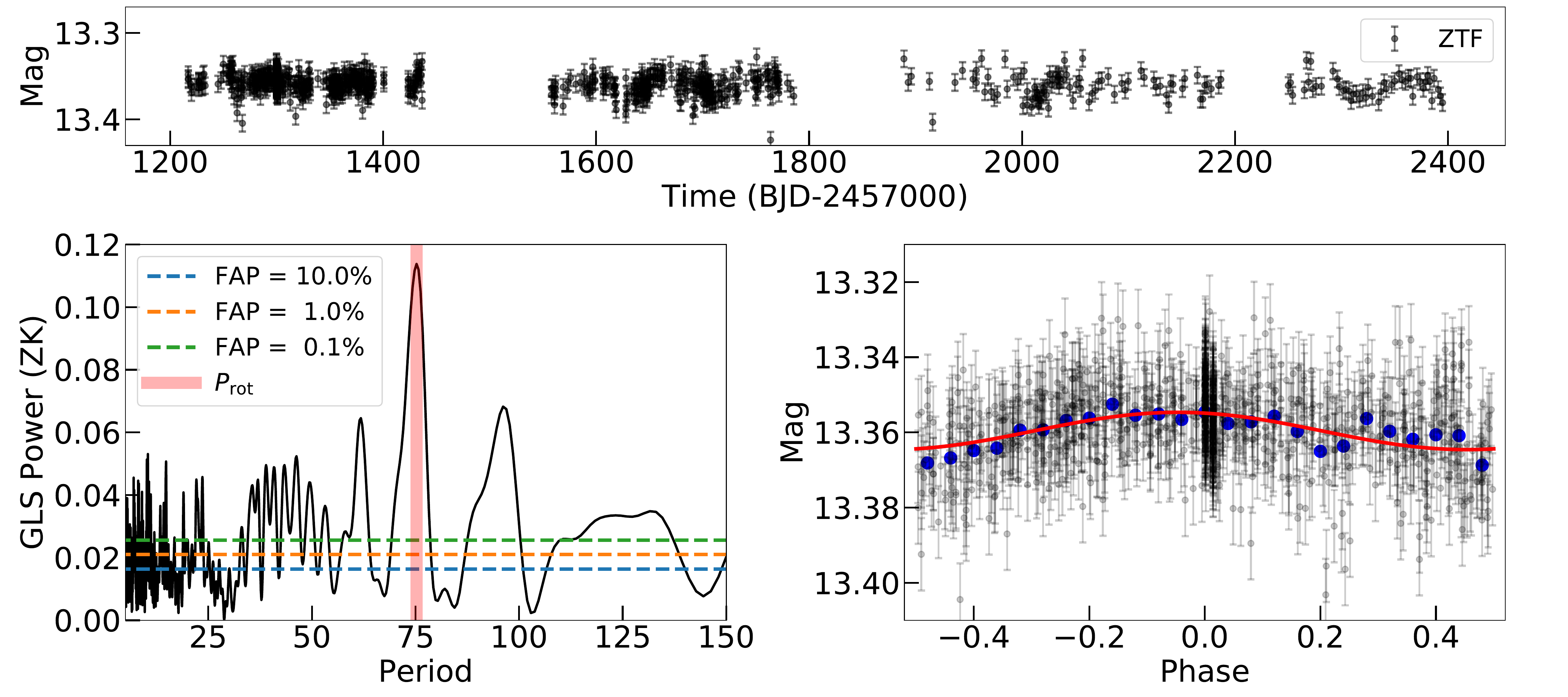}
\caption{{\it Top panel:} The ground-based long-term light curve of \tar\ taken by ZTF. {\it Bottom left panel:} The GLS periodogram of the ZTF photometry. The vertical red line represents the $\sim$75 days rotational signal of \tar. The theoretical FAP levels of 10\%, 1\% and 0.1\% are marked as horizontal lines with different colors. {\it Bottom right panel:} The phase-folded ZTF light curve at 75 days along with the best-fit sinusoidal model, shown as a red solid curve. The blue dots are the binned data. }
\label{ztf}
\end{figure*}

Stellar activity, often manifesting as stellar rotation signals, is expected to affect the RV measurements and make it challenge to accurately determine the planet mass \citep{Queloz2009,Howard2013,Pepe2013}, especially when its timescale is close to the planet orbital period \citep{Gan2021}. In order to evaluate the effect of the stellar activity on the RVs, we first search for the periodic signals in the \tess\ PDCSAP light curve after masking the known in-transit data using the generalized Lomb-Scargle (GLS) periodogram \citep{Zechmeister2009}, and we find no signs of stellar variation. Therefore, we do not present the periodograms here. However, we note that the PDCSAP photometry from \tess\ flattens variability on timescales greater than about 15 day and \tess\ is insensitive to long-term stellar rotational features due to its sector-by-sector observational strategy. Thus, we further examine the archival long-term photometric time-series data from ground-based surveys. We look into the rotational modulation of \tar\ in the publicly available light curve taken by the Zwicky Transient Facility (ZTF; \citealt{Masci2019}). A total of 1054 measurements were acquired in $r-$ band spanning 1112 days. After removing the observations flagged as bad-quality, we have 1011 measurements left with a standard deviation of 0.011 mag. We compute the GLS periodogram and find a clear peak at $75\pm5$ days (see Figure \ref{ztf}). This is consistent with the estimation of $P_{\rm rot}\sim82.97$ days derived by \cite{Newton2016} using the data from MEarth \citep{Nutzman2008}. Additionally, the lack of significant flaring activity existing in the \tess\ light curves also suggests that the host star is quiet and inactive. We thus attribute this $75\pm5$ days signal to the stellar rotation. Adopting the empirical relations from \cite{Engle2018}, we estimate that \tar\ has an age of $4.6\pm 1.0$ Gyr, consistent with our thin-disk population conclusion. 


\section{Analysis and results}\label{analysis}
In this section, we outline our data analysis steps including modeling the space and ground light curves as well as the SPIRou radial velocities, which mainly follows \cite{Gantoi530}. In short, we begin with fitting the \tess-only photometry and then take the posterior information as a prior to detrend the ground-based light curves (see Section~\ref{Photometric_Analysis}). We next perform a pre-analysis to the RVs and test the significance of eccentricity (see Section~\ref{RV_Analysis}), and we carry out a joint-fit of all data to obtain the best-fit physical parameters of \hbox{TOI-2136\,b} (see Section~\ref{Joint_fit}). Finally, we conduct a transit timing variation (TTV) analysis to look for potential evidence of another non-transiting planet (see Section~\ref{Transit_timing_variation}).

\subsection{Photometric Analysis}\label{Photometric_Analysis}

\subsubsection{TESS only}\label{tess_only}

We first employ the \code{juliet} package \citep{juliet} to fit the detrended \tess\ light curve, which makes use of \code{batman} \citep{Kreidberg2015} to build the transit model and \code{dynesty} \citep{Higson2019,Speagle2019} to carry out dynamic nested sampling and determine the Bayesian posteriors of system parameters. Instead of fitting the planet-to-star radius ratio ($p=R_{p}/R_{\ast}$) and the impact parameter $b=a\cos i/R_{\ast}$ directly, \code{juliet} utilizes the new parametrizations $r_{1}$ and $r_{2}$ to make the sampling more efficient as it focuses on physically meaningful values of a transiting system with $0 < b < 1 + p$ \citep{Espinoza2018}. We carry out a circular-orbit fit with $e=0$. Consequently, the left degrees of freedom are $r_{1}$, $r_{2}$, mid-transit epoch $T_0$, orbital period $P_{b}$ and stellar density $\rho_{\ast}$. We place uniform priors on both $T_{0}$ and $P_{b}$ according to the outputs from our TLS analysis, and allow $r_{1}$ as well as $r_{2}$ to vary uniformly between 0 and 1. Regarding the stellar density, we impose a non-informative log-uniform prior. We fit two limb-darkening coefficients under the triangular sampling scheme (i.e., $q_{1}$ and $q_{2}$; \citealt{Kipping2013}), and adopt uniform priors on both of them. In addition, we also include an extra flux jitter term to account for additional systematics, on which we set a wide log-uniform prior. We do not take light contamination into account as the \tess\ PDCSAP light curve has already been corrected for the dilution effect. The prior settings and the median along with $1\sigma$ credible intervals of transit parameter's posteriors are given in Table \ref{tess_only_fit_priors}. 

In order to investigate the potential evidence of orbital eccentricity from the photometric-only data, we rerun our fit with free $e$ and $w$ and compare the Bayesian model log-evidence ($\ln Z$) difference between the circular and eccentric models. Basically, \code{juliet} considers that a model is significantly favored if it has a $\ln Z$ improvement over 5 and moderately supported if $\Delta \ln Z>2.5$ based on the criteria described in \citep{Trotta2008}. We find that the circular orbit model is slightly preferred with a Bayesian evidence improvement of $\Delta \ln Z=\ln Z_{\rm Circular}-\ln Z_{\rm Eccentric}=1.1$. Therefore, we conclude that no significant orbital eccentricity preference is shown in the \tess\ data. 

\subsubsection{Ground-based photometry}\label{ground_based_photometry}

Since part of ground light curves show obvious linear coherence between the flux and time, we perform a uniform detrending using Gaussian process regressors with the \code{celerite} Mat\'{e}rn-3/2 kernel to remove their systematic trends. Rather than mask out the in-transit data and interpolate to renormalize the light curve as stated in Section \ref{tess_data}, here we perform a simultaneous GP+transit fit to all ground data given their short out-of-transit span. We take the posteriors from the previous circular orbit fit, and put informative priors on $P_{b}$, $T_{0}$, $r_1$, $r_2$ and $\rho_{\ast}$. We show in Table \ref{ground_only_fit_priors} our prior adoption and present our raw and reprocessed ground light curves in Figure \ref{ground_transit_detrend}.

\subsection{RV Analysis}\label{RV_Analysis}
We perform an RV-only fit using \code{juliet}, which employs the \code{radvel} package \citep{Fulton2018} to model the Keplerian RV signals. Since the expected RVs caused by the planet perturbation is expected to be small, we choose to fix the orbital period $P_{b}$ and mid-transit epoch $T_{0,b}$ at the best-fit transit ephemeris derived from the \tess\ only fit to reduce introducing additional uncertainties. As the \tess\ photometric data do not show evidence for eccentricity, we assume a circular orbit and fix eccentricity $e$ at 0, and the argument of periastron $\omega$ at $90^{\circ}$. Moreover, we do not take the RV slope $\dot{\gamma}$ or the quadratic trend $\ddot{\gamma}$ into consideration and fix them at zero due to the short time span of our RV data. We include a simple jitter term $\sigma_{\rm RV}$ that is added in quadrature to the error bars of each data point to account for the white noise. We set uniform priors on both the RV semi-amplitude $K_{b}$ and the systemic velocity $\mu$ but a log-uniform prior on $\sigma_{RV}$. We obtain $K_{b}=4.1\pm1.5$ m/s, which is consistent with the expected value $\sim3.7$ m/s supposing a planet mass estimated using the mass-radius relation from \cite{Chen2017}.

We next construct a Keplerian model with free $e$ and $\omega$ to look for the significance of the eccentricity in the RV data. We find that the circular orbit model is slightly preferred with a Bayesian evidence improvement of $\Delta \ln Z=\ln Z_{\rm Circular}-\ln Z_{\rm Eccentric}=1.2$, agreeing with our findings in the \tess\ photometric data (see Section \ref{tess_only}).

\subsection{Joint-fit}\label{Joint_fit}
\begin{figure}
\centering
\includegraphics[width=0.49\textwidth]{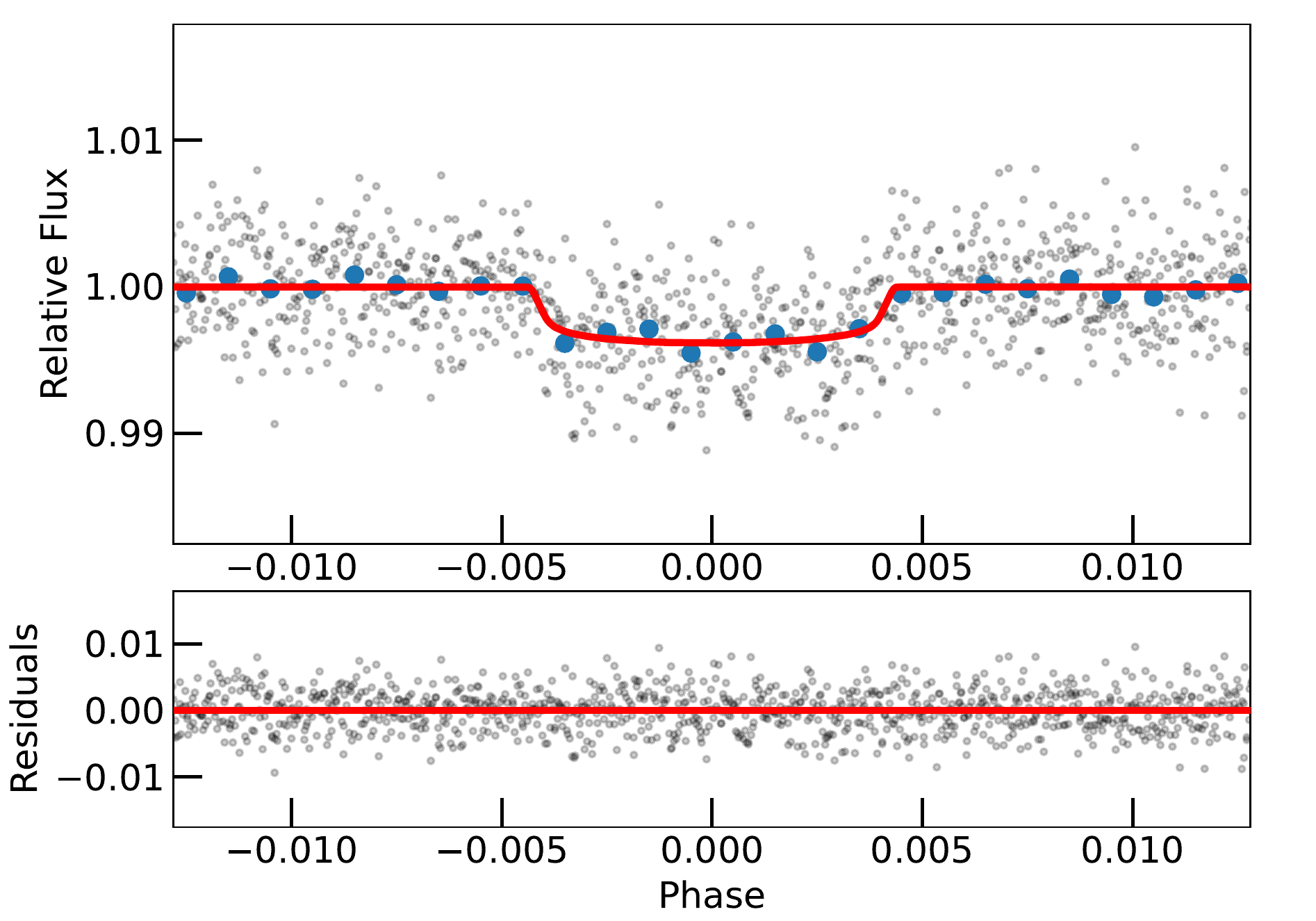}
\caption{The \tess\ light curve folded in phase with the orbital period of \hbox{TOI-2136\,b}. The red solid line represents the best-fit transit model from the final joint-fit (see Section \ref{Joint_fit}). The blue dots are the binned data every phase interval of 0.001. The residuals are plotted below.} 
\label{TESS_phase_folded}
\end{figure}

\begin{figure}
\centering
\includegraphics[width=0.49\textwidth]{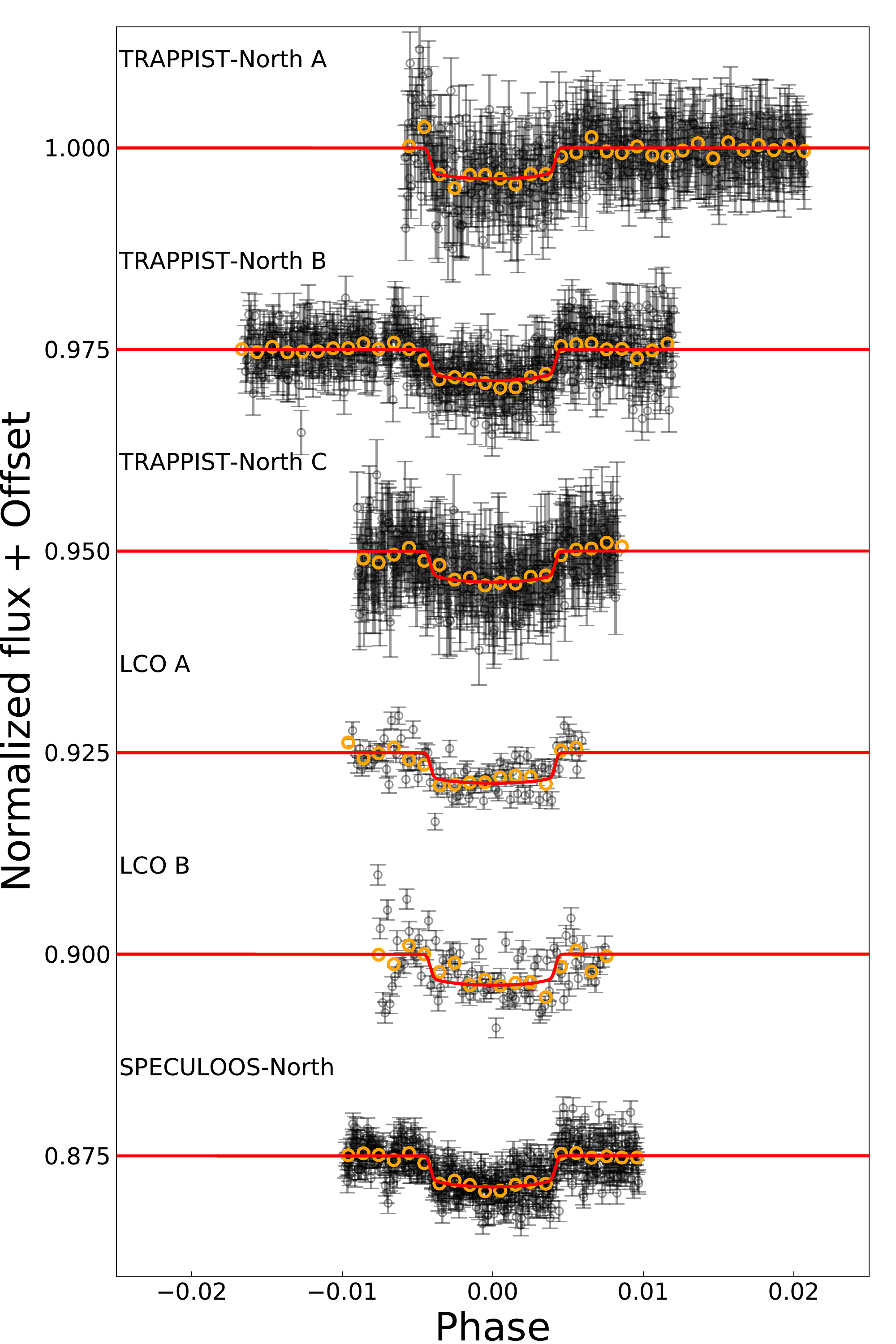}
\caption{All ground-based photometry phase-folded on the best-fit orbital period of \hbox{TOI-2136\,b} with arbitrary offsets. The red solid lines are the median transit models from the final joint-fit (see Section \ref{Joint_fit}). The over-plotted orange circles are the binned data every phase interval of 0.001.} 
\label{ground_phase_folded}
\end{figure}

\begin{figure*}
\centering
\includegraphics[width=\textwidth]{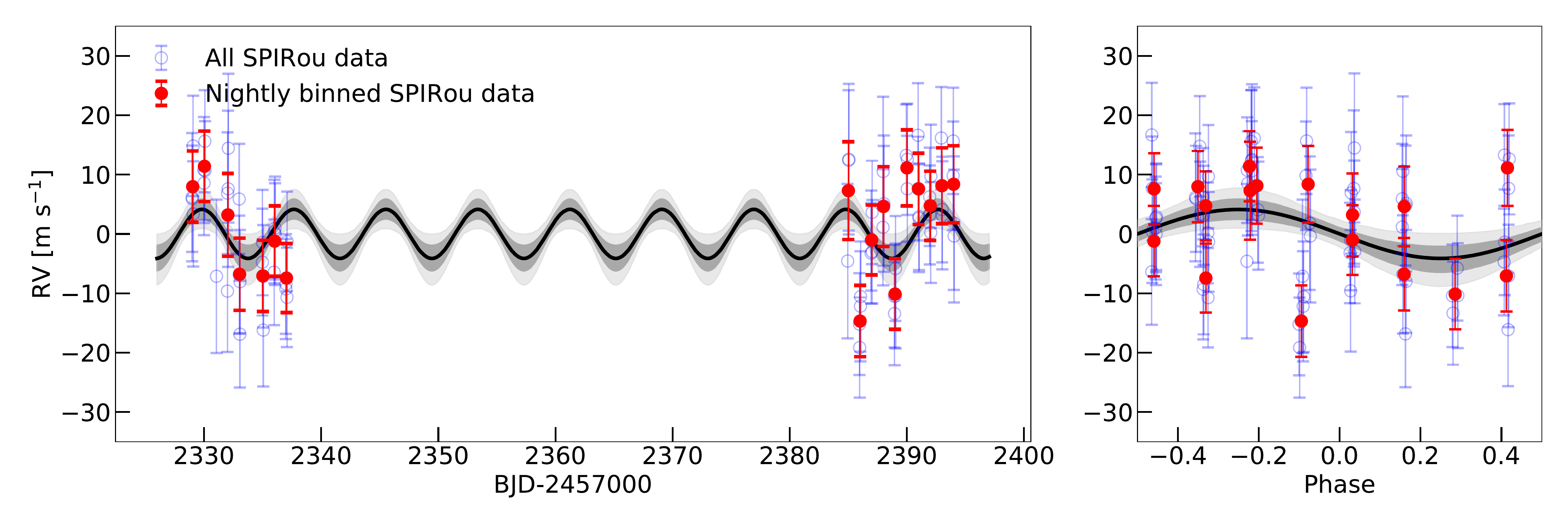}
\caption{{\it Left panel}: Time series of the SPIRou RVs after subtracting the best-fit systemic velocity. The blue circles are all original SPIRou data taken every night while the red dots are the nightly binned RVs. The black solid line is the median RV model from the final joint-fit (see Section \ref{Joint_fit}). The grey shaded areas denote the one and two sigma credible intervals of the RV model. {\it Right panel}: Phase-folded SPIRou RVs. The presented RV error bars in both panels are the the quadrature sum of the fitted instrument jitter and the measurement uncertainties.} 
\label{rvtotal}
\end{figure*}

Building on the results from the independent transit and RV fits, we finally carry out a joint-fit using \code{juliet} to simultaneously model all detrended space and ground light curves together with the SPIRou RVs to infer the properties of the \hbox{TOI-2136\,b}. We place the same priors as we did in Section \ref{tess_only} except that we adopt Gaussian priors for the linear limb darkening coefficients of the ground-based light curves, centered at the estimates from the \code{LDTK} package \citep{Husser2013,Parviainen2015} with a $1\sigma$ value of 0.1. As there is less contamination in the ground data, we fix all dilution factors $D_{i}$ to 1. For the RV part, we use the same priors as in Section \ref{RV_Analysis}. The phase-folded \tess\ and ground light curves along with the best-fit transit models are shown in Figures \ref{TESS_phase_folded} and \ref{ground_phase_folded}. The RV timeseries and the best-fit RV model are presented in Figure \ref{rvtotal}. The fitted RV semi-amplitude is $4.2\pm1.4$ m/s, a detection close to a $3\sigma$ significance. Our joint-fit model reveals that the planet has a radius of $2.19\pm0.17\ R_{\oplus}$ with a mass of $6.37\pm2.45\ M_{\oplus}$. All priors and the median of the posterior distributions for each fitted parameter are summarized in Table \ref{allpriors}. We also run a separate joint-fit using \code{EXOFASTv2} \citep{Eastman:2019}, and we verify that similar results were obtained within $1\sigma$. 

\begin{table*}
    {\renewcommand{\arraystretch}{1.2}
    \caption{Parameter priors and the best-fit values along with the 68\% credibility intervals in the final joint fit for \tar. $\mathcal{N}$($\mu\ ,\ \sigma^{2}$) means a normal prior with mean $\mu$ and standard deviation $\sigma$. $\mathcal{U}$(a\ , \ b) stands for a uniform prior ranging from a to b. $\mathcal{J}$(a\ , \ b) stands for a Jeffrey's prior ranging from a to b.}
    \begin{tabular}{lccr}
        \hline\hline
        Parameter       &Prior &Best-fit    &Description\\\hline
        \it{Planetary parameters}\\
        $P_{b}$ (days)  &$\mathcal{U}$ ($7.6$\ ,\ $8.0$)  &$7.851928^{+0.000018}_{-0.000016}$
        &Orbital period of \hbox{TOI-2136\,b}.\\
        $T_{0,b}$ (BJD-2457000)  &$\mathcal{U}$ ($2014$\ ,\ $2020$)   &$2017.7043^{+0.0009}_{-0.0007}$
        &Mid-transit time of \hbox{TOI-2136\,b}.\\
        $r_{1,b}$  &$\mathcal{U}$ (0\ ,\ 1)   &$0.57^{+0.13}_{-0.12}$
        &Parametrisation for {\it p} and {\it b}.\\
        $r_{2,b}$  &$\mathcal{U}$ (0\ ,\ 1)   &$0.0591^{+0.0010}_{-0.0009}$
        &Parametrisation for {\it p} and {\it b}.\\
        $e_{b}$                     &0  &Fixed  &Orbital eccentricity of \hbox{TOI-2136\,b}.\\
        $\omega_{b}$ (deg)          &90 &Fixed  &Argument of periapsis of \hbox{TOI-2136\,b}.\\
        \it{Photometry parameters}\\
        $D_{\rm all}$  &Fixed    &$1$      &Photometric dilution factors.\\
        $M_{\rm TESS}$  &$\mathcal{N}$ (0\ ,\ $0.1^{2}$)   &$-0.000001^{+0.000016}_{-0.000017}$      &Mean out-of-transit flux of \tess\ photometry.\\
        $\sigma_{\rm TESS}$ (ppm) &$\mathcal{J}$ ($10^{-6}$\ ,\ $10^{6}$)  &$0.02^{+2.32}_{-0.01}$
          &\tess\ additive photometric jitter term.\\
        $q_{1}$       &$\mathcal{U}$ (0\ ,\ 1)           &$0.27^{+0.22}_{-0.16}$   &Quadratic limb darkening coefficient of \tess\ photometry.\\
        $q_{2}$       &$\mathcal{U}$ (0\ ,\ 1)                 &$0.28^{+0.29}_{-0.18}$  &Quadratic limb darkening coefficient of \tess\ photometry.\\
        
        $M_{\rm TRAPPIST-North,\ A}$  &$\mathcal{N}$ (0\ ,\ $0.1^{2}$)    &$-0.00009^{+0.00018}_{-0.00017}$
        &Mean out-of-transit flux of TRAPPIST-North-A photometry.\\
        $\sigma_{\rm TRAPPIST-North,\ A}$ (ppm) &$\mathcal{J}$ ($0.1$\ ,\ $10^{5}$)  &$7.5^{+75.7}_{-6.9}$
          &Additive photometric jitter term of TRAPPIST-North-A photometry.\\
        $q_{\rm TRAPPIST-North,\ A}$   &$\mathcal{N}$ ($0.31$\ ,\ $0.1^{2}$)                &$0.31^{+0.08}_{-0.07}$     &Linear limb darkening coefficient of TRAPPIST-North-A photometry.\\
        
        $M_{\rm TRAPPIST-North,\ B}$  &$\mathcal{N}$ (0\ ,\ $0.1^{2}$)    &$-0.00002^{+0.00011}_{-0.00011}$
        &Mean out-of-transit flux of TRAPPIST-North-B photometry.\\
        $\sigma_{\rm TRAPPIST-North,\ B}$ (ppm) &$\mathcal{J}$ ($0.1$\ ,\ $10^{5}$)  &$8.5^{+122.9}_{-7.9}$
          &Additive photometric jitter term of TRAPPIST-North-B photometry.\\
        $q_{\rm TRAPPIST-North,\ B}$   &$\mathcal{N}$ ($0.31$\ ,\ $0.1^{2}$)                &$0.34^{+0.09}_{-0.07}$     &Linear limb darkening coefficient of TRAPPIST-North-B photometry.\\
        
        $M_{\rm TRAPPIST-North,\ C}$  &$\mathcal{N}$ (0\ ,\ $0.1^{2}$)    &$-0.00004^{+0.00023}_{-0.00023}$
        &Mean out-of-transit flux of TRAPPIST-North-C photometry.\\
        $\sigma_{\rm TRAPPIST-North,\ C}$ (ppm) &$\mathcal{J}$ ($0.1$\ ,\ $10^{5}$)  &$7.7^{+78.2}_{-7.1}$
          &Additive photometric jitter term of TRAPPIST-North-C photometry.\\
        $q_{\rm TRAPPIST-North,\ C}$   &$\mathcal{N}$ ($0.31$\ ,\ $0.1^{2}$)                &$0.32^{+0.08}_{-0.07}$     &Linear limb darkening coefficient of TRAPPIST-North-C photometry.\\
        
        $M_{\rm LCO-CTIO,\ A}$  &$\mathcal{N}$ (0\ ,\ $0.1^{2}$)    &$-0.00006^{+0.00017}_{-0.00017}$
        &Mean out-of-transit flux of LCO-CTIO-A photometry.\\
        $\sigma_{\rm LCO-CTIO,\ A}$ (ppm) &$\mathcal{J}$ ($0.1$\ ,\ $10^{5}$)  &$1533.9^{+147.9}_{-137.7}$
          &Additive photometric jitter term of LCO-CTIO-A photometry.\\
        $q_{\rm LCO-CTIO,\ A}$   &$\mathcal{N}$ ($0.31$\ ,\ $0.1^{2}$)                &$0.26^{+0.07}_{-0.07}$     &Linear limb darkening coefficient of LCO-CTIO-A photometry.\\
        
        $M_{\rm LCO-CTIO,\ B}$  &$\mathcal{N}$ (0\ ,\ $0.1^{2}$)    &$-0.00005^{+0.00026}_{-0.00028}$
        &Mean out-of-transit flux of LCO-CTIO-B photometry.\\
        $\sigma_{\rm LCO-CTIO,\ B}$ (ppm) &$\mathcal{J}$ ($0.1$\ ,\ $10^{5}$)  &$2700.2^{+226.3}_{-212.4}$
          &Additive photometric jitter term of LCO-CTIO-B photometry.\\
        $q_{\rm LCO-CTIO,\ B}$   &$\mathcal{N}$ ($0.31$\ ,\ $0.1^{2}$)                &$0.32^{+0.09}_{-0.09}$     &Linear limb darkening coefficient of LCO-CTIO-B photometry.\\
        
        $M_{\rm SPECULOOS-North}$  &$\mathcal{N}$ (0\ ,\ $0.1^{2}$)    &$-0.00002^{+0.00009}_{-0.00009}$
        &Mean out-of-transit flux of SPECULOOS-North photometry.\\
        $\sigma_{\rm SPECULOOS-North}$ (ppm) &$\mathcal{J}$ ($0.1$\ ,\ $10^{5}$)  &$1532.7^{+73.7}_{-66.7}$
          &Additive photometric jitter term of SPECULOOS-North photometry.\\
        $q_{\rm SPECULOOS-North}$   &$\mathcal{N}$ ($0.31$\ ,\ $0.1^{2}$)                &$0.35^{+0.06}_{-0.07}$     &Linear limb darkening coefficient of SPECULOOS-North photometry.\\
        
        \it{Stellar parameters}\\
        ${\rho}_{\ast}$ ($\rm kg\ m^{-3}$) &$\mathcal{J}$ ($10^{3}$\ ,\ $\rm 10^{5}$)   &$14023^{+2462}_{-4219}$
         &Stellar density.\\
        \it{RV parameters}\\
        $K_{b}$ ($\rm m\ s^{-1}$)  &$\mathcal{U}$ ($0$\ ,\ $30$)      &$4.2^{+1.4}_{-1.4}$
        &RV semi-amplitude of \hbox{TOI-2136\,b}.\\
        $\rm \mu_{SPIRou}$ ($\rm m\ s^{-1}$) &$\mathcal{U}$ ($-29100$\ ,\ $-29000$)   &$-29067.3^{+1.1}_{-1.2}$
        &Systemic velocity for SPIRou.\\
        $\rm \sigma_{SPIRou}$ ($\rm m\ s^{-1}$) &$\mathcal{J}$ ($0.1$\ ,\ $100$)   &$4.4^{+1.4}_{-1.6}$
        &Extra jitter term for SPIRou.\\ \hline
        \it{Derived parameters}\\
        $R_{p}/R_{\ast}$  & &$0.0591^{+0.0010}_{-0.0009}$ &Planet radius in units of stellar radius.\\
        $R_{p}$ ($R_{\oplus}$) & &$2.19^{+0.17}_{-0.17}$ &Planet radius.\\
        $M_{P}$ ($M_{\oplus}$) & &$6.37^{+2.45}_{-2.29}$ &Planet mass.\\
        $\rho_{p}$ ($\rm g\ cm^{-3}$) & &$3.34^{+2.55}_{-1.63}$ &Planet density.\\
        ${b}$   &   &$0.35^{+0.20}_{-0.18}$ &Impact parameter.\\
        $a/R_{\ast}$   &  &$35.75^{+1.98}_{-2.01}$ &Semi-major axis in units of stellar radii.\\
        $a$ (au)   &    &$0.057^{+0.006}_{-0.006}$ &Semi-major axis.\\
        $i$ (deg)  &   &$89.4^{+0.3}_{-0.4}$ &Inclination angle.\\
        $S_{p}$ ($S_{\oplus}$) & &$4.0^{+2.1}_{-1.5}$ &Insolation flux relative to the Earth.\\
        $T_{\rm eq}^{[1]}$ (K)  &   &$395^{+24}_{-22}$ &Equilibrium temperature.\\
        \hline\hline
    \label{allpriors}    
    \end{tabular}}
    \begin{tablenotes}
       \item[1]  [1]\ We set an albedo $A_{B}=0$ here and assume there is no heat distribution between the dayside and nightside. 
    \end{tablenotes}
\end{table*}

\subsection{Transit timing variation}\label{Transit_timing_variation}

\begin{figure}
\centering
\includegraphics[width=0.49\textwidth]{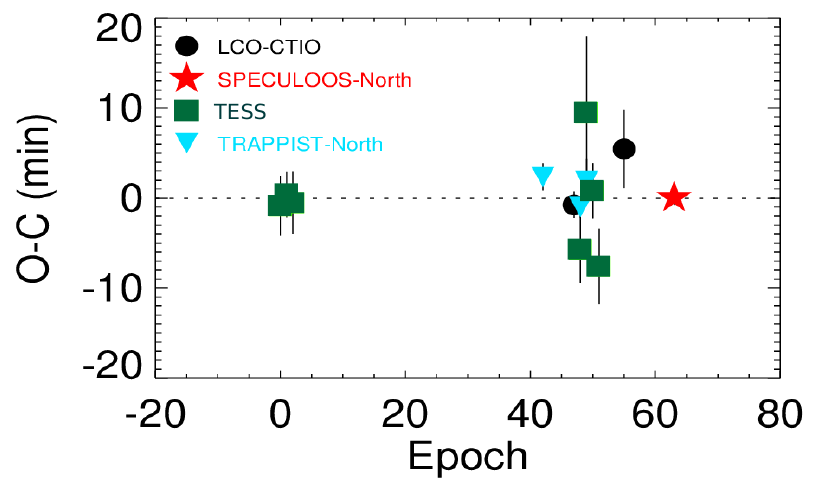}
\caption{The Transit Timing Variations of \hbox{TOI-2136}. Each symbol is a different telescope, and they are plotted as a function of epoch number. No significant TTV signal was detected.} 
\label{ttv}
\end{figure}

We search for the transit timing variations (TTVs) with all the photometric datasets (\tess, LCOGT, TRAPPIST-North, SPECULOOS-North) using \code{EXOFASTv2}. \code{EXOFASTv2} uses the Differential Evolution Markov chain Monte Carlo method to derive the values and their uncertainties of the stellar and planetary parameters of the system. It fits a linear ephemeris to the transit times and adds a penalty for the deviation of the step's linear ephemeris from the best-fit linear ephemeris of the transit times. For the TTV analysis of \hbox{TOI-2136\,b}, we fix the stellar parameters to the values as in Table \ref{starparam} and orbital parameters to the results obtained from the joint-fit performed. 
The results of the analysis showing the difference between the observed transit times and the calculated linear ephemeris from all the transits is presented in Figure \ref{ttv}. We find no evidence of a significant TTV signal in the current photometric data.

\subsection{Statistical validation}
Since the mass constraint on the planet has a significance slightly below $3\sigma$, we make use of the \code{TRICERATOPS} package \citep{Giacalone2021} to vet and statistically validate the planetary nature of \hbox{TOI-2136\,b}. \code{TRICERATOPS} is a Bayesian tool that takes host and nearby stars into consideration and calculate the probabilities of different transit-producing scenarios. The output false positive probability (FPP) value quantifies the possibility that the transit signal is not due to a planet around the host star. We first apply \code{TRICERATOPS} to the \tess\ light curve along with the contrast curve obtained by the `Alopeke speckle imaging (832 nm). The resulting FPP value 0.014 is close to the normal FPP threshold of 0.015 (1.5\%) to classify a validated planet \citep{Giacalone2021}. \cite{Wells2021} found that ground light curves sometimes put a better photometric constraint than the \tess\ data. We thus rerun the pipeline using the same contrast curve but the SPECULOOS-North/Artemis time-series, which yields a FPP value of $4\times10^{-3}$. Therefore, we consider this TOI to be a validated planet.

\section{Discussion}\label{discussion}

\subsection{Composition of \tar\,b}

\begin{figure}
\centering
\includegraphics[width=0.49\textwidth]{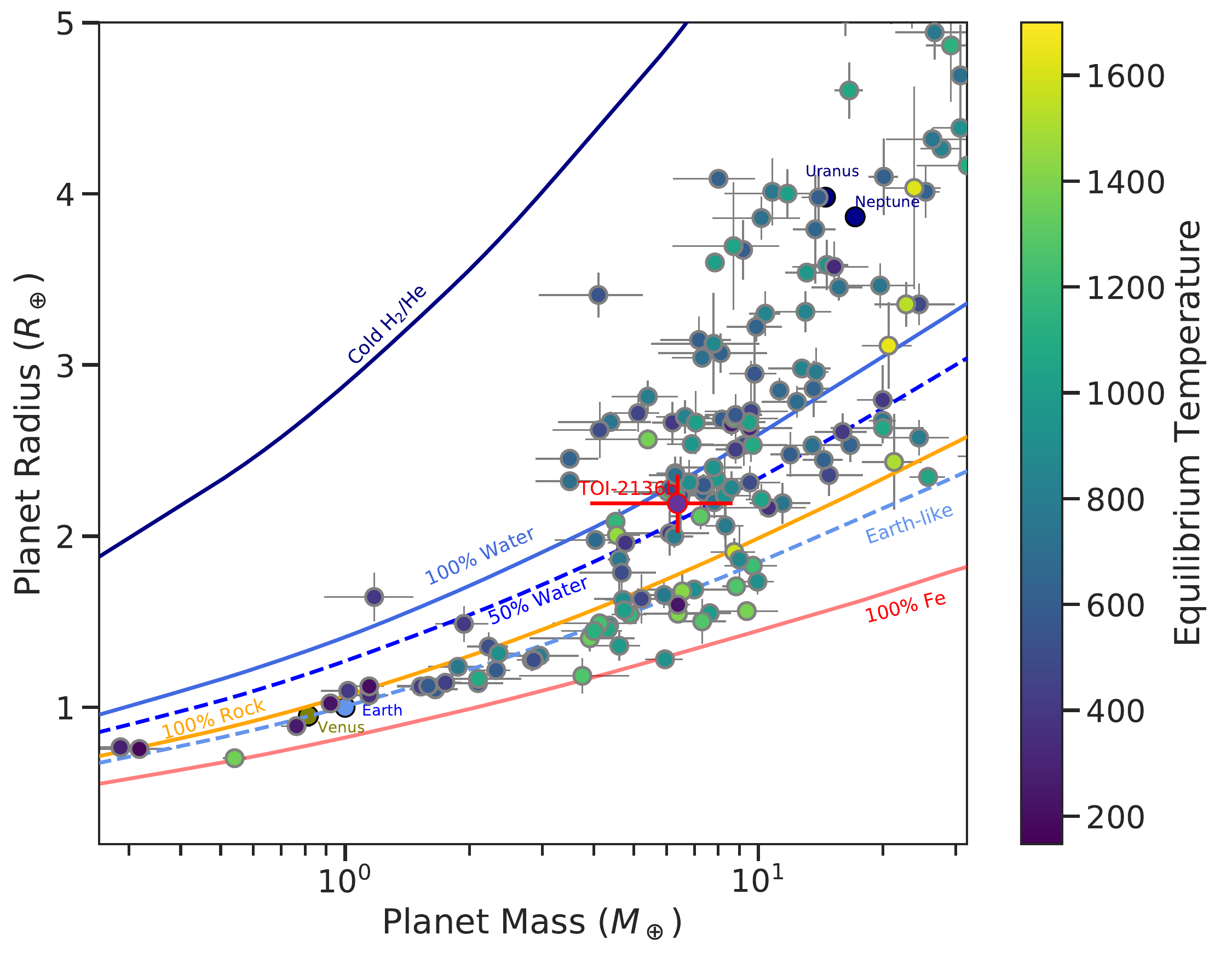}
\caption{Mass-radius curves with planets color-coded by their surface temperatures, indicating the potential bulk compositions of \hbox{TOI-2136\,b}. Data are taken from the TEPCat database of well-characterized planets. Theoretical models for the planet’s internal composition are taken from \citet{Zeng:2016}.}
\label{MR}
\end{figure}

We use the radius constraint on \hbox{TOI-2136\,b} derived from the transit photometry and the measured mass from the SPIRou RV data to investigate the location of this planet in the mass-radius diagram. Figure \ref{MR} shows the mass and radius distribution of a sample of well-characterized planets with the precisions on both measurements better than $30\%$ taken from the TEPcat database \citep{Southworth:2011}. The composition curves are retrieved from \cite{Zeng:2016}. The mass and radius of \hbox{TOI-2136\,b} are compared to the two-layer internal structure models of \citep{Zeng:2016}. As can be seen from Figure \ref{MR}, \hbox{TOI-2136\,b} appears to have a composition consistent with a pure water-ice world or a rocky planet with moderate atmosphere. 

We further investigate the composition of \hbox{TOI-2136\,b} using the Exoplanet Composition Interpolator\footnote{\url{https://tools.emac.gsfc.nasa.gov/ECI/}}. The algorithm takes the planet evolution models proposed by \cite{Lopez2014} and interpolates between the grid of these pre-computed models to explore the interiors and compositions of the planets. Taking the planet mass, radius, insolation flux as well as stellar age as inputs, we find the rocky core and gaseous envelope of \hbox{TOI-2136\,b} have mass fractions of $98.7^{+1.0}_{-1.5}\%$ and $1.3^{+1.5}_{-1.0}\%$, respectively. 

\subsection{\hbox{\tar\,b} and radius valley}

The bimodality of radius distribution shown in small planets around FGK stars, which splits them into super-Earths and sub-Neptunes, is known as a transition between planets with and without extended gaseous envelopes \citep{Fulton2017,Fulton2018gap}. \cite{Martinez2019} found that this transition radius is orbital period dependent, following a power law of $r_{\rm p,valley}\propto P^{-0.11}$. This finding approximately agrees with the prediction from the thermally driven atmospheric mass-loss scenarios including photoevaporation and core-powered envelope escape ($r_{\rm p, valley}\propto P^{-0.15}$; \citealt{Lopez2018}). However, observational results from \cite{Cloutier2020} suggested that the transition radius of small planets around low mass stars is likely in accordance with the gas poor formation model \citep{Lee2014,Lee2016}, following $r_{\rm p, valley}\propto P^{0.11}$. For early type M dwarfs with mass around $0.64\ M_{\odot}$, \cite{Cloutier2020_1235b} found that the thermally driven atmospheric mass-loss scenario remains efficient at sculpting their close-in planets. Recent work from \cite{Luque2021}, instead, tentatively reached a different conclusion. They proposed that the planetary radius valley for stars within a mass range between $0.54$ and $0.64\ M_{\odot}$ probably results from gas poor formation. The relative dominance of these two kinds of competing physical processes at the low stellar mass end remains unclear. Thus, populating the number of small planets with known bulk composition is crucial to solve the puzzles. 

\begin{figure}
\centering
\includegraphics[width=0.49\textwidth]{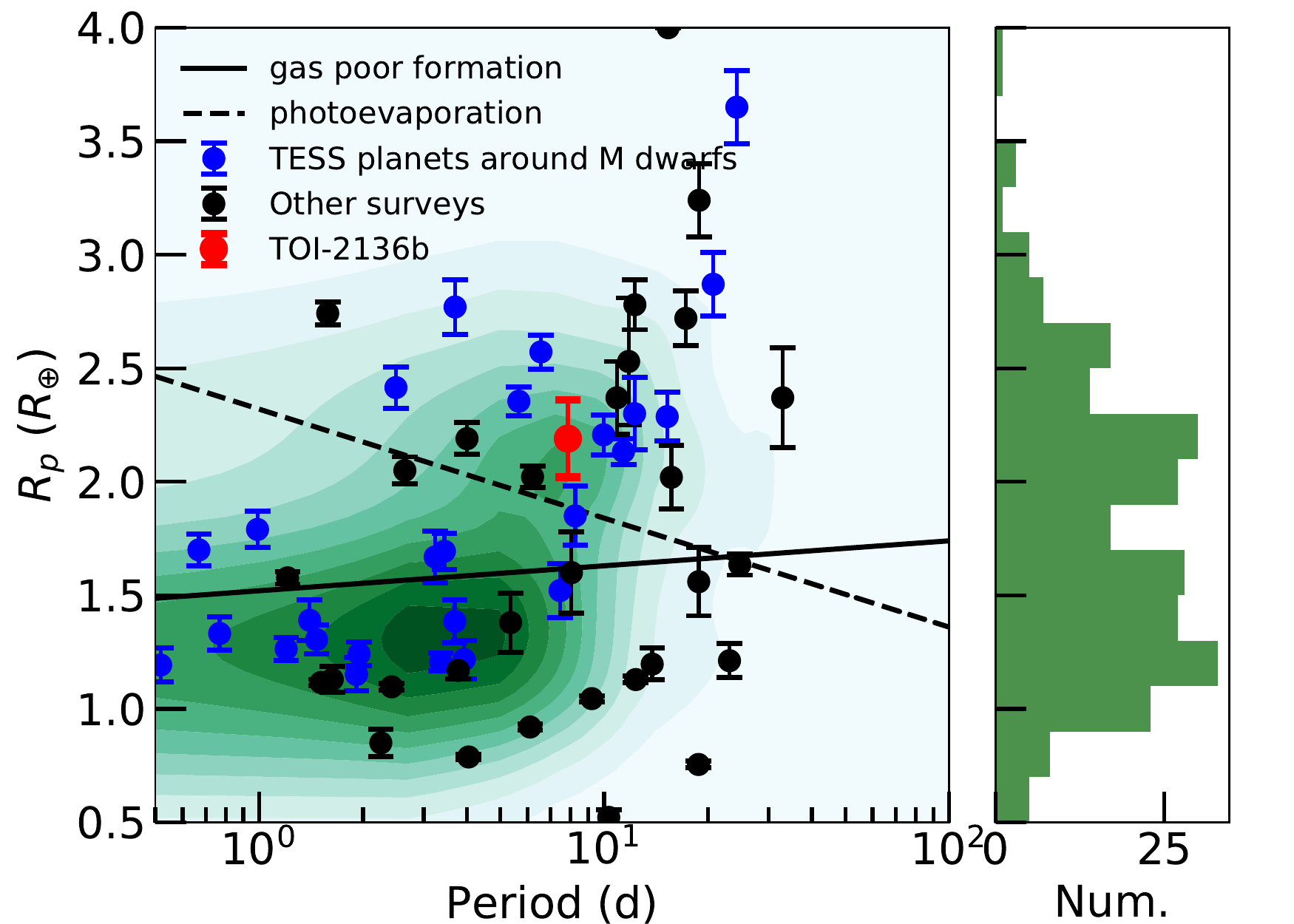}
\caption{The planet radius and orbital period diagram of all confirmed small planets hosted by low mass stars ($M_{\ast}\lesssim0.65\ M_{\odot}$). The green contours are the density distribution of planets without mass measurements. The 1d radius distribution is shown on the right. The colored points are the planets with mass constraint from TTV or RV. Specially, the blue dots are planets detected by the TESS mission. The solid and dashed lines depict the locations of radius valley for low mass stars predicted by the gas poor and photoevaporation models, taken from \citet{Cloutier2020}. \hbox{TOI-2136\,b} is marked as a red dot.} 
\label{PR}
\end{figure}

Figure \ref{PR} shows the orbital period and radius of planets with mass determination around M dwarfs ($M_{\ast}\lesssim0.65\ M_{\odot}$). We can see that \tess\ has doubled the number of small planets with known density, making them important for further investigating the strength of the two aforementioned envelope escape physical processes. With a period of $P_{b}=7.85$ days and a radius of $R_{p}=2.19\pm0.17\ R_{\oplus}$, \hbox{TOI-2136\,b} is located slightly above the radius valley for low mass dwarfs predicted by the thermally driven atmospheric mass loss model (See Figure \ref{PR}). Theoretical studies infer that \hbox{TOI-2136\,b} should be predominantly gaseous. Indeed, our previous analysis shows that \hbox{TOI-2136\,b} likely retains a H/He envelope with a small mass fraction. Given an estimated stellar age of $4.6\pm1.0$ Gyr, \tar\ has, in principle, finished the photoevaporation stage, which has a timescale of hundreds of Myrs \citep{Owen2013,Owen2017}. However, it is possible that \tar\ is still undergoing the mass loss process following the core powered mechanism that has a Gyr timescale \citep{Ginzburg2018}. 

Another interesting question is the behaviour of the radius valley as a function of stellar mass. Both competing physical processes predict a positive correlation between the center of radius valley and stellar mass, although different models show a difference in the slope at each mass bin \citep{Lopez2018,Gupta2019,Wu2019}. Consequently, comparing the theoretical predictions with the observational findings may rule out certain models. \cite{Cloutier2020} obtained a similar positive trend using a sub-sample of \kepler\ and {\it K2} planets. However, the sample size is small, especially for planets around mid-to-late M dwarfs, leading to a relatively large statistical uncertainty. Nevertheless, \hbox{TOI-2136\,b} joins the small but growing sample of planets around mid-M dwarfs that may help understand evolution of the transition radius with stellar mass in the future.

\subsection{Prospects for future observations}
\begin{figure}
\centering
\includegraphics[width=0.49\textwidth]{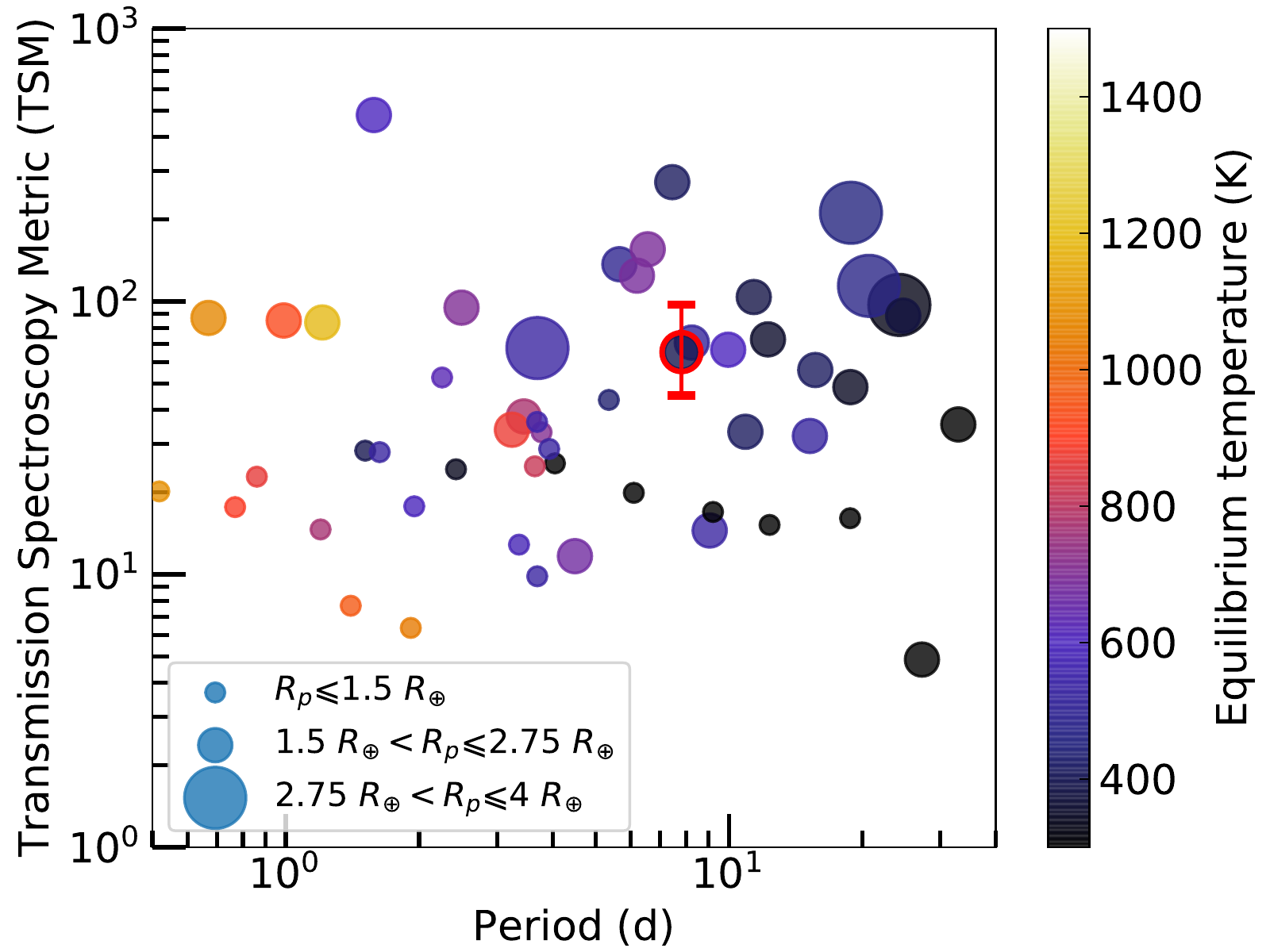}
\caption{The transmission spectroscopy metric as a function of orbital period for small planets around low mass stars ($M_{\ast}\lesssim0.65\ M_{\odot}$), colored by the planet equilibrium temperature. \hbox{TOI-2136\,b} is shown as a dot surrounded by a red circle with error bars. The size of each point is proportional to the planet radius.} 
\label{tsm}
\end{figure}

Given the proximity, small size and brightness in the near infrared, \tar\ is a promising star for atmospheric studies of its planet.  Following the criteria proposed in \cite{Kempton2018}, we compute the Transmission Spectroscopy Metric (TSM) of \hbox{TOI-2136\,b} to examine its potential opportunities for atmospheric characterization with the {\it James Webb Space Telescope} (\jwst, \citealt{Gardner2006}). We derive a TSM of $65^{+20}_{-32}$ for \hbox{TOI-2136\,b}. We compare the TSM factor of \hbox{TOI-2136\,b} with other small planets ($R_{p}\leq4\ R_{\oplus}$) harbored around low mass stars ($M_{\ast}\leq0.65\ M_{\odot}$) with mass measurements from RVs or TTVs in Figure \ref{tsm}. \cite{Kempton2018} quantified $\rm TSM=90$ as a recommended threshold for planets with $1.5<R_{p}<10\ R_{\oplus}$ to be high-quality atmospheric characterization targets. Thus, \hbox{TOI-2136\,b} is located close to the first rank of targets with a relatively low equilibrium temperature $T_{\rm eq}$. In addition, we also estimate the signal amplitude of \hbox{TOI-2136\,b} in the transit transmission spectroscopy following the approach described in \cite{Gillon2016}:
\begin{equation}
    {\it S} = \frac{2R_{\rm p}h_{\rm eff}}{R_{\ast}^{2}},
\end{equation}
where $R_{p}$ and $R_{\ast}$ are the planet and its host star radius, $h_{\rm eff}$ is the effective atmospheric height. We calculate the signal amplitude under the typical case that $h_{\rm eff}/H=7$, where $H=kT/\mu g$ is the atmospheric scale height. We find a $\it S$ of $382\pm196$ ppm, assuming a bond albedo of 0 and a mean molecular mass $\mu$ of 2.3 amu for sub-Neptunes \citep{Demory2020}. The large uncertainty mainly comes from the loose constraint on the planet mass. \cite{Schlawin2020} reported a noise floor level 10 ppm for \jwst\ for NIRSpec ($ \lambda= 5.0 - 11\ \mu {\rm m}$). Supposing the lower limit on $\it S$ measurement of \hbox{TOI-2136\,b}, it would be between 18.6 times and the higher limit, 57.8 times the 10 ppm uncertainty. Taking two aspects into consideration, we suggest that \hbox{TOI-2136\,b} is an exciting target for further atmospheric researches. A number of studies on the diversity of sub-Neptunian atmospheres have already been made \citep[e.g.,][]{Lavvas2019,Chouqar2020}. 

As noted above, the mean molecular mass $\mu$ is degenerated with the surface gravity of the planet (i.e., planet mass $M_{p}$). Thus, a well-measured planet mass is required to fully understand the compositions of the planet atmosphere. Otherwise, the accuracy and precision of the retrieved atmospheric parameters will be largely limited \citep{Batalha2019}. Since the current SPIRou RVs only provide a $2\sigma$ mass constraint and \tar\ is a quiet M dwarf without strong stellar activity, future subsequent spectroscopy observations are encouraged to determine the planet mass at the $3\sigma$ confidence level and look for other potential non-transiting planets. Due to the faintness of \tar\ (Vmag=14.3), it challenges most optical spectroscopy instruments on the ground. However, it is still accessible by NIR facilities like InfrRed Doppler spectrograph (IRD; \citealt{Kotani2018}) and Habitable-zone Planet Finder (HPF; \citealt{Mahadevan2014}) or red-optical spectrographs on large telescopes like MAROON-X \citep{Seifahrt2018}, which is dedicated to conducting RV measurements for mid-to-late M dwarfs. 

\subsection{Detection limits}
\begin{figure}
\centering
\includegraphics[width=0.49\textwidth]{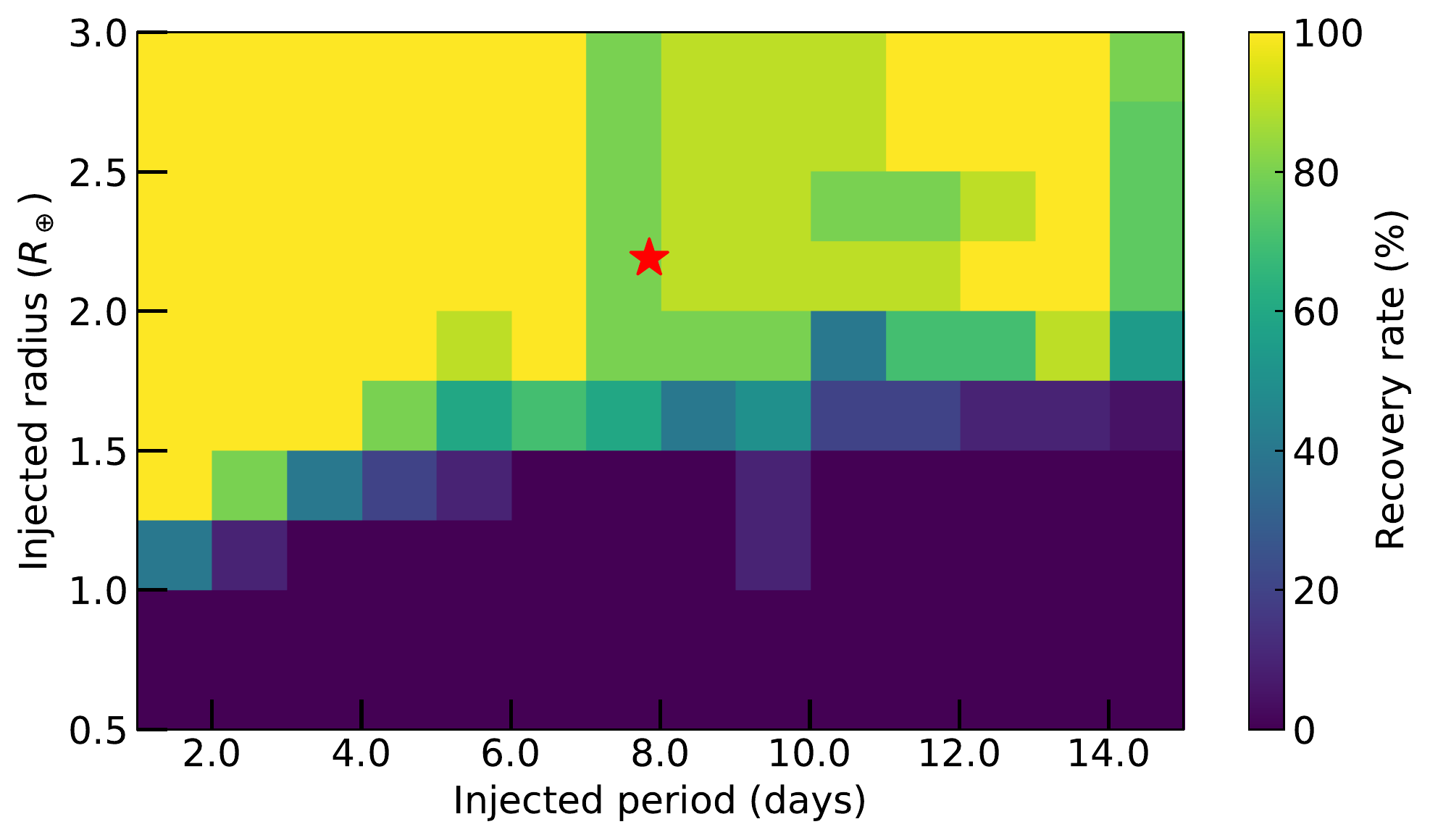}
\caption{The results of injection-and-recovery test on the \tess\ light curve of \tar. We search the $P$-$R_{p}$ space with 10 random generated mid-transit times for each grid and we explore a total of 1680 scenarios. Different colors represent different recovery rates. The yellow and green regions are the planetary parameter space with high recovery rate while the planets located in the dark regions may be missed. The red star marks the position of \hbox{TOI-2136\,b}.} 
\label{inj_recov}
\end{figure}

Based on the results from the \kepler\ survey, \cite{Muirhead2015} found that $21^{+7}_{-5}\%$ of mid-M dwarf stars like \tar\ host compact multiple planets with periods all shorter than 10 days. This rate is not very different from that of early-type M dwarfs but much higher than solar-like stars. Therefore, we perform an injection-and-recovery test using \code{MATRIX ToolKit}\footnote{\url{https://github.com/PlanetHunters/tkmatrix}} \citep{Pozuelos2020,Demory2020} to explore the detection limits of the current \tess\ data and determine the type of planets we perhaps miss. We make use of all available PDC-SAP light curves of \tar\ after removing the known transits of \hbox{TOI-2136\,b}. We explore a period-radius space of $1\sim15\ {\rm days}$ and $0.5\sim3.0\ R_{\oplus}$ with step sizes of 1 day and 0.25 $R_{\oplus}$. During the injection, we assume that the synthetic ``planet'' has an inclination of $i=90^{\circ}$ on a circular orbit, and randomly generate ten light curves with different $T_{0}$ for each grid. We thus examine a total of 1680 scenarios. For each light curve, we use a biweight filter with a window size of 0.5 day to remove the systematic trends. \code{MATRIX ToolKit} defines a successful recovery if the the detected period is within 5\% of the injected period and the transit duration is within 1 hr when compared with the set value. Figure \ref{inj_recov} depicts our test results. We find that: (1) most planets smaller than super Earths ($R_{p}\lesssim1.5\ R_{\oplus}$) across the period range we searched are likely to be still buried in the light curve and remain undetectable \citep{Brady2021}; (2) planets that have $R_{p}\gtrsim2.0\ R_{\oplus}$ with periods up to 15 days can be ruled out with a recovery rate $\geq 80\%$. Since the current \tess\ dataset is insensitive to planets with period larger than 15 days, here we note that future \tess\ observations to be done in Sector 53 and 54 between 13$^{\rm th}$ Jun. 2022 and 5$^{\rm th}$ Aug. 2022 during the Extended Mission would help better understand the architecture of this system. 

\section{Conclusion}\label{conclusion}
In this study, we report the discovery and characterization of the \tar\ system, a sub-Neptune around a faint M4.5 dwarf ($V=14.1$ mag), detected by the \tess\ mission. We confirm the planetary nature of \hbox{TOI-2136\,b} through a combination of 2-min cadence \tess\ observations, ground-based photometry, high angular resolution imaging and SPIRou spectroscopic observations. The transit and RV joint-fit model gives a planet radius of $R_{p}=2.19^{+0.17}_{-0.17}\ R_{\oplus}$, a mass of $M_{p}=6.37^{+2.45}_{-2.29}\ M_{\oplus}$ and an equilibrium temperature of $T_{\rm eq}=395^{+24}_{-22}$ K. The bulk density $\rho_{p}=3.34^{+2.55}_{-1.63}\ g\ cm^{-3}$ of \hbox{TOI-2136\,b} is consistent with a water world or a rocky planet with moderate atmosphere. Planetary structure models of \hbox{TOI-2136\,b} suggests that it may contains a rocky core with a H/He envelope with a mass fraction of $1.3^{+1.5}_{-1.0}\%$. Given the period and radius of \hbox{TOI-2136\,b}, it falls close to the location of radius valley predicted by the the thermally driven envelope escape model for M dwarfs, making it a great laboratory to investigate the formation and evolution models of small planets around low-mass stars. The small size and quiet nature of the host star as well as its brightness in the NIR make \hbox{TOI-2136\,b} amenable to be further observed by most JWST modes for studying atmospheric compositions. 

\section*{Affiliations}
$^{1}$Department of Astronomy, Tsinghua University, Beijing 100084, People's Republic of China\\
$^{2}$Oukaimeden Observatory, High Energy Physics and Astrophysics
Laboratory, Cadi Ayyad University, Marrakech, Morocco\\
$^{3}$Departamento de Fisica e Astronomia, Faculdade de Ciencias, Universidade do Porto, Rua do Campo Alegre, 4169-007 porto, Portugal\\
$^{4}$Instituto de Astrofisica e Ciencias do Espaco, Universidade do porto, CAUP, Rua das Estrelas, 150-762 Porto, Portugal\\
$^{5}$National Astronomical Observatories, Chinese Academy of Sciences, 20A Datun Road, Chaoyang District, Beijing 100012, People's Republic of China\\
$^{6}$Universit\'e de Montr\'eal, D\'epartement de Physique, IREX, Montr\'eal, QC H3C 3J7, Canada\\
$^{7}$Observatoire du Mont-M\'egantic, Universit\'e de Montr\'eal, Montr\'eal, QC H3C 3J7, Canada\\
$^{8}$Canada-France-Hawaii Telescope, CNRS, Kamuela, HI 96743, USA\\
$^{9}$Univ. de Toulouse, CNRS, IRAP, 14 Avenue Belin, 31400 Toulouse, France\\
$^{10}$Department of Astronomy, University of California Berkeley, Berkeley, CA 94720, USA\\
$^{11}$Center for Astrophysics and Space Sciences, University of California, San Diego, 9500 Gilman Dr, La Jolla, CA 92093, USA\\
$^{12}$Center for Astrophysics \textbar \ Harvard \& Smithsonian, 60 Garden Street, Cambridge, MA 02138, USA\\
$^{13}$Department of Physics and Kavli Institute for Astrophysics and Space Research, Massachusetts Institute of Technology, Cambridge, MA 02139, USA\\
$^{14}$Astrobiology Research Unit, Universit\'e de Li\`ege, All\'ee du 6 Ao\^ut 19C, B-4000 Li\`ege, Belgium\\
$^{15}$Department of Earth, Atmospheric and Planetary Science, Massachusetts Institute of Technology, 77 Massachusetts Avenue, Cambridge, MA 02139, USA\\
$^{16}$Instituto de Astrof\'isica de Canarias (IAC), Calle V\'ia L\'actea s/n, 38200, La Laguna, Tenerife, Spain\\
$^{17}$NASA Ames Research Center, Moffett Field, CA 94035, USA\\
$^{18}$George Mason University, 4400 University Drive, Fairfax, VA 22030, USA\\
$^{19}$Space Sciences, Technologies and Astrophysics Research (STAR) Institute, Universit\' de Li\`ege, All\'ee du 6 Ao\^ut 19C, B-4000 Li\`ege, Belgium\\
$^{20}$Center for Space and Habitability, University of Bern, Gesellschaftsstrasse 6, CH-3012, Bern, Switzerland\\
$^{21}$School of Physics \& Astronomy, University of Birmingham, Edgbaston, Birmimgham B15 2TT, UK\\
$^{22}$CEA, Universit\'e Paris-Saclay, Universit\'e de Paris, F-91191 Gif-sur-Yvette, France\\
$^{23}$School of Aerospace Engineering, Tsinghua University, Beijing 100084, People’s Republic of China\\
$^{24}$NASA Exoplanet Science Institute, Caltech/IPAC, Mail Code 100-22, 1200 E. California Blvd., Pasadena, CA 91125, USA\\
$^{25}$Instituto de Astronom\'ia, Universidad Nacional Aut\'onoma de M\'exico, Ciudad Universitaria, Ciudad de M\'exico, 04510, M\'exico\\
$^{26}$European Space Research and Technology Centre (ESTEC), European Space Agency (ESA), Keplerlaan 1, 2201 AZ Noordwijk, The Netherlands\\
$^{27}$University of Southern Queensland, Centre for Astrophysics, West Street, Toowoomba, QLD 4350 Australia\\
$^{28}$Department of Physics \& Astronomy, Swarthmore College, Swarthmore PA 19081, USA\\
$^{29}$Department of Physics, Tsinghua University, Beijing 100084, People's Republic of China\\
$^{30}$Department of Physics, University of Warwick, Coventry, CV4 7AL, UK\\
$^{31}$Cavendish Laboratory, JJ Thomson Avenue, Cambridge, CB3 0H3, UK\\
$^{32}$ETH Zurich, Department of Physics, Wolfgang-Pauli-Strasse 2, CH-8093 Zurich, Switzerland\\
$^{33}$Department of Astronomy, University of Maryland, College Park, College Park, MD 20742 USA\\
$^{34}$Patashnick Voorheesville Observatory, Voorheesville, NY 12186, USA\\
$^{35}$Department of Aeronautics and Astronautics, MIT, 77 Massachusetts Avenue, Cambridge, MA 02139, USA\\
$^{36}$Department of Astrophysical Sciences, Princeton University, 4 Ivy Lane, Princeton, NJ 08544, USA\\

\section*{Acknowledgments}
We are grateful to Coel Hellier for the insights regarding the WASP data. We thank Ryan Cloutier for useful discussions. We also thank Elise Furlan for the contributions to the speckle data.
This work is partly supported by the National Science Foundation of China (Grant No. 11390372, 11761131004 and 12133005 to SM and TG).
A. Soubkiou is partly supported by a grant from the MOBILE 2 BE project, coordinated by the University of Porto in the framework of the European Programme Erasmus plus.
B.V.R. thanks the Heising-Simons Foundation for support.
YGMC is supported by UNAM-PAPIIT-IG101321. CXH's work is surpported by ARC DECRA Grant. 
This research uses data obtained through the China's Telescope Access Program (TAP), which has been funded by the TAP member institutes.
This publication benefits from the support of the French Community of Belgium in the context of the FRIA Doctoral Grant awarded to Mathilde Timmermans.
The research leading to these results has received funding from the ARC grant for Concerted Research Actions, financed by the Wallonia-Brussels Federation. TRAPPIST is funded by the Belgian Fund for Scientific Research (Fond National de la Recherche Scientifique, FNRS) under the grant PDR T.0120.21. TRAPPIST-North is a project funded by the University of Liege (Belgium), in collaboration with Cadi Ayyad University of Marrakech (Morocco) MG and EJ are F.R.S.-FNRS Senior Research Associate.
CL is supported by the National Science Foundation Graduate Research Fellowship under Grant No. DGE1745303.
Some of the observations in the paper made use of the High-Resolution Imaging instrument ‘Alopeke obtained under Gemini LLP Proposal Number: GN/S-2021A-LP-105. ‘Alopeke was funded by the NASA Exoplanet Exploration Program and built at the NASA Ames Research Center by Steve B. Howell, Nic Scott, Elliott P. Horch, and Emmett Quigley. Alopeke was mounted on the Gemini North (and/or South) telescope of the international Gemini Observatory, a program of NSF’s OIR Lab, which is managed by the Association of Universities for Research in Astronomy (AURA) under a cooperative agreement with the National Science Foundation. on behalf of the Gemini partnership: the National Science Foundation (United States), National Research Council (Canada), Agencia Nacional de Investigación y Desarrollo (Chile), Ministerio de Ciencia, Tecnología e Innovación (Argentina), Ministério da Ciência, Tecnologia, Inovações e Comunicações (Brazil), and Korea Astronomy and Space Science Institute (Republic of Korea).
This work makes use of observations from the LCOGT network. Part of the LCOGT telescope time was granted by NOIRLab through the Mid-Scale Innovations Program (MSIP). MSIP is funded by NSF.
The ULiege's contribution to SPECULOOS has received funding from the European Research Council under the European Union's Seventh Framework Programme (FP/2007-2013) (grant Agreement n$^\circ$ 336480/SPECULOOS), from the Balzan Prize Foundation, from the Belgian Scientific Research Foundation (F.R.S.-FNRS; grant n$^\circ$ T.0109.20), from the University of Liege, and from the ARC grant for Concerted Research Actions financed by the Wallonia-Brussels Federation. MG and EJ are F.R.S-FNRS Senior Research Associates. VVG is F.R.S-FNRS Research Associate. This work is supported by a grant from the Simons Foundation (PI Queloz, grant number 327127). J.d.W. and MIT gratefully acknowledge financial support from the Heising-Simons Foundation, Dr. and Mrs. Colin Masson and Dr. Peter A. Gilman for Artemis, the first telescope of the SPECULOOS network situated in Tenerife, Spain. MNG acknowledges support from the European Space Agency (ESA) as an ESA Research Fellow. This work is supported by the Swiss National Science Foundation (PP00P2-163967, PP00P2-190080 and the National Centre for Competence in Research PlanetS). This work has received fund from the European Research Council (ERC) under the European Union's Horizon 2020 research and innovation programme (grant agreement n$^\circ$ 803193/BEBOP), from the MERAC foundation, and from the Science and Technology Facilities Council (STFC; grant n$^\circ$ ST/S00193X/1).
The National Geographic Society - Palomar Observatory Sky Atlas (POSS-I) was made by the California Institute of Technology with grants from the National Geographic Society.
Funding for the TESS mission is provided by NASA's Science Mission directorate. 
We acknowledge the use of \tess\ public data from pipelines at the \tess\ Science Office and at the \tess\ Science Processing Operations Center. 
Resources supporting this work were provided by the NASA High-End Computing (HEC) Program through the NASA Advanced Supercomputing (NAS) Division at Ames Research Center for the production of the SPOC data products.
This research has made use of the Exoplanet Follow-up Observation Program website, which is operated by the California Institute of Technology, under contract with the National Aeronautics and Space Administration under the Exoplanet Exploration Program. 
This paper includes data collected by the \tess\ mission, which are publicly available from the Mikulski Archive for Space Telescopes\ (MAST). 
This work has made use of data from the European Space Agency (ESA) mission
{\it Gaia} (\url{https://www.cosmos.esa.int/gaia}), processed by the {\it Gaia} Data Processing and Analysis Consortium (DPAC,
\url{https://www.cosmos.esa.int/web/gaia/dpac/consortium}). Funding for the DPAC has been provided by national institutions, in particular the institutions participating in the {\it Gaia} Multilateral Agreement.
This work made use of \texttt{tpfplotter} by J. Lillo-Box (publicly available in www.github.com/jlillo/tpfplotter), which also made use of the python packages \texttt{astropy}, \texttt{lightkurve}, \texttt{matplotlib} and \texttt{numpy}.

\section*{Data Availability}
This paper includes photometric data collected by the \tess\ mission and ground instruments, which are publicly available in ExoFOP, at \url{https://exofop.ipac.caltech.edu/tess/target.php?id=336128819}. All spectroscopy data underlying this article are listed in the appendix. All of the high-resolution speckle imaging data is available at the NASA exoplanet Archive with no proprietary period.



\bibliographystyle{mnras}
\bibliography{planet} 

\begin{thebibliography}{}
\makeatletter
\relax
\def\mn@urlcharsother{\let\do\@makeother \do\$\do\&\do\#\do\^\do\_\do\%\do\~}
\def\mn@doi{\begingroup\mn@urlcharsother \@ifnextchar [ {\mn@doi@}
  {\mn@doi@[]}}
\def\mn@doi@[#1]#2{\def\@tempa{#1}\ifx\@tempa\@empty \href
  {http://dx.doi.org/#2} {doi:#2}\else \href {http://dx.doi.org/#2} {#1}\fi
  \endgroup}
\def\mn@eprint#1#2{\mn@eprint@#1:#2::\@nil}
\def\mn@eprint@arXiv#1{\href {http://arxiv.org/abs/#1} {{\tt arXiv:#1}}}
\def\mn@eprint@dblp#1{\href {http://dblp.uni-trier.de/rec/bibtex/#1.xml}
  {dblp:#1}}
\def\mn@eprint@#1:#2:#3:#4\@nil{\def\@tempa {#1}\def\@tempb {#2}\def\@tempc
  {#3}\ifx \@tempc \@empty \let \@tempc \@tempb \let \@tempb \@tempa \fi \ifx
  \@tempb \@empty \def\@tempb {arXiv}\fi \@ifundefined
  {mn@eprint@\@tempb}{\@tempb:\@tempc}{\expandafter \expandafter \csname
  mn@eprint@\@tempb\endcsname \expandafter{\@tempc}}}

\bibitem[\protect\citeauthoryear{{Agol} et~al.,}{{Agol}
  et~al.}{2021}]{Agol2021}
{Agol} E.,  et~al., 2021, \mn@doi [Planetary Science Journal]
  {10.3847/PSJ/abd022}, \href
  {https://ui.adsabs.harvard.edu/abs/2021PSJ.....2....1A} {2, 1}

\bibitem[\protect\citeauthoryear{{Aller}, {Lillo-Box}, {Jones}, {Miranda}  \&
  {Barcel{\'o} Forteza}}{{Aller} et~al.}{2020}]{Aller2020}
{Aller} A.,  {Lillo-Box} J.,  {Jones} D.,  {Miranda} L.~F.,   {Barcel{\'o}
  Forteza} S.,  2020, \mn@doi [\aap] {10.1051/0004-6361/201937118}, \href
  {https://ui.adsabs.harvard.edu/abs/2020A&A...635A.128A} {635, A128}

\bibitem[\protect\citeauthoryear{{Artigau} et~al.,}{{Artigau}
  et~al.}{2014}]{Artigau2014PCA}
{Artigau} {\'E}.,  et~al., 2014, in {Peck} A.~B.,  {Benn} C.~R.,   {Seaman}
  R.~L.,  eds,  Society of Photo-Optical Instrumentation Engineers (SPIE)
  Conference Series Vol. 9149, Observatory Operations: Strategies, Processes,
  and Systems V. p. 914905 (\mn@eprint {arXiv} {1406.6927}),
  \mn@doi{10.1117/12.2056385}

\bibitem[\protect\citeauthoryear{{Artigau} et~al.,}{{Artigau}
  et~al.}{2021}]{Artigau2021}
{Artigau} {\'E}.,  et~al., 2021, \mn@doi [\aj] {10.3847/1538-3881/ac096d},
  \href {https://ui.adsabs.harvard.edu/abs/2021AJ....162..144A} {162, 144}

\bibitem[\protect\citeauthoryear{{Baranec} et~al.,}{{Baranec}
  et~al.}{2014}]{Baranec2014}
{Baranec} C.,  et~al., 2014, \mn@doi [\apjl] {10.1088/2041-8205/790/1/L8},
  \href {https://ui.adsabs.harvard.edu/abs/2014ApJ...790L...8B} {790, L8}

\bibitem[\protect\citeauthoryear{{Barkaoui} et~al.,}{{Barkaoui}
  et~al.}{2019}]{Barkaoui:2019}
{Barkaoui} K.,  et~al., 2019, \mn@doi [\aj] {10.3847/1538-3881/aaf422}, \href
  {https://ui.adsabs.harvard.edu/abs/2019AJ....157...43B} {157, 43}

\bibitem[\protect\citeauthoryear{{Batalha}, {Lewis}, {Fortney}, {Batalha},
  {Kempton}, {Lewis}  \& {Line}}{{Batalha} et~al.}{2019}]{Batalha2019}
{Batalha} N.~E.,  {Lewis} T.,  {Fortney} J.~J.,  {Batalha} N.~M.,  {Kempton}
  E.,  {Lewis} N.~K.,   {Line} M.~R.,  2019, \mn@doi [\apjl]
  {10.3847/2041-8213/ab4909}, \href
  {https://ui.adsabs.harvard.edu/abs/2019ApJ...885L..25B} {885, L25}

\bibitem[\protect\citeauthoryear{{Benedict} et~al.,}{{Benedict}
  et~al.}{2016}]{Benedict2016}
{Benedict} G.~F.,  et~al., 2016, \mn@doi [\aj] {10.3847/0004-6256/152/5/141},
  \href {http://adsabs.harvard.edu/abs/2016AJ....152..141B} {152, 141}

\bibitem[\protect\citeauthoryear{{Bensby}, {Feltzing}  \&
  {Lundstr{\"o}m}}{{Bensby} et~al.}{2003}]{Bensby2003}
{Bensby} T.,  {Feltzing} S.,   {Lundstr{\"o}m} I.,  2003, \mn@doi [\aap]
  {10.1051/0004-6361:20031213}, \href
  {https://ui.adsabs.harvard.edu/abs/2003A%26A...410..527B} {410, 527}

\bibitem[\protect\citeauthoryear{{Bensby}, {Feltzing}  \& {Oey}}{{Bensby}
  et~al.}{2014}]{Bensby2014}
{Bensby} T.,  {Feltzing} S.,   {Oey} M.~S.,  2014, \mn@doi [\aap]
  {10.1051/0004-6361/201322631}, \href
  {http://adsabs.harvard.edu/abs/2014A%26A...562A..71B} {562, A71}

\bibitem[\protect\citeauthoryear{{Borucki} et~al.,}{{Borucki}
  et~al.}{2010}]{Borucki2010}
{Borucki} W.~J.,  et~al., 2010, \mn@doi [Science] {10.1126/science.1185402},
  \href {http://adsabs.harvard.edu/abs/2010Sci...327..977B} {327, 977}

\bibitem[\protect\citeauthoryear{Bouchy, Pepe  \& Queloz}{Bouchy
  et~al.}{2001}]{bouchy_fundamental_2001}
Bouchy F.,  Pepe F.,   Queloz D.,  2001, \mn@doi [Astronomy \& Astrophysics]
  {10.1051/0004-6361:20010730}, 374, 733

\bibitem[\protect\citeauthoryear{{Bovy}}{{Bovy}}{2015}]{Bovy2015}
{Bovy} J.,  2015, \mn@doi [\apjs] {10.1088/0067-0049/216/2/29}, \href
  {http://adsabs.harvard.edu/abs/2015ApJS..216...29B} {216, 29}

\bibitem[\protect\citeauthoryear{{Boyajian}, {van Belle}  \& {von
  Braun}}{{Boyajian} et~al.}{2014}]{Boyajian2014}
{Boyajian} T.~S.,  {van Belle} G.,   {von Braun} K.,  2014, \mn@doi [\aj]
  {10.1088/0004-6256/147/3/47}, \href
  {https://ui.adsabs.harvard.edu/abs/2014AJ....147...47B} {147, 47}

\bibitem[\protect\citeauthoryear{{Brady} \& {Bean}}{{Brady} \&
  {Bean}}{2021}]{Brady2021}
{Brady} M.,  {Bean} J.,  2021, arXiv e-prints, \href
  {https://ui.adsabs.harvard.edu/abs/2021arXiv211208337B} {p. arXiv:2112.08337}

\bibitem[\protect\citeauthoryear{{Brown} et~al.,}{{Brown}
  et~al.}{2013}]{Brown2013}
{Brown} T.~M.,  et~al., 2013, \mn@doi [\pasp] {10.1086/673168}, \href
  {http://adsabs.harvard.edu/abs/2013PASP..125.1031B} {125, 1031}

\bibitem[\protect\citeauthoryear{{Cersullo}, {Wildi}, {Chazelas}  \&
  {Pepe}}{{Cersullo} et~al.}{2017}]{Cersullo2017}
{Cersullo} F.,  {Wildi} F.,  {Chazelas} B.,   {Pepe} F.,  2017, \mn@doi [\aap]
  {10.1051/0004-6361/201629972}, \href
  {https://ui.adsabs.harvard.edu/abs/2017A&A...601A.102C} {601, A102}

\bibitem[\protect\citeauthoryear{{Charbonneau} et~al.,}{{Charbonneau}
  et~al.}{2009}]{Charbonneau2009}
{Charbonneau} D.,  et~al., 2009, \mn@doi [\nat] {10.1038/nature08679}, \href
  {https://ui.adsabs.harvard.edu/abs/2009Natur.462..891C} {462, 891}

\bibitem[\protect\citeauthoryear{{Chen} \& {Kipping}}{{Chen} \&
  {Kipping}}{2017}]{Chen2017}
{Chen} J.,  {Kipping} D.,  2017, \mn@doi [\apj] {10.3847/1538-4357/834/1/17},
  \href {https://ui.adsabs.harvard.edu/abs/2017ApJ...834...17C} {834, 17}

\bibitem[\protect\citeauthoryear{{Chen} \& {Rogers}}{{Chen} \&
  {Rogers}}{2016}]{Chen2016}
{Chen} H.,  {Rogers} L.~A.,  2016, \mn@doi [\apj]
  {10.3847/0004-637X/831/2/180}, \href
  {https://ui.adsabs.harvard.edu/abs/2016ApJ...831..180C} {831, 180}

\bibitem[\protect\citeauthoryear{{Choi}, {Dotter}, {Conroy}, {Cantiello},
  {Paxton}  \& {Johnson}}{{Choi} et~al.}{2016}]{choi:2016}
{Choi} J.,  {Dotter} A.,  {Conroy} C.,  {Cantiello} M.,  {Paxton} B.,
  {Johnson} B.~D.,  2016, \mn@doi [\apj] {10.3847/0004-637X/823/2/102}, \href
  {https://ui.adsabs.harvard.edu/abs/2016ApJ...823..102C} {823, 102}

\bibitem[\protect\citeauthoryear{{Chouqar}, {Benkhaldoun}, {Jabiri},
  {Lustig-Yaeger}, {Soubkiou}  \& {Szentgyorgyi}}{{Chouqar}
  et~al.}{2020}]{Chouqar2020}
{Chouqar} J.,  {Benkhaldoun} Z.,  {Jabiri} A.,  {Lustig-Yaeger} J.,  {Soubkiou}
  A.,   {Szentgyorgyi} A.,  2020, \mn@doi [\mnras] {10.1093/mnras/staa1198},
  \href {https://ui.adsabs.harvard.edu/abs/2020MNRAS.495..962C} {495, 962}

\bibitem[\protect\citeauthoryear{{Ciardi}, {Beichman}, {Horch}  \&
  {Howell}}{{Ciardi} et~al.}{2015}]{Ciardi2015}
{Ciardi} D.~R.,  {Beichman} C.~A.,  {Horch} E.~P.,   {Howell} S.~B.,  2015,
  \mn@doi [\apj] {10.1088/0004-637X/805/1/16}, \href
  {https://ui.adsabs.harvard.edu/abs/2015ApJ...805...16C} {805, 16}

\bibitem[\protect\citeauthoryear{{Cloutier} \& {Menou}}{{Cloutier} \&
  {Menou}}{2020}]{Cloutier2020}
{Cloutier} R.,  {Menou} K.,  2020, \mn@doi [\aj] {10.3847/1538-3881/ab8237},
  \href {https://ui.adsabs.harvard.edu/abs/2020AJ....159..211C} {159, 211}

\bibitem[\protect\citeauthoryear{{Cloutier} et~al.,}{{Cloutier}
  et~al.}{2020}]{Cloutier2020_1235b}
{Cloutier} R.,  et~al., 2020, \mn@doi [\aj] {10.3847/1538-3881/ab9534}, \href
  {https://ui.adsabs.harvard.edu/abs/2020AJ....160...22C} {160, 22}

\bibitem[\protect\citeauthoryear{{Collins}, {Kielkopf}, {Stassun}  \&
  {Hessman}}{{Collins} et~al.}{2017}]{Collins2017}
{Collins} K.~A.,  {Kielkopf} J.~F.,  {Stassun} K.~G.,   {Hessman} F.~V.,  2017,
  \mn@doi [\aj] {10.3847/1538-3881/153/2/77}, \href
  {http://adsabs.harvard.edu/abs/2017AJ....153...77C} {153, 77}

\bibitem[\protect\citeauthoryear{Cristofari et~al.,}{Cristofari
  et~al.}{2021}]{cristofari_estimating_2021}
Cristofari P.~I.,  et~al., 2021, \mn@doi [Monthly Notices of the Royal
  Astronomical Society] {10.1093/mnras/stab3679}, p. stab3679

\bibitem[\protect\citeauthoryear{{Cushing}, {Vacca}  \& {Rayner}}{{Cushing}
  et~al.}{2004}]{Cushing2004}
{Cushing} M.~C.,  {Vacca} W.~D.,   {Rayner} J.~T.,  2004, \mn@doi [\pasp]
  {10.1086/382907}, \href
  {https://ui.adsabs.harvard.edu/abs/2004PASP..116..362C} {116, 362}

\bibitem[\protect\citeauthoryear{{Cushing}, {Rayner}  \& {Vacca}}{{Cushing}
  et~al.}{2005}]{Cushing2005}
{Cushing} M.~C.,  {Rayner} J.~T.,   {Vacca} W.~D.,  2005, \mn@doi [\apj]
  {10.1086/428040}, \href
  {https://ui.adsabs.harvard.edu/abs/2005ApJ...623.1115C} {623, 1115}

\bibitem[\protect\citeauthoryear{{Cutri} et~al.,}{{Cutri}
  et~al.}{2003}]{Cutri:2003}
{Cutri} R.~M.,  et~al., 2003, {2MASS All Sky Catalog of point sources.}

\bibitem[\protect\citeauthoryear{{Delrez} et~al.,}{{Delrez}
  et~al.}{2018}]{Delrez2018}
{Delrez} L.,  et~al., 2018, in {Marshall} H.~K.,  {Spyromilio} J.,  eds,
  Society of Photo-Optical Instrumentation Engineers (SPIE) Conference Series
  Vol. 10700, Ground-based and Airborne Telescopes VII. p. 107001I (\mn@eprint
  {arXiv} {1806.11205}), \mn@doi{10.1117/12.2312475}

\bibitem[\protect\citeauthoryear{{Demory} et~al.,}{{Demory}
  et~al.}{2020}]{Demory2020}
{Demory} B.~O.,  et~al., 2020, \mn@doi [\aap] {10.1051/0004-6361/202038616},
  \href {https://ui.adsabs.harvard.edu/abs/2020A&A...642A..49D} {642, A49}

\bibitem[\protect\citeauthoryear{{Donati} et~al.,}{{Donati}
  et~al.}{2020}]{Donati2020}
{Donati} J.~F.,  et~al., 2020, \mn@doi [\mnras] {10.1093/mnras/staa2569}, \href
  {https://ui.adsabs.harvard.edu/abs/2020MNRAS.498.5684D} {498, 5684}

\bibitem[\protect\citeauthoryear{{Dotter}}{{Dotter}}{2016}]{Dotter:2016}
{Dotter} A.,  2016, \mn@doi [\apjs] {10.3847/0067-0049/222/1/8}, \href
  {http://adsabs.harvard.edu/abs/2016ApJS..222....8D} {222, 8}

\bibitem[\protect\citeauthoryear{{Eastman} et~al.,}{{Eastman}
  et~al.}{2019}]{Eastman:2019}
{Eastman} J.~D.,  et~al., 2019, arXiv e-prints, \href
  {https://ui.adsabs.harvard.edu/abs/2019arXiv190709480E} {p. arXiv:1907.09480}

\bibitem[\protect\citeauthoryear{{Engle} \& {Guinan}}{{Engle} \&
  {Guinan}}{2018}]{Engle2018}
{Engle} S.~G.,  {Guinan} E.~F.,  2018, \mn@doi [Research Notes of the American
  Astronomical Society] {10.3847/2515-5172/aab1f8}, \href
  {https://ui.adsabs.harvard.edu/abs/2018RNAAS...2...34E} {2, 34}

\bibitem[\protect\citeauthoryear{{Espinoza}}{{Espinoza}}{2018}]{Espinoza2018}
{Espinoza} N.,  2018, \mn@doi [Research Notes of the American Astronomical
  Society] {10.3847/2515-5172/aaef38}, \href
  {https://ui.adsabs.harvard.edu/abs/2018RNAAS...2..209E} {2, 209}

\bibitem[\protect\citeauthoryear{{Espinoza}, {Kossakowski}  \&
  {Brahm}}{{Espinoza} et~al.}{2019}]{juliet}
{Espinoza} N.,  {Kossakowski} D.,   {Brahm} R.,  2019, \mn@doi [\mnras]
  {10.1093/mnras/stz2688}, \href
  {https://ui.adsabs.harvard.edu/abs/2019MNRAS.490.2262E} {490, 2262}

\bibitem[\protect\citeauthoryear{{Foreman-Mackey}, {Agol}, {Ambikasaran}  \&
  {Angus}}{{Foreman-Mackey} et~al.}{2017}]{Foreman2017}
{Foreman-Mackey} D.,  {Agol} E.,  {Ambikasaran} S.,   {Angus} R.,  2017,
  \mn@doi [\aj] {10.3847/1538-3881/aa9332}, \href
  {https://ui.adsabs.harvard.edu/abs/2017AJ....154..220F} {154, 220}

\bibitem[\protect\citeauthoryear{{Fressin} et~al.,}{{Fressin}
  et~al.}{2013}]{Fressin2013}
{Fressin} F.,  et~al., 2013, \mn@doi [\apj] {10.1088/0004-637X/766/2/81}, \href
  {https://ui.adsabs.harvard.edu/abs/2013ApJ...766...81F} {766, 81}

\bibitem[\protect\citeauthoryear{{Fukui} et~al.,}{{Fukui}
  et~al.}{2021}]{Fukui2021}
{Fukui} A.,  et~al., 2021, \mn@doi [\aj] {10.3847/1538-3881/ac13a5}, \href
  {https://ui.adsabs.harvard.edu/abs/2021AJ....162..167F} {162, 167}

\bibitem[\protect\citeauthoryear{{Fulton} \& {Petigura}}{{Fulton} \&
  {Petigura}}{2018}]{Fulton2018gap}
{Fulton} B.~J.,  {Petigura} E.~A.,  2018, \mn@doi [\aj]
  {10.3847/1538-3881/aae828}, \href
  {https://ui.adsabs.harvard.edu/abs/2018AJ....156..264F} {156, 264}

\bibitem[\protect\citeauthoryear{{Fulton} et~al.,}{{Fulton}
  et~al.}{2017}]{Fulton2017}
{Fulton} B.~J.,  et~al., 2017, \mn@doi [\aj] {10.3847/1538-3881/aa80eb}, \href
  {http://adsabs.harvard.edu/abs/2017AJ....154..109F} {154, 109}

\bibitem[\protect\citeauthoryear{{Fulton}, {Petigura}, {Blunt}  \&
  {Sinukoff}}{{Fulton} et~al.}{2018}]{Fulton2018}
{Fulton} B.~J.,  {Petigura} E.~A.,  {Blunt} S.,   {Sinukoff} E.,  2018, \mn@doi
  [\pasp] {10.1088/1538-3873/aaaaa8}, \href
  {https://ui.adsabs.harvard.edu/abs/2018PASP..130d4504F} {130, 044504}

\bibitem[\protect\citeauthoryear{{Furlan} \& {Howell}}{{Furlan} \&
  {Howell}}{2017}]{Furlan2017b}
{Furlan} E.,  {Howell} S.~B.,  2017, \mn@doi [\aj] {10.3847/1538-3881/aa7b70},
  \href {https://ui.adsabs.harvard.edu/abs/2017AJ....154...66F} {154, 66}

\bibitem[\protect\citeauthoryear{{Furlan} \& {Howell}}{{Furlan} \&
  {Howell}}{2020}]{Furlan2020}
{Furlan} E.,  {Howell} S.~B.,  2020, \mn@doi [\apj] {10.3847/1538-4357/ab9c9c},
  \href {https://ui.adsabs.harvard.edu/abs/2020ApJ...898...47F} {898, 47}

\bibitem[\protect\citeauthoryear{{Gaia Collab.} et~al.,}{{Gaia Collab.}
  et~al.}{2018}]{Gaia:2018}
{Gaia Collab.} et~al., 2018, \mn@doi [\aap] {10.1051/0004-6361/201833051},
  \href {http://adsabs.harvard.edu/abs/2018A%26A...616A...1G} {616, A1}

\bibitem[\protect\citeauthoryear{{Gaia Collaboration} et~al.,}{{Gaia
  Collaboration} et~al.}{2021}]{GaiaEDR3}
{Gaia Collaboration} et~al., 2021, \mn@doi [\aap]
  {10.1051/0004-6361/202039657}, \href
  {https://ui.adsabs.harvard.edu/abs/2021A&A...649A...1G} {649, A1}

\bibitem[\protect\citeauthoryear{{Gan} et~al.,}{{Gan} et~al.}{2020}]{Gan2020}
{Gan} T.,  et~al., 2020, \mn@doi [\aj] {10.3847/1538-3881/ab775a}, \href
  {https://ui.adsabs.harvard.edu/abs/2020AJ....159..160G} {159, 160}

\bibitem[\protect\citeauthoryear{{Gan} et~al.,}{{Gan} et~al.}{2021}]{Gan2021}
{Gan} T.,  et~al., 2021, \mn@doi [\mnras] {10.1093/mnras/staa3886}, \href
  {https://ui.adsabs.harvard.edu/abs/2021MNRAS.501.6042G} {501, 6042}

\bibitem[\protect\citeauthoryear{{Gan} et~al.,}{{Gan} et~al.}{2022}]{Gantoi530}
{Gan} T.,  et~al., 2022, \mn@doi [\mnras] {10.1093/mnras/stab3708}, \href
  {https://ui.adsabs.harvard.edu/abs/2022MNRAS.511...83G} {511, 83}

\bibitem[\protect\citeauthoryear{{Garcia}, {Timmermans}, {Pozuelos}, {Ducrot},
  {Gillon}, {Delrez}, {Wells}  \& {Jehin}}{{Garcia} et~al.}{2022}]{garcia2021}
{Garcia} L.~J.,  {Timmermans} M.,  {Pozuelos} F.~J.,  {Ducrot} E.,  {Gillon}
  M.,  {Delrez} L.,  {Wells} R.~D.,   {Jehin} E.,  2022, \mn@doi [\mnras]
  {10.1093/mnras/stab3113}, \href
  {https://ui.adsabs.harvard.edu/abs/2022MNRAS.509.4817G} {509, 4817}

\bibitem[\protect\citeauthoryear{{Gardner} et~al.,}{{Gardner}
  et~al.}{2006}]{Gardner2006}
{Gardner} J.~P.,  et~al., 2006, \mn@doi [\ssr] {10.1007/s11214-006-8315-7},
  \href {https://ui.adsabs.harvard.edu/abs/2006SSRv..123..485G} {123, 485}

\bibitem[\protect\citeauthoryear{{Gavel} et~al.,}{{Gavel}
  et~al.}{2014}]{2014SPIE.9148E..05G}
{Gavel} D.,  et~al., 2014, in {Marchetti} E.,  {Close} L.~M.,   {Vran} J.-P.,
  eds,  Society of Photo-Optical Instrumentation Engineers (SPIE) Conference
  Series Vol. 9148, Adaptive Optics Systems IV. p. 914805 (\mn@eprint {arXiv}
  {1407.8207}), \mn@doi{10.1117/12.2055256}

\bibitem[\protect\citeauthoryear{{Giacalone} et~al.,}{{Giacalone}
  et~al.}{2021}]{Giacalone2021}
{Giacalone} S.,  et~al., 2021, \mn@doi [\aj] {10.3847/1538-3881/abc6af}, \href
  {https://ui.adsabs.harvard.edu/abs/2021AJ....161...24G} {161, 24}

\bibitem[\protect\citeauthoryear{{Gillon}, {Jehin}, {Magain}, {Chantry},
  {Hutsem{\'e}kers}, {Manfroid}, {Queloz}  \& {Udry}}{{Gillon}
  et~al.}{2011}]{Gillon:2011}
{Gillon} M.,  {Jehin} E.,  {Magain} P.,  {Chantry} V.,  {Hutsem{\'e}kers} D.,
  {Manfroid} J.,  {Queloz} D.,   {Udry} S.,  2011, in European Physical Journal
  Web of Conferences. p. 06002 (\mn@eprint {arXiv} {1101.5807}),
  \mn@doi{10.1051/epjconf/20101106002}

\bibitem[\protect\citeauthoryear{{Gillon} et~al.,}{{Gillon}
  et~al.}{2016}]{Gillon2016}
{Gillon} M.,  et~al., 2016, \mn@doi [\nat] {10.1038/nature17448}, \href
  {https://ui.adsabs.harvard.edu/abs/2016Natur.533..221G} {533, 221}

\bibitem[\protect\citeauthoryear{{Ginzburg}, {Schlichting}  \&
  {Sari}}{{Ginzburg} et~al.}{2018}]{Ginzburg2018}
{Ginzburg} S.,  {Schlichting} H.~E.,   {Sari} R.,  2018, \mn@doi [\mnras]
  {10.1093/mnras/sty290}, \href
  {https://ui.adsabs.harvard.edu/abs/2018MNRAS.476..759G} {476, 759}

\bibitem[\protect\citeauthoryear{{Gupta} \& {Schlichting}}{{Gupta} \&
  {Schlichting}}{2019}]{Gupta2019}
{Gupta} A.,  {Schlichting} H.~E.,  2019, \mn@doi [\mnras]
  {10.1093/mnras/stz1230}, \href
  {https://ui.adsabs.harvard.edu/abs/2019MNRAS.487...24G} {487, 24}

\bibitem[\protect\citeauthoryear{{Gupta} \& {Schlichting}}{{Gupta} \&
  {Schlichting}}{2020}]{Gupta2020}
{Gupta} A.,  {Schlichting} H.~E.,  2020, \mn@doi [\mnras]
  {10.1093/mnras/staa315}, \href
  {https://ui.adsabs.harvard.edu/abs/2020MNRAS.493..792G} {493, 792}

\bibitem[\protect\citeauthoryear{{Gupta} \& {Schlichting}}{{Gupta} \&
  {Schlichting}}{2021}]{Gupta2021}
{Gupta} A.,  {Schlichting} H.~E.,  2021, \mn@doi [\mnras]
  {10.1093/mnras/stab1128}, \href
  {https://ui.adsabs.harvard.edu/abs/2021MNRAS.504.4634G} {504, 4634}

\bibitem[\protect\citeauthoryear{{Higson}, {Handley}, {Hobson}  \&
  {Lasenby}}{{Higson} et~al.}{2019}]{Higson2019}
{Higson} E.,  {Handley} W.,  {Hobson} M.,   {Lasenby} A.,  2019, \mn@doi
  [Statistics and Computing] {10.1007/s11222-018-9844-0}, \href
  {https://ui.adsabs.harvard.edu/abs/2019S&C....29..891H} {29, 891}

\bibitem[\protect\citeauthoryear{{Hippke} \& {Heller}}{{Hippke} \&
  {Heller}}{2019}]{Hippke2019}
{Hippke} M.,  {Heller} R.,  2019, \mn@doi [\aap] {10.1051/0004-6361/201834672},
  \href {https://ui.adsabs.harvard.edu/abs/2019A&A...623A..39H} {623, A39}

\bibitem[\protect\citeauthoryear{Hobson et~al.,}{Hobson
  et~al.}{2021}]{hobson_spirou_2021}
Hobson M.~J.,  et~al., 2021, \mn@doi [Astronomy \& Astrophysics]
  {10.1051/0004-6361/202038413}, 648, A48

\bibitem[\protect\citeauthoryear{{Howard} et~al.,}{{Howard}
  et~al.}{2012}]{Howard2012}
{Howard} A.~W.,  et~al., 2012, \mn@doi [\apjs] {10.1088/0067-0049/201/2/15},
  \href {https://ui.adsabs.harvard.edu/abs/2012ApJS..201...15H} {201, 15}

\bibitem[\protect\citeauthoryear{{Howard} et~al.,}{{Howard}
  et~al.}{2013}]{Howard2013}
{Howard} A.~W.,  et~al., 2013, \mn@doi [\nat] {10.1038/nature12767}, \href
  {https://ui.adsabs.harvard.edu/abs/2013Natur.503..381H} {503, 381}

\bibitem[\protect\citeauthoryear{{Howell}, {Everett}, {Sherry}, {Horch}  \&
  {Ciardi}}{{Howell} et~al.}{2011}]{Howell2011}
{Howell} S.~B.,  {Everett} M.~E.,  {Sherry} W.,  {Horch} E.,   {Ciardi} D.~R.,
  2011, \mn@doi [\aj] {10.1088/0004-6256/142/1/19}, \href
  {https://ui.adsabs.harvard.edu/abs/2011AJ....142...19H} {142, 19}

\bibitem[\protect\citeauthoryear{{Howell} et~al.,}{{Howell}
  et~al.}{2014}]{Howell2014}
{Howell} S.~B.,  et~al., 2014, \mn@doi [\pasp] {10.1086/676406}, \href
  {https://ui.adsabs.harvard.edu/abs/2014PASP..126..398H} {126, 398}

\bibitem[\protect\citeauthoryear{{Howell}, {Everett}, {Horch}, {Winters},
  {Hirsch}, {Nusdeo}  \& {Scott}}{{Howell} et~al.}{2016}]{Howell2016}
{Howell} S.~B.,  {Everett} M.~E.,  {Horch} E.~P.,  {Winters} J.~G.,  {Hirsch}
  L.,  {Nusdeo} D.,   {Scott} N.~J.,  2016, \mn@doi [\apjl]
  {10.3847/2041-8205/829/1/L2}, \href
  {https://ui.adsabs.harvard.edu/abs/2016ApJ...829L...2H} {829, L2}

\bibitem[\protect\citeauthoryear{{Howell}, {Matson}, {Ciardi}, {Everett},
  {Livingston}, {Scott}, {Horch}  \& {Winn}}{{Howell}
  et~al.}{2021}]{Howell2021}
{Howell} S.~B.,  {Matson} R.~A.,  {Ciardi} D.~R.,  {Everett} M.~E.,
  {Livingston} J.~H.,  {Scott} N.~J.,  {Horch} E.~P.,   {Winn} J.~N.,  2021,
  \mn@doi [\aj] {10.3847/1538-3881/abdec6}, \href
  {https://ui.adsabs.harvard.edu/abs/2021AJ....161..164H} {161, 164}

\bibitem[\protect\citeauthoryear{Husser, {Wende-von Berg}, Dreizler, Homeier,
  Reiners, Barman  \& Hauschildt}{Husser et~al.}{2013}]{Husser2013}
Husser T.-O.,  {Wende-von Berg} S.,  Dreizler S.,  Homeier D.,  Reiners A.,
  Barman T.,   Hauschildt P.~H.,  2013, \mn@doi [A{\&}A]
  {10.1051/0004-6361/201219058}, 553, A6

\bibitem[\protect\citeauthoryear{{Jehin} et~al.,}{{Jehin}
  et~al.}{2011}]{Jehin:2011}
{Jehin} E.,  et~al., 2011, The Messenger, \href
  {https://ui.adsabs.harvard.edu/abs/2011Msngr.145....2J} {145, 2}

\bibitem[\protect\citeauthoryear{{Jenkins}}{{Jenkins}}{2002}]{Jenkins2002}
{Jenkins} J.~M.,  2002, \mn@doi [\apj] {10.1086/341136}, \href
  {https://ui.adsabs.harvard.edu/abs/2002ApJ...575..493J} {575, 493}

\bibitem[\protect\citeauthoryear{{Jenkins} et~al.,}{{Jenkins}
  et~al.}{2016}]{Jenkins2016}
{Jenkins} J.~M.,  et~al., 2016, in Software and Cyberinfrastructure for
  Astronomy IV. p. 99133E, \mn@doi{10.1117/12.2233418}

\bibitem[\protect\citeauthoryear{{Jenkins}, {Tenenbaum}, {Seader}, {Burke},
  {McCauliff}, {Smith}, {Twicken}  \& {Chandrasekaran}}{{Jenkins}
  et~al.}{2020}]{Jenkins2020}
{Jenkins} J.~M.,  {Tenenbaum} P.,  {Seader} S.,  {Burke} C.~J.,  {McCauliff}
  S.~D.,  {Smith} J.~C.,  {Twicken} J.~D.,   {Chandrasekaran} H.,  2020,
  {Kepler Data Processing Handbook: Transiting Planet Search}, Kepler Science
  Document KSCI-19081-003

\bibitem[\protect\citeauthoryear{{Jensen}}{{Jensen}}{2013}]{Jensen2013}
{Jensen} E.,  2013, {Tapir: A web interface for transit/eclipse observability}
  (\mn@eprint {ascl} {1306.007})

\bibitem[\protect\citeauthoryear{{Jensen-Clem}, {Duev}, {Riddle}, {Salama},
  {Baranec}, {Law}, {Kulkarni}  \& {Ramprakash}}{{Jensen-Clem}
  et~al.}{2018}]{Jensen2018}
{Jensen-Clem} R.,  {Duev} D.~A.,  {Riddle} R.,  {Salama} M.,  {Baranec} C.,
  {Law} N.~M.,  {Kulkarni} S.~R.,   {Ramprakash} A.~N.,  2018, \mn@doi [\aj]
  {10.3847/1538-3881/aa9be6}, \href
  {https://ui.adsabs.harvard.edu/abs/2018AJ....155...32J} {155, 32}

\bibitem[\protect\citeauthoryear{{Jin}, {Mordasini}, {Parmentier}, {van
  Boekel}, {Henning}  \& {Ji}}{{Jin} et~al.}{2014}]{Jin2014}
{Jin} S.,  {Mordasini} C.,  {Parmentier} V.,  {van Boekel} R.,  {Henning} T.,
  {Ji} J.,  2014, \mn@doi [\apj] {10.1088/0004-637X/795/1/65}, \href
  {https://ui.adsabs.harvard.edu/abs/2014ApJ...795...65J} {795, 65}

\bibitem[\protect\citeauthoryear{{Johnson} \& {Soderblom}}{{Johnson} \&
  {Soderblom}}{1987}]{Johnson1987}
{Johnson} D.~R.~H.,  {Soderblom} D.~R.,  1987, \mn@doi [\aj] {10.1086/114370},
  \href {http://adsabs.harvard.edu/abs/1987AJ.....93..864J} {93, 864}

\bibitem[\protect\citeauthoryear{{Johnson} et~al.,}{{Johnson}
  et~al.}{2017}]{Johnson2017}
{Johnson} J.~A.,  et~al., 2017, \mn@doi [\aj] {10.3847/1538-3881/aa80e7}, \href
  {https://ui.adsabs.harvard.edu/abs/2017AJ....154..108J} {154, 108}

\bibitem[\protect\citeauthoryear{{Kempton} et~al.,}{{Kempton}
  et~al.}{2018}]{Kempton2018}
{Kempton} E. M.~R.,  et~al., 2018, \mn@doi [\pasp] {10.1088/1538-3873/aadf6f},
  \href {https://ui.adsabs.harvard.edu/abs/2018PASP..130k4401K} {130, 114401}

\bibitem[\protect\citeauthoryear{{Kipping}}{{Kipping}}{2013}]{Kipping2013}
{Kipping} D.~M.,  2013, \mn@doi [\mnras] {10.1093/mnras/stt1435}, \href
  {https://ui.adsabs.harvard.edu/abs/2013MNRAS.435.2152K} {435, 2152}

\bibitem[\protect\citeauthoryear{{Klein} et~al.,}{{Klein}
  et~al.}{2021}]{Klein2021}
{Klein} B.,  et~al., 2021, \mn@doi [\mnras] {10.1093/mnras/staa3702}, \href
  {https://ui.adsabs.harvard.edu/abs/2021MNRAS.502..188K} {502, 188}

\bibitem[\protect\citeauthoryear{{Kotani} et~al.,}{{Kotani}
  et~al.}{2018}]{Kotani2018}
{Kotani} T.,  et~al., 2018, in {Evans} C.~J.,  {Simard} L.,   {Takami} H.,
  eds,  Society of Photo-Optical Instrumentation Engineers (SPIE) Conference
  Series Vol. 10702, Ground-based and Airborne Instrumentation for Astronomy
  VII. p. 1070211, \mn@doi{10.1117/12.2311836}

\bibitem[\protect\citeauthoryear{{Kov{\'a}cs}, {Zucker}  \&
  {Mazeh}}{{Kov{\'a}cs} et~al.}{2002}]{Kovacs2002}
{Kov{\'a}cs} G.,  {Zucker} S.,   {Mazeh} T.,  2002, \mn@doi [\aap]
  {10.1051/0004-6361:20020802}, \href
  {http://adsabs.harvard.edu/abs/2002A%26A...391..369K} {391, 369}

\bibitem[\protect\citeauthoryear{{Kreidberg}}{{Kreidberg}}{2015}]{Kreidberg2015}
{Kreidberg} L.,  2015, \mn@doi [\pasp] {10.1086/683602}, \href
  {http://adsabs.harvard.edu/abs/2015PASP..127.1161K} {127, 1161}

\bibitem[\protect\citeauthoryear{{Kupke} et~al.,}{{Kupke}
  et~al.}{2012}]{2012SPIE.8447E..3GK}
{Kupke} R.,  et~al., 2012, in {Ellerbroek} B.~L.,  {Marchetti} E.,
  {V{\'e}ran} J.-P.,  eds,  Society of Photo-Optical Instrumentation Engineers
  (SPIE) Conference Series Vol. 8447, Adaptive Optics Systems III. p. 84473G,
  \mn@doi{10.1117/12.926470}

\bibitem[\protect\citeauthoryear{{Lamman} et~al.,}{{Lamman}
  et~al.}{2020}]{Lamman2020}
{Lamman} C.,  et~al., 2020, \mn@doi [\aj] {10.3847/1538-3881/ab6ef1}, \href
  {https://ui.adsabs.harvard.edu/abs/2020AJ....159..139L} {159, 139}

\bibitem[\protect\citeauthoryear{{Lavvas}, {Koskinen}, {Steinrueck},
  {Garc{\'\i}a Mu{\~n}oz}  \& {Showman}}{{Lavvas} et~al.}{2019}]{Lavvas2019}
{Lavvas} P.,  {Koskinen} T.,  {Steinrueck} M.~E.,  {Garc{\'\i}a Mu{\~n}oz} A.,
   {Showman} A.~P.,  2019, \mn@doi [\apj] {10.3847/1538-4357/ab204e}, \href
  {https://ui.adsabs.harvard.edu/abs/2019ApJ...878..118L} {878, 118}

\bibitem[\protect\citeauthoryear{{Lee} \& {Chiang}}{{Lee} \&
  {Chiang}}{2016}]{Lee2016}
{Lee} E.~J.,  {Chiang} E.,  2016, \mn@doi [\apj] {10.3847/0004-637X/817/2/90},
  \href {https://ui.adsabs.harvard.edu/abs/2016ApJ...817...90L} {817, 90}

\bibitem[\protect\citeauthoryear{{Lee}, {Chiang}  \& {Ormel}}{{Lee}
  et~al.}{2014}]{Lee2014}
{Lee} E.~J.,  {Chiang} E.,   {Ormel} C.~W.,  2014, \mn@doi [\apj]
  {10.1088/0004-637X/797/2/95}, \href
  {https://ui.adsabs.harvard.edu/abs/2014ApJ...797...95L} {797, 95}

\bibitem[\protect\citeauthoryear{{Lester} et~al.,}{{Lester}
  et~al.}{2021}]{Lester2021}
{Lester} K.~V.,  et~al., 2021, \mn@doi [\aj] {10.3847/1538-3881/ac0d06}, \href
  {https://ui.adsabs.harvard.edu/abs/2021AJ....162...75L} {162, 75}

\bibitem[\protect\citeauthoryear{{Li}, {Tenenbaum}, {Twicken}, {Burke},
  {Jenkins}, {Quintana}, {Rowe}  \& {Seader}}{{Li} et~al.}{2019}]{Li2019}
{Li} J.,  {Tenenbaum} P.,  {Twicken} J.~D.,  {Burke} C.~J.,  {Jenkins} J.~M.,
  {Quintana} E.~V.,  {Rowe} J.~F.,   {Seader} S.~E.,  2019, \mn@doi [\pasp]
  {10.1088/1538-3873/aaf44d}, \href
  {https://ui.adsabs.harvard.edu/abs/2019PASP..131b4506L} {131, 024506}

\bibitem[\protect\citeauthoryear{{Lopez} \& {Fortney}}{{Lopez} \&
  {Fortney}}{2014}]{Lopez2014}
{Lopez} E.~D.,  {Fortney} J.~J.,  2014, \mn@doi [\apj]
  {10.1088/0004-637X/792/1/1}, \href
  {http://adsabs.harvard.edu/abs/2014ApJ...792....1L} {792, 1}

\bibitem[\protect\citeauthoryear{{Lopez} \& {Rice}}{{Lopez} \&
  {Rice}}{2018}]{Lopez2018}
{Lopez} E.~D.,  {Rice} K.,  2018, \mn@doi [\mnras] {10.1093/mnras/sty1707},
  \href {https://ui.adsabs.harvard.edu/abs/2018MNRAS.479.5303L} {479, 5303}

\bibitem[\protect\citeauthoryear{{Luque} et~al.,}{{Luque}
  et~al.}{2019}]{Luque2019}
{Luque} R.,  et~al., 2019, \mn@doi [\aap] {10.1051/0004-6361/201935801}, \href
  {https://ui.adsabs.harvard.edu/abs/2019A&A...628A..39L} {628, A39}

\bibitem[\protect\citeauthoryear{{Luque} et~al.,}{{Luque}
  et~al.}{2021}]{Luque2021}
{Luque} R.,  et~al., 2021, \mn@doi [\aap] {10.1051/0004-6361/202039455}, \href
  {https://ui.adsabs.harvard.edu/abs/2021A&A...645A..41L} {645, A41}

\bibitem[\protect\citeauthoryear{{Mahadevan} et~al.,}{{Mahadevan}
  et~al.}{2014}]{Mahadevan2014}
{Mahadevan} S.,  et~al., 2014, in {Ramsay} S.~K.,  {McLean} I.~S.,   {Takami}
  H.,  eds,  Society of Photo-Optical Instrumentation Engineers (SPIE)
  Conference Series Vol. 9147, Ground-based and Airborne Instrumentation for
  Astronomy V. p. 91471G, \mn@doi{10.1117/12.2056417}

\bibitem[\protect\citeauthoryear{{Mann}, {Brewer}, {Gaidos}, {L{\'e}pine}  \&
  {Hilton}}{{Mann} et~al.}{2013}]{Mann2013}
{Mann} A.~W.,  {Brewer} J.~M.,  {Gaidos} E.,  {L{\'e}pine} S.,   {Hilton}
  E.~J.,  2013, \mn@doi [\aj] {10.1088/0004-6256/145/2/52}, \href
  {https://ui.adsabs.harvard.edu/abs/2013AJ....145...52M} {145, 52}

\bibitem[\protect\citeauthoryear{{Mann}, {Feiden}, {Gaidos}, {Boyajian}  \&
  {von Braun}}{{Mann} et~al.}{2015}]{Mann2015}
{Mann} A.~W.,  {Feiden} G.~A.,  {Gaidos} E.,  {Boyajian} T.,   {von Braun} K.,
  2015, \mn@doi [\apj] {10.1088/0004-637X/804/1/64}, \href
  {http://adsabs.harvard.edu/abs/2015ApJ...804...64M} {804, 64}

\bibitem[\protect\citeauthoryear{{Mann} et~al.,}{{Mann}
  et~al.}{2019}]{Mann2019}
{Mann} A.~W.,  et~al., 2019, \mn@doi [\apj] {10.3847/1538-4357/aaf3bc}, \href
  {https://ui.adsabs.harvard.edu/abs/2019ApJ...871...63M} {871, 63}

\bibitem[\protect\citeauthoryear{{Martinez}, {Cunha}, {Ghezzi}  \&
  {Smith}}{{Martinez} et~al.}{2019}]{Martinez2019}
{Martinez} C.~F.,  {Cunha} K.,  {Ghezzi} L.,   {Smith} V.~V.,  2019, \mn@doi
  [\apj] {10.3847/1538-4357/ab0d93}, \href
  {https://ui.adsabs.harvard.edu/abs/2019ApJ...875...29M} {875, 29}

\bibitem[\protect\citeauthoryear{{Martioli} et~al.,}{{Martioli}
  et~al.}{2022}]{Martioli2022}
{Martioli} E.,  et~al., 2022, arXiv e-prints, \href
  {https://ui.adsabs.harvard.edu/abs/2022arXiv220201259M} {p. arXiv:2202.01259}

\bibitem[\protect\citeauthoryear{{Masci} et~al.,}{{Masci}
  et~al.}{2019}]{Masci2019}
{Masci} F.~J.,  et~al., 2019, \mn@doi [\pasp] {10.1088/1538-3873/aae8ac}, \href
  {https://ui.adsabs.harvard.edu/abs/2019PASP..131a8003M} {131, 018003}

\bibitem[\protect\citeauthoryear{{McCully}, {Volgenau}, {Harbeck}, {Lister},
  {Saunders}, {Turner}, {Siiverd}  \& {Bowman}}{{McCully}
  et~al.}{2018}]{McCully2018}
{McCully} C.,  {Volgenau} N.~H.,  {Harbeck} D.-R.,  {Lister} T.~A.,  {Saunders}
  E.~S.,  {Turner} M.~L.,  {Siiverd} R.~J.,   {Bowman} M.,  2018, in Software
  and Cyberinfrastructure for Astronomy V. p. 107070K (\mn@eprint {arXiv}
  {1811.04163}), \mn@doi{10.1117/12.2314340}

\bibitem[\protect\citeauthoryear{{McGurk} et~al.,}{{McGurk}
  et~al.}{2014}]{2014SPIE.9148E..3AM}
{McGurk} R.,  et~al., 2014, in {Marchetti} E.,  {Close} L.~M.,   {Vran} J.-P.,
  eds,  Society of Photo-Optical Instrumentation Engineers (SPIE) Conference
  Series Vol. 9148, Adaptive Optics Systems IV. p. 91483A (\mn@eprint {arXiv}
  {1407.8205}), \mn@doi{10.1117/12.2057027}

\bibitem[\protect\citeauthoryear{{Ment} et~al.,}{{Ment}
  et~al.}{2019}]{Ment2019}
{Ment} K.,  et~al., 2019, \mn@doi [\aj] {10.3847/1538-3881/aaf1b1}, \href
  {https://ui.adsabs.harvard.edu/abs/2019AJ....157...32M} {157, 32}

\bibitem[\protect\citeauthoryear{{Morris}, {Twicken}, {Smith}, {Clarke},
  {Jenkins}, {Bryson}, {Girouard}  \& {Klaus}}{{Morris}
  et~al.}{2020}]{Morris2020}
{Morris} R.~L.,  {Twicken} J.~D.,  {Smith} J.~C.,  {Clarke} B.~D.,  {Jenkins}
  J.~M.,  {Bryson} S.~T.,  {Girouard} F.,   {Klaus} T.~C.,  2020, {Kepler Data
  Processing Handbook: Photometric Analysis}, Kepler Science Document
  KSCI-19081-003

\bibitem[\protect\citeauthoryear{{Moutou} et~al.,}{{Moutou}
  et~al.}{2020}]{Moutou2020}
{Moutou} C.,  et~al., 2020, \mn@doi [\aap] {10.1051/0004-6361/202038108}, \href
  {https://ui.adsabs.harvard.edu/abs/2020A&A...642A..72M} {642, A72}

\bibitem[\protect\citeauthoryear{{Muirhead} et~al.,}{{Muirhead}
  et~al.}{2015}]{Muirhead2015}
{Muirhead} P.~S.,  et~al., 2015, \mn@doi [\apj] {10.1088/0004-637X/801/1/18},
  \href {https://ui.adsabs.harvard.edu/abs/2015ApJ...801...18M} {801, 18}

\bibitem[\protect\citeauthoryear{{Newton}, {Charbonneau}, {Irwin},
  {Berta-Thompson}, {Rojas-Ayala}, {Covey}  \& {Lloyd}}{{Newton}
  et~al.}{2014}]{2014AJ....147...20N}
{Newton} E.~R.,  {Charbonneau} D.,  {Irwin} J.,  {Berta-Thompson} Z.~K.,
  {Rojas-Ayala} B.,  {Covey} K.,   {Lloyd} J.~P.,  2014, \mn@doi [\aj]
  {10.1088/0004-6256/147/1/20}, \href
  {https://ui.adsabs.harvard.edu/abs/2014AJ....147...20N} {147, 20}

\bibitem[\protect\citeauthoryear{{Newton}, {Irwin}, {Charbonneau},
  {Berta-Thompson}, {Dittmann}  \& {West}}{{Newton} et~al.}{2016}]{Newton2016}
{Newton} E.~R.,  {Irwin} J.,  {Charbonneau} D.,  {Berta-Thompson} Z.~K.,
  {Dittmann} J.~A.,   {West} A.~A.,  2016, \mn@doi [\apj]
  {10.3847/0004-637X/821/2/93}, \href
  {https://ui.adsabs.harvard.edu/abs/2016ApJ...821...93N} {821, 93}

\bibitem[\protect\citeauthoryear{{Niraula} et~al.,}{{Niraula}
  et~al.}{2020}]{Niraula2020}
{Niraula} P.,  et~al., 2020, \mn@doi [\aj] {10.3847/1538-3881/aba95f}, \href
  {https://ui.adsabs.harvard.edu/abs/2020AJ....160..172N} {160, 172}

\bibitem[\protect\citeauthoryear{{Nutzman} \& {Charbonneau}}{{Nutzman} \&
  {Charbonneau}}{2008}]{Nutzman2008}
{Nutzman} P.,  {Charbonneau} D.,  2008, \mn@doi [\pasp] {10.1086/533420}, \href
  {https://ui.adsabs.harvard.edu/abs/2008PASP..120..317N} {120, 317}

\bibitem[\protect\citeauthoryear{{Owen} \& {Wu}}{{Owen} \&
  {Wu}}{2013}]{Owen2013}
{Owen} J.~E.,  {Wu} Y.,  2013, \mn@doi [\apj] {10.1088/0004-637X/775/2/105},
  \href {http://adsabs.harvard.edu/abs/2013ApJ...775..105O} {775, 105}

\bibitem[\protect\citeauthoryear{{Owen} \& {Wu}}{{Owen} \&
  {Wu}}{2017}]{Owen2017}
{Owen} J.~E.,  {Wu} Y.,  2017, \mn@doi [\apj] {10.3847/1538-4357/aa890a}, \href
  {https://ui.adsabs.harvard.edu/abs/2017ApJ...847...29O} {847, 29}

\bibitem[\protect\citeauthoryear{Parviainen \& Aigrain}{Parviainen \&
  Aigrain}{2015}]{Parviainen2015}
Parviainen H.,  Aigrain S.,  2015, \mn@doi [MNRAS] {10.1093/mnras/stv1857},
  453, 3821

\bibitem[\protect\citeauthoryear{{Pecaut} \& {Mamajek}}{{Pecaut} \&
  {Mamajek}}{2013}]{Pecaut2013}
{Pecaut} M.~J.,  {Mamajek} E.~E.,  2013, \mn@doi [\apjs]
  {10.1088/0067-0049/208/1/9}, \href
  {http://adsabs.harvard.edu/abs/2013ApJS..208....9P} {208, 9}

\bibitem[\protect\citeauthoryear{{Pepe} et~al.,}{{Pepe}
  et~al.}{2013}]{Pepe2013}
{Pepe} F.,  et~al., 2013, \mn@doi [\nat] {10.1038/nature12768}, \href
  {https://ui.adsabs.harvard.edu/abs/2013Natur.503..377P} {503, 377}

\bibitem[\protect\citeauthoryear{{Petigura}, {Howard}  \& {Marcy}}{{Petigura}
  et~al.}{2013}]{Petigura2013}
{Petigura} E.~A.,  {Howard} A.~W.,   {Marcy} G.~W.,  2013, \mn@doi [Proceedings
  of the National Academy of Science] {10.1073/pnas.1319909110}, \href
  {https://ui.adsabs.harvard.edu/abs/2013PNAS..11019273P} {110, 19273}

\bibitem[\protect\citeauthoryear{{Petigura} et~al.,}{{Petigura}
  et~al.}{2017}]{Petigura2017}
{Petigura} E.~A.,  et~al., 2017, \mn@doi [\aj] {10.3847/1538-3881/aa80de},
  \href {https://ui.adsabs.harvard.edu/abs/2017AJ....154..107P} {154, 107}

\bibitem[\protect\citeauthoryear{{Pozuelos} et~al.,}{{Pozuelos}
  et~al.}{2020}]{Pozuelos2020}
{Pozuelos} F.~J.,  et~al., 2020, \mn@doi [\aap] {10.1051/0004-6361/202038047},
  \href {https://ui.adsabs.harvard.edu/abs/2020A&A...641A..23P} {641, A23}

\bibitem[\protect\citeauthoryear{{Queloz} et~al.,}{{Queloz}
  et~al.}{2009}]{Queloz2009}
{Queloz} D.,  et~al., 2009, \mn@doi [\aap] {10.1051/0004-6361/200913096}, \href
  {https://ui.adsabs.harvard.edu/abs/2009A&A...506..303Q} {506, 303}

\bibitem[\protect\citeauthoryear{{Rayner}, {Toomey}, {Onaka}, {Denault},
  {Stahlberger}, {Vacca}, {Cushing}  \& {Wang}}{{Rayner}
  et~al.}{2003}]{Rayner2003}
{Rayner} J.~T.,  {Toomey} D.~W.,  {Onaka} P.~M.,  {Denault} A.~J.,
  {Stahlberger} W.~E.,  {Vacca} W.~D.,  {Cushing} M.~C.,   {Wang} S.,  2003,
  \mn@doi [\pasp] {10.1086/367745}, \href
  {https://ui.adsabs.harvard.edu/abs/2003PASP..115..362R} {115, 362}

\bibitem[\protect\citeauthoryear{{Rayner}, {Cushing}  \& {Vacca}}{{Rayner}
  et~al.}{2009}]{Rayner2009}
{Rayner} J.~T.,  {Cushing} M.~C.,   {Vacca} W.~D.,  2009, \mn@doi [\apjs]
  {10.1088/0067-0049/185/2/289}, \href
  {https://ui.adsabs.harvard.edu/abs/2009ApJS..185..289R} {185, 289}

\bibitem[\protect\citeauthoryear{{Ricker} et~al.,}{{Ricker}
  et~al.}{2015}]{Ricker2015}
{Ricker} G.~R.,  et~al., 2015, \mn@doi [Journal of Astronomical Telescopes,
  Instruments, and Systems] {10.1117/1.JATIS.1.1.014003}, \href
  {https://ui.adsabs.harvard.edu/abs/2015JATIS...1a4003R} {1, 014003}

\bibitem[\protect\citeauthoryear{{Savel}, {Dressing}, {Hirsch}, {Ciardi},
  {Fleming}, {Giacalone}, {Mayo}  \& {Christiansen}}{{Savel}
  et~al.}{2020}]{2020AJ....160..287S}
{Savel} A.~B.,  {Dressing} C.~D.,  {Hirsch} L.~A.,  {Ciardi} D.~R.,  {Fleming}
  J. P.~C.,  {Giacalone} S.~A.,  {Mayo} A.~W.,   {Christiansen} J.~L.,  2020,
  \mn@doi [\aj] {10.3847/1538-3881/abc47d}, \href
  {https://ui.adsabs.harvard.edu/abs/2020AJ....160..287S} {160, 287}

\bibitem[\protect\citeauthoryear{{Schlafly} \& {Finkbeiner}}{{Schlafly} \&
  {Finkbeiner}}{2011}]{Schlafly:2011}
{Schlafly} E.~F.,  {Finkbeiner} D.~P.,  2011, \mn@doi [\apj]
  {10.1088/0004-637X/737/2/103}, \href
  {http://adsabs.harvard.edu/abs/2011ApJ...737..103S} {737, 103}

\bibitem[\protect\citeauthoryear{{Schlawin}, {Leisenring}, {Misselt}, {Greene},
  {McElwain}, {Beatty}  \& {Rieke}}{{Schlawin} et~al.}{2020}]{Schlawin2020}
{Schlawin} E.,  {Leisenring} J.,  {Misselt} K.,  {Greene} T.~P.,  {McElwain}
  M.~W.,  {Beatty} T.,   {Rieke} M.,  2020, \mn@doi [\aj]
  {10.3847/1538-3881/abb811}, \href
  {https://ui.adsabs.harvard.edu/abs/2020AJ....160..231S} {160, 231}

\bibitem[\protect\citeauthoryear{{Scott} et~al.,}{{Scott}
  et~al.}{2021}]{Scott2021}
{Scott} N.~J.,  et~al., 2021, \mn@doi [Frontiers in Astronomy and Space
  Sciences] {10.3389/fspas.2021.716560}, \href
  {https://ui.adsabs.harvard.edu/abs/2021FrASS...8..138S} {8, 138}

\bibitem[\protect\citeauthoryear{{Sebastian} et~al.,}{{Sebastian}
  et~al.}{2021}]{Sebastian_2021AA}
{Sebastian} D.,  et~al., 2021, \mn@doi [\aap] {10.1051/0004-6361/202038827},
  \href {https://ui.adsabs.harvard.edu/abs/2021A&A...645A.100S} {645, A100}

\bibitem[\protect\citeauthoryear{{Seifahrt}, {St{\"u}rmer}, {Bean}  \&
  {Schwab}}{{Seifahrt} et~al.}{2018}]{Seifahrt2018}
{Seifahrt} A.,  {St{\"u}rmer} J.,  {Bean} J.~L.,   {Schwab} C.,  2018, in
  {Evans} C.~J.,  {Simard} L.,   {Takami} H.,  eds,  Society of Photo-Optical
  Instrumentation Engineers (SPIE) Conference Series Vol. 10702, Ground-based
  and Airborne Instrumentation for Astronomy VII. p. 107026D (\mn@eprint
  {arXiv} {1805.09276}), \mn@doi{10.1117/12.2312936}

\bibitem[\protect\citeauthoryear{{Shporer} et~al.,}{{Shporer}
  et~al.}{2020}]{Shporer2020}
{Shporer} A.,  et~al., 2020, \mn@doi [\apjl] {10.3847/2041-8213/ab7020}, \href
  {https://ui.adsabs.harvard.edu/abs/2020ApJ...890L...7S} {890, L7}

\bibitem[\protect\citeauthoryear{{Smith} et~al.,}{{Smith}
  et~al.}{2012}]{Smith2012}
{Smith} J.~C.,  et~al., 2012, \mn@doi [\pasp] {10.1086/667697}, \href
  {https://ui.adsabs.harvard.edu/abs/2012PASP..124.1000S} {124, 1000}

\bibitem[\protect\citeauthoryear{{Soto} et~al.,}{{Soto}
  et~al.}{2021}]{Soto2021}
{Soto} M.~G.,  et~al., 2021, \mn@doi [\aap] {10.1051/0004-6361/202140618},
  \href {https://ui.adsabs.harvard.edu/abs/2021A&A...649A.144S} {649, A144}

\bibitem[\protect\citeauthoryear{{Southworth}}{{Southworth}}{2011}]{Southworth:2011}
{Southworth} J.,  2011, \mn@doi [\mnras] {10.1111/j.1365-2966.2011.19399.x},
  \href {https://ui.adsabs.harvard.edu/abs/2011MNRAS.417.2166S} {417, 2166}

\bibitem[\protect\citeauthoryear{{Speagle}}{{Speagle}}{2020}]{Speagle2019}
{Speagle} J.~S.,  2020, \mn@doi [\mnras] {10.1093/mnras/staa278}, \href
  {https://ui.adsabs.harvard.edu/abs/2020MNRAS.tmp..280S} {}

\bibitem[\protect\citeauthoryear{{Stassun} et~al.,}{{Stassun}
  et~al.}{2018}]{Stassun2017tic}
{Stassun} K.~G.,  et~al., 2018, \mn@doi [\aj] {10.3847/1538-3881/aad050}, \href
  {https://ui.adsabs.harvard.edu/abs/2018AJ....156..102S} {156, 102}

\bibitem[\protect\citeauthoryear{{Stassun} et~al.,}{{Stassun}
  et~al.}{2019}]{Stassun2019tic}
{Stassun} K.~G.,  et~al., 2019, \mn@doi [\aj] {10.3847/1538-3881/ab3467}, \href
  {https://ui.adsabs.harvard.edu/abs/2019AJ....158..138S} {158, 138}

\bibitem[\protect\citeauthoryear{{Stumpe} et~al.,}{{Stumpe}
  et~al.}{2012}]{Stumpe2012}
{Stumpe} M.~C.,  et~al., 2012, \mn@doi [\pasp] {10.1086/667698}, \href
  {https://ui.adsabs.harvard.edu/abs/2012PASP..124..985S} {124, 985}

\bibitem[\protect\citeauthoryear{{Stumpe}, {Smith}, {Catanzarite}, {Van Cleve},
  {Jenkins}, {Twicken}  \& {Girouard}}{{Stumpe} et~al.}{2014}]{Stumpe2014}
{Stumpe} M.~C.,  {Smith} J.~C.,  {Catanzarite} J.~H.,  {Van Cleve} J.~E.,
  {Jenkins} J.~M.,  {Twicken} J.~D.,   {Girouard} F.~R.,  2014, \mn@doi [\pasp]
  {10.1086/674989}, \href
  {https://ui.adsabs.harvard.edu/abs/2014PASP..126..100S} {126, 100}

\bibitem[\protect\citeauthoryear{{Trotta}}{{Trotta}}{2008}]{Trotta2008}
{Trotta} R.,  2008, \mn@doi [Contemporary Physics] {10.1080/00107510802066753},
  \href {https://ui.adsabs.harvard.edu/abs/2008ConPh..49...71T} {49, 71}

\bibitem[\protect\citeauthoryear{{Twicken}, {Clarke}, {Bryson}, {Tenenbaum},
  {Wu}, {Jenkins}, {Girouard}  \& {Klaus}}{{Twicken}
  et~al.}{2010}]{Twicken2010}
{Twicken} J.~D.,  {Clarke} B.~D.,  {Bryson} S.~T.,  {Tenenbaum} P.,  {Wu} H.,
  {Jenkins} J.~M.,  {Girouard} F.,   {Klaus} T.~C.,  2010, in {Radziwill}
  N.~M.,  {Bridger} A.,  eds,  Society of Photo-Optical Instrumentation
  Engineers (SPIE) Conference Series Vol. 7740, Software and
  Cyberinfrastructure for Astronomy. p. 774023, \mn@doi{10.1117/12.856790}

\bibitem[\protect\citeauthoryear{{Twicken} et~al.,}{{Twicken}
  et~al.}{2018}]{Twicken2018}
{Twicken} J.~D.,  et~al., 2018, \mn@doi [\pasp] {10.1088/1538-3873/aab694},
  \href {https://ui.adsabs.harvard.edu/abs/2018PASP..130f4502T} {130, 064502}

\bibitem[\protect\citeauthoryear{{Van Eylen}, {Agentoft}, {Lundkvist},
  {Kjeldsen}, {Owen}, {Fulton}, {Petigura}  \& {Snellen}}{{Van Eylen}
  et~al.}{2018}]{Van_Eylen2018}
{Van Eylen} V.,  {Agentoft} C.,  {Lundkvist} M.~S.,  {Kjeldsen} H.,  {Owen}
  J.~E.,  {Fulton} B.~J.,  {Petigura} E.,   {Snellen} I.,  2018, \mn@doi
  [\mnras] {10.1093/mnras/sty1783}, \href
  {https://ui.adsabs.harvard.edu/abs/2018MNRAS.479.4786V} {479, 4786}

\bibitem[\protect\citeauthoryear{{Vanderspek} et~al.,}{{Vanderspek}
  et~al.}{2019}]{Vanderspek2019}
{Vanderspek} R.,  et~al., 2019, \mn@doi [\apjl] {10.3847/2041-8213/aafb7a},
  \href {https://ui.adsabs.harvard.edu/abs/2019ApJ...871L..24V} {871, L24}

\bibitem[\protect\citeauthoryear{{Wells} et~al.,}{{Wells}
  et~al.}{2021}]{Wells2021}
{Wells} R.~D.,  et~al., 2021, \mn@doi [\aap] {10.1051/0004-6361/202141277},
  \href {https://ui.adsabs.harvard.edu/abs/2021A&A...653A..97W} {653, A97}

\bibitem[\protect\citeauthoryear{{Wright} et~al.,}{{Wright}
  et~al.}{2010}]{Wright:2010}
{Wright} E.~L.,  et~al., 2010, \mn@doi [\aj] {10.1088/0004-6256/140/6/1868},
  \href {http://adsabs.harvard.edu/abs/2010AJ....140.1868W} {140, 1868}

\bibitem[\protect\citeauthoryear{{Wu}}{{Wu}}{2019}]{Wu2019}
{Wu} Y.,  2019, \mn@doi [\apj] {10.3847/1538-4357/ab06f8}, \href
  {https://ui.adsabs.harvard.edu/abs/2019ApJ...874...91W} {874, 91}

\bibitem[\protect\citeauthoryear{{Zechmeister} \& {K{\"u}rster}}{{Zechmeister}
  \& {K{\"u}rster}}{2009}]{Zechmeister2009}
{Zechmeister} M.,  {K{\"u}rster} M.,  2009, \mn@doi [\aap]
  {10.1051/0004-6361:200811296}, \href
  {http://adsabs.harvard.edu/abs/2009A%26A...496..577Z} {496, 577}

\bibitem[\protect\citeauthoryear{{Zeng}, {Sasselov}  \& {Jacobsen}}{{Zeng}
  et~al.}{2016}]{Zeng:2016}
{Zeng} L.,  {Sasselov} D.~D.,   {Jacobsen} S.~B.,  2016, \mn@doi [\apj]
  {10.3847/0004-637X/819/2/127}, \href
  {https://ui.adsabs.harvard.edu/abs/2016ApJ...819..127Z} {819, 127}

\bibitem[\protect\citeauthoryear{{Ziegler}, {Tokovinin}, {Brice{\~n}o}, {Mang},
  {Law}  \& {Mann}}{{Ziegler} et~al.}{2020}]{Ziegler2020}
{Ziegler} C.,  {Tokovinin} A.,  {Brice{\~n}o} C.,  {Mang} J.,  {Law} N.,
  {Mann} A.~W.,  2020, \mn@doi [\aj] {10.3847/1538-3881/ab55e9}, \href
  {https://ui.adsabs.harvard.edu/abs/2020AJ....159...19Z} {159, 19}

\makeatother
\end{thebibliography}




\appendix
\section{SPIRou RVs}
\begin{table}
     \centering
     \caption{SPIRou RV measurements of \tar. Each observation took an exposure time of 900s. The data points marked with $\ast$ are outliers, which were removed during the RV analysis.}
     \begin{tabular}{ccc}
         \hline\hline
         BJD$_\mathrm{TDB}$       &RV\ (m~s$^{-1}$) &$\sigma_{\rm RV}$\ (m~s$^{-1}$) \\\hline
         2459328.981 &-29061.14 &9.74\\
2459328.992 &-29061.40 &7.65\\
2459329.065 &-29052.51 &7.08\\
2459329.076 &-29063.97 &7.55\\
2459329.987 &-29056.56 &7.60\\
2459329.997 &-29058.77 &7.42\\
2459330.065 &-29051.71 &7.27\\
2459330.075 &-29056.66 &6.92\\
2459331.060 &-29074.46 &12.06\\
2459331.995 &-29076.96 &9.11\\
2459332.006 &-29060.47 &9.12\\
2459332.057 &-29059.73 &12.31\\
2459332.068 &-29052.90 &11.65\\
2459332.997 &-29061.45 &8.02\\
2459333.008 &-29074.18 &8.81\\
2459333.059 &-29084.22 &7.70\\
2459333.070 &-29075.37 &7.33\\
2459334.975 &-29072.09 &7.69\\
2459334.986 &-29068.77 &7.58\\
2459335.051 &-29083.52 &8.29\\
2459335.062 &-29074.47 &7.25\\
2459335.985 &-29073.77 &7.59\\
2459335.995 &-29066.93 &7.38\\
2459336.058 &-29066.35 &7.32\\
2459336.069 &-29067.37 &7.19\\
2459336.983 &-29076.59 &7.03\\
2459336.993 &-29075.78 &7.01\\
2459337.077 &-29078.00 &6.97\\
2459337.088 &-29068.59 &6.97\\
2459384.944 &-29071.89 &12.14\\
2459384.955$^{\ast}$ &-29028.67 &12.07\\
2459385.032 &-29054.93 &10.89\\
2459385.042 &-29054.76 &11.79\\
2459385.956 &-29082.54 &7.20\\
2459385.967 &-29086.46 &7.04\\
2459386.035 &-29079.50 &8.03\\
2459386.046 &-29077.88 &8.04\\
2459386.947 &-29070.58 &7.05\\
2459386.958 &-29068.43 &7.02\\
2459387.028 &-29063.71 &7.36\\
2459387.039 &-29070.33 &7.40\\
2459387.961 &-29066.21 &8.62\\
2459387.971 &-29056.82 &11.73\\
2459388.026 &-29062.60 &8.94\\
2459388.037 &-29062.24 &10.48\\
2459388.935 &-29077.83 &7.29\\
2459388.946 &-29080.74 &7.34\\
2459389.029 &-29073.15 &7.51\\
2459389.040 &-29077.78 &7.51\\
2459389.933$^{\ast}$ &-29035.46 &7.34\\
2459389.944 &-29054.10 &7.19\\
2459390.026 &-29059.72 &7.61\\
2459390.037 &-29054.76 &8.11\\
2459390.949 &-29050.71 &7.46\\
2459390.959 &-29059.70 &7.41\\
2459391.028 &-29064.68 &7.78\\
2459391.039 &-29064.43 &7.71\\
2459391.944 &-29061.08 &6.85\\
2459391.955 &-29064.11 &6.92\\
2459392.037 &-29067.10 &7.07\\
          \hline
     \end{tabular}
     \label{spirourv}
\end{table}

\begin{table}
 \contcaption{SPIRou RV measurements of \tar. Each observation took an exposure time of 900s. The data points marked with $\ast$ are outliers, which were removed during the RV analysis.}
 \label{tab:continued}
 \begin{tabular}{ccc}
  \hline\hline
 BJD$_\mathrm{TDB}$       &RV\ (m~s$^{-1}$) &$\sigma_{\rm RV}$\ (m~s$^{-1}$) \\\hline
 2459392.048 &-29057.61 &7.34\\
2459392.929$^{\ast}$ &-29037.53 &7.37\\
2459392.940 &-29051.17 &7.23\\
2459393.009 &-29063.23 &7.58\\
2459393.020 &-29064.21 &7.82\\
2459393.943 &-29057.52 &7.87\\
2459393.954 &-29051.67 &7.65\\
2459394.016 &-29065.51 &10.25\\
2459394.027 &-29067.68 &10.18\\
   \hline\hline
 \end{tabular}
\end{table}

\section{Prior settings for TESS-only fit and ground photometric data detrending.}

\begin{table*}
    \centering
    {\renewcommand{\arraystretch}{1.3}
    \caption{Prior settings and posterior values for the fit to the \tess\ only data.}
    \begin{tabular}{lccr}
        \hline\hline
        Parameter       &Best-fit Value       &Prior     &Description\\\hline
        \it{Planetary parameters}\\
        $P_{b}$ (days)   &$7.85192^{+0.0005}_{-0.0005}$  
        &$\mathcal{U}$ ($7.6$\ ,\ $8.0$)
        &Orbital period of \hbox{TOI-2136\,b}.\\
        $T_{0,b}$ (BJD-2457000)    &$2017.7039^{+0.0013}_{-0.0015}$ 
        &$\mathcal{U}$ ($2014$\ ,\ $2020$) 
        &Mid-transit time of \hbox{TOI-2136\,b}.\\
        $r_{1,b}$    &$0.599^{+0.083}_{-0.067}$ 
        &$\mathcal{U}$ (0\ ,\ 1)
        &Parametrisation for {\it p} and {\it b}.\\
        $r_{2,b}$    &$0.058^{+0.002}_{-0.002}$ 
        &$\mathcal{U}$ (0\ ,\ 1)
        &Parametrisation for {\it p} and {\it b}.\\
        $e_{b}$                     &0  &Fixed  &Orbital eccentricity of \hbox{TOI-2136\,b}.\\
        $\omega_{b}$ (deg)          &90 &Fixed  &Argument of periapsis of \hbox{TOI-2136\,b}.\\
        \it{Stellar parameters}\\
        ${\rho}_{\ast}$ ($\rm kg\ m^{-3}$)   &$12721^{+1573}_{-1789}$
        &$\mathcal{J}$ ($10^{3}$\ ,\ $\rm 10^{5}$) &Stellar density.\\
        \it{\tess\ photometry parameters}\\
        $D_{\rm TESS}$     &$1$ 
        &Fixed      &\tess\ photometric dilution factor.\\
        $M_{\rm TESS}$    &$0.00002^{+0.00001}_{-0.00001}$
        &$\mathcal{N}$ (0\ ,\ $0.1^{2}$)      &Mean out-of-transit flux of \tess\ photometry.\\
        $\sigma_{\rm TESS}$ (ppm) &$0.03^{+4.62}_{-0.02}$
        &$\mathcal{J}$ ($10^{-6}$\ ,\ $10^{6}$)      &\tess\ additive photometric jitter term.\\
        $q_{1}$                &$0.37^{+0.36}_{-0.23}$       &$\mathcal{U}$ (0\ ,\ 1)  &Quadratic limb darkening coefficient.\\
        $q_{2}$                &$0.31^{+0.32}_{-0.20}$       &$\mathcal{U}$ (0\ ,\ 1)  &Quadratic limb darkening coefficient.\\
        
        \hline\hline 
    \end{tabular}\label{tess_only_fit_priors}}
\end{table*}

\begin{table*}
    \centering
    {\renewcommand{\arraystretch}{1.2}
    \caption{Prior settings for detrending the ground data.}
    \begin{tabular}{lcr}
        \hline\hline
        Parameter          &Prior     &Description\\\hline
        \it{Planetary parameters}\\
        $P_{b}$ (days)     
        &$\mathcal{U}$ (7.84\ ,\ 7.86)
        &Orbital period of \hbox{TOI-2136\,b}.\\
        $T_{0,b}$ (BJD-2457000)   
        &$\mathcal{U}$ ($2017.699$\ ,\ $2017.709$) 
        &Mid-transit time of \hbox{TOI-2136\,b}.\\
        $r_{1,b}$     
        &$\mathcal{U}$ (0.4\ ,\ 0.8)
        &Parametrisation for {\it p} and {\it b}.\\
        $r_{2,b}$    
        &$\mathcal{U}$ (0.05\ ,\ 0.07)
        &Parametrisation for {\it p} and {\it b}.\\
        $e_{b}$                       &0 (Fixed)  &Orbital eccentricity of \hbox{TOI-2136\,b}.\\
        $\omega_{b}$ (deg)            &90 (Fixed)  &Argument of periapsis of \hbox{TOI-2136\,b}.\\
        \it{Stellar parameters}\\
        ${\rho}_{\ast}$ ($\rm kg\ m^{-3}$)   
        &$\mathcal{N}$ ($12721$\ ,\ $\rm 1789^{2}$) &Stellar density.\\
        \it{Photometry parameters for each ground light curve}\\
        $D_{i}$      
        &1 (Fixed)      &Photometric dilution factor.\\
        $M_{i}$    
        &$\mathcal{N}$ (0\ ,\ $0.1^{2}$)      &Mean out-of-transit flux of ground photometry.\\
        $\sigma_{i}$ (ppm) 
        &$\mathcal{J}$ ($10^{-1}$\ ,\ $10^{5}$)      &Ground additive photometric jitter term.\\
        $q_{i}$                       &$\mathcal{U}$ (0\ ,\ 1)  &Linear limb darkening coefficient.\\
        
        \hline\hline 
    \end{tabular}\label{ground_only_fit_priors}}
\end{table*}


\bsp	
\label{lastpage}
\end{document}